\documentclass{cernyrep}
\usepackage{texnames}
\usepackage[T1]{fontenc}
\begin{document}
\title{Quantum Chromodynamics}

\author{Hsiang-nan Li}

\institute{Institute of Physics, Academia Sinica, Taipei,
Taiwan 115, Republic of China}

\maketitle 

\begin{abstract}
I review the basics of perturbative QCD, including infrared divergences
and safety, collinear and $k_T$ factorization theorems, and various
evolution equations and resummation techniques for single- and
double-logarithmic corrections. I then elaborate its
applications to studies of jet substructures and hadronic two-body
heavy-quark decays.
\end{abstract}

\section{Introduction}

One of the important missions of the Large Hadron Collider (LHC) is to
search for new physics beyond the standard model. The identification of
new physics signals usually requires precise understanding of
standard-model background, whose contributions mainly arise from quantum
chromodynamics (QCD). Many theoretical approaches have been developed
based on QCD, which are appropriate for studies of processes in different
kinematic regions and involving different hadronic systems. The theoretical
framework for high-energy hadron collisions is known as the perturbative QCD
(pQCD). I will focus on pQCD below, introducing its fundamental
ingredients and applications to LHC physics. Supplementary material can be
found in \cite{S95}.

The simple QCD Lagrangian reveals rich dynamics. It exhibits the confinement
at low energy, which accounts for the existence of various hadronic bound
states, such as pions, protons, $B$ mesons, and etc.. This nonperturbative
dynamics is manifested by infrared divergences in perturbative
calculations of bound-state properties like parton distribution functions
and fragmentation functions. On the other hand, the asymptotic freedom
at high energy leads to a small coupling constant, that allows
formulation of pQCD. Therefore, it is possible to test QCD in high-energy
scattering, which is, however, nontrivial due to bound-state properties of
involved hadrons. That is, high-energy QCD processes still involve both
perturbative and nonperturbative dynamics. A sophisticated theoretical
framework needs to be established in order to realize the goal of pQCD:
it is the factorization theorem \cite{CSS81}, in which infrared divergences are
factorized out of a process, and the remaining piece goes to a hard kernel.
The point is to prove the universality of the infrared divergences, namely,
the independence of processes the same hadron participates in. Then the
infrared divergences are absorbed into a parton distribution function (PDF)
for the hadron, which just needs to be determined once, either from
experimental data or by nonperturbative methods. The universality of a
PDF guarantees the infrared finiteness of hard kernels for all processes
involving the same hadron. Convoluting these hard kernels with the
determined PDF, one can make predictions. In other words, the universality
of a PDF warrants the predictive power of the factorization theorem.

Though infrared divergences are factorized into a PDF, the associated logarithmic
terms may appear in a process, that is not fully inclusive. To improve
perturbative expansion, these logarithmic corrections should be organized by
evolution equations or resummation techniques. For the summation of different
single logarithms, the Dokshitzer-Gribov-Lipatov-Altarelli-Parisi (DGLAP) equation
\cite{AP} and the Balitsky-Fadin-Kuraev-Lipatov (BFKL) equation \cite{BFKL}
have been proposed. For different double logarithms, the threshold
resummation \cite{S,CT,KM} and the $k_T$ resummation \cite{CS,Collins:1984kg} have been
developed. Besides, an attempt has been made to combine the DGLAP and BFKL equations,
leading to the Ciafaloni-Catani-Fiorani-Marchesini (CCFM) equation \cite{CCFM}.
Similarly, the threshold and $k_T$ resummations has been unified under the joint
resummation \cite{Li99,LSV00}, which is applicable to processes in a wider kinematic 
range. A simple framework for understanding all the above evolution equations
and resummation techniques will be provided.

After being equipped with the pQCD formalism, we are ready to learn
its applications to various processes, for which I will introduce jet
substructures and hadronic two-body heavy-quark decays. It will be demonstrated
that jet substructures, information which is crucial for particle identification
at the LHC and usually acquired from event generators \cite{S13}, are actually
calculable using the resummation technique. Among jet substructures investigated
in the literature, the distribution in jet invariant mass and the energy profile
within a jet cone will be elaborated. For the latter, it will be shown that the
factorization theorem goes beyond the conventional naive factorization assumption
\cite{BSW}, and provides valuable predictions for branching ratios and CP
asymmetries of hadronic two-body heavy-quark decays, that can be confronted by
LHCb data. Specifically, I will concentrate on three major approaches, the
QCD-improved factorization \cite{BBNS}, the perturbative QCD \cite{LY1,CL,KLS,LUY},
and the soft-collinear-effective theory \cite{bfl,bfps,cbis,bpssoft}. Some
long-standing puzzles in $B$ meson decays and their plausible resolutions are
reviewed. For more details on this subject, refer to \cite{L03}.

\section{Factorization Theorem}

The QCD lagrangian is written as
\begin{eqnarray}
{\cal L}_{QCD}=\bar\psi(i\not\! D_aT_a-m)\psi-\frac{1}{4}F^{\mu\nu}_aF_{\mu\nu a},
\label{qcdl}
\end{eqnarray}
with the quark field $\psi$, the quark mass $m$,
and the covariant derivative and the gauge field tensor
\begin{eqnarray}
D^\mu_a&=&\partial^\mu+igA^\mu_a,\nonumber\\
F^{\mu\nu}_a&=&\partial^\mu A^\nu_a-\partial^\nu A^\mu_a-gf_{abc}A^\mu_bA^\nu_c,
\end{eqnarray}
respectively. The color matrices $T_a$ and the structure constants $f_{abc}$ obey
\begin{eqnarray}
[T^{(F)}_a,T^{(F)}_b]=if_{abc}T^{(F)}_c,\;\;\;\;(T_a^{(A)})_{bc}=-if_{abc},
\end{eqnarray}
where $F$ $(A)$ denotes the fundamental (adjoint) representation.
Adding the gauge-fixing term in the path-integral quantization to remove
spurious degrees of freedom, Eq.~(\ref{qcdl}) becomes
\begin{eqnarray}
{\cal L}_{QCD}&=&\bar\psi(i\not\! D_aT_a-m)\psi-\frac{1}{4}F^{\mu\nu}_aF_{\mu\nu a}
-\frac{1}{2}\lambda(\partial_\mu A^\mu_a)^2
+\partial_\mu\eta^\dagger_a(\partial^\mu+gf_{abc}A^\mu_c)\eta_b,\label{qcd2}
\end{eqnarray}
with the gauge parameter $\lambda$, and the ghost field $\eta$.
The last term in the above expression comes from the Jacobian for
the variable change, as fixing the gauge.

The Feynman rules for QCD can be derived from Eq.~(\ref{qcd2}) following the
standard procedures \cite{Sterman}. The quark and gluon propagators with the
momentum $p$ are given by $i\not\!p/(p^2+i\epsilon)$ and $-ig^{\mu\nu}/p^2$ in
the Feynman gauge, respectively. The quark-gluon-quark vertex and the ghost-gluon-ghost
vertex are written as $-ig\gamma_\mu T_a$ and $gf_{abc}p'_\mu$, respectively, where
the subscripts $\mu$ and $a$ are associated with the gluon, $p'$ is the momentum
of the outgoing ghost, and $b$ ($c$) is associated with the outgoing (incoming) ghost.
The three-gluon vertex and the four-gluon vertex are given by
\begin{eqnarray}
\Gamma_{3g}&=&-gf_{a_1a_2a_3}[g^{\nu_1\nu_2}(p_1-p_2)^{\nu_3}+g^{\nu_2\nu_3}(p_2-p_3)^{\nu_1}
+g^{\nu_3\nu_1}(p_3-p_1)^{\nu_2},\nonumber\\
\Gamma_{4g}&=&-ig^2[f_{ea_1a_2}f_{ea_3a_4}(g^{\nu_1\nu_3}
g^{\nu_2\nu_4}-g^{\nu_1\nu_4}g^{\nu_2\nu_3})
+f_{ea_1a_3}f_{ea_4a_2}(g^{\nu_1\nu_4}g^{\nu_3\nu_2}-g^{\nu_1\nu_2}g^{\nu_3\nu_4})
\nonumber\\
& &\;\;\;\;+f_{ea_1a_4}f_{ea_2a_3}(g^{\nu_1\nu_2}g^{\nu_4\nu_3}-g^{\nu_1\nu_3}g^{\nu_4\nu_2})],
\end{eqnarray}
respectively, where the subscripts $a_1$, $a_2$, $\cdots$ and $\nu_1$, $\nu_2$, $\cdots$
are assigned to gluons counterclockwise. The particle momenta flow
into the vertices in all the above Feynman rules.

\subsection{Infrared Divergences and Safety}

The first step to establish the factorization theorem is to identify
infrared divergences in Feynman diagrams for a QCD process at
quark-gluon level. We start with the vertex correction to the amplitude
$\gamma^*(q)\to q(p_1)\bar q(p_2)$,
in which a virtual photon of momentum $q=p_1+p_1$ splits into a quark
of momentum $p_1$ and an anti-quark of momentum $p_2$. Given the Feynman
rules, one has the loop integral
\begin{eqnarray}
\int\frac{d^4l}{(2\pi)^4}(-ig\gamma^\nu T_a)
\frac{i(\not\! p_1-\not\! l)}{(p_1-l)^2+i\epsilon}(-ie\gamma_\mu)
\frac{-i(\not\! p_2-\not\! l)}{(p_2-l)^2+i\epsilon}(-ig\gamma_\nu T_a)
\frac{-i}{l^2+i\epsilon},\label{ver1}
\end{eqnarray}
where $l$ is the loop momentum carried by the gluon, and the inclusion of
the corresponding counterterm for the regularization of a ultraviolet
divergence is understood. The appearance of infrared divergences becomes
more transparent, as performing the contour integration in the light-cone
frame, in which the coordinates $l^\mu=(l^+,l^-,{\bf l}_T)$ are defined by
\begin{eqnarray}
l^{\pm}=\frac{l^0\pm l^z}{\sqrt{2}},\;\;\;\;{\bf l}_T=(l^x,l^y).
\end{eqnarray}
When an on-shell particle moves along the light cone, only one
component of its momentum is large in this frame.
For example, the above quark momenta can be chosen as
$p_1^\mu=(p_1^+,0,{\bf 0}_T)$ and $p_2^\mu=(0,p_2^-,{\bf 0}_T)$.

In terms of the light-cone coordinates, Eq.~(\ref{ver1}) is reexpressed as
\begin{eqnarray}
\int\frac{dl^+dl^-d^2l_T}{(2\pi)^4}\frac{1}{2(l^+-p_1^+)l^--l_T^2+i\epsilon}
\frac{1}{2l^+(l^--p_2^-)-l_T^2+i\epsilon}\frac{1}{2l^+l^--l_T^2+i\epsilon},\label{ver}
\end{eqnarray}
where only the denominators are shown, since infrared divergences are mainly
determined by pole structures. The poles of $l^-$ are located, for $0<l^+<p_1^+$, at
\begin{eqnarray}
& &l^-=\frac{l_T^2}{2(l^+-p_1^+)}+i\epsilon,\;\;
l^-=p_2^-+\frac{l_T^2}{2l^+}-i\epsilon,\;\;l^-=\frac{l_T^2}{2l^+}-i\epsilon.\label{pole}
\end{eqnarray}
As $l^+\sim O(p_1^+)$, the contour of $l^-$ is pinched at $l^-\sim O(l_T^2/p_1^+)$
by the first and third poles, defining the collinear region.
As $l^+\sim O(l_T)$, the contour of $l^-$ is pinched at $l^-\sim O(l_T)$,
defining the soft region. That is, the collinear (soft)
region corresponds to the configuration of
$l^\mu\sim(E, \Lambda^2/E,\Lambda)$ ($l^\mu\sim(\Lambda, \Lambda,\Lambda)$),
where $E$ and $\Lambda$ denote a large scale and a small scale, respectively.
Another leading configuration arises from the hard region characterized by
$l^\mu\sim(E, E,E)$. A simple power counting implies that all the above three
regions give logarithmic divergences.
Picking up the first pole in Eq.~(\ref{pole}), Eq.~(\ref{ver}) becomes
\begin{eqnarray}
\frac{-i}{2p_1^+}\int\frac{dl^+d^2l_T}{(2\pi)^3}\frac{p_1^+-l^+}
{2p_2^-l^+(p_1^+-l^+)+p_1^+l_T^2} \frac{1}{l_T^2}\approx
\frac{-i}{4p_1\cdot p_2}\frac{1}{(2\pi)^3}\int \frac{dl^+}
{l^+} \int\frac{d^2l_T}{l_T^2},
\end{eqnarray}
which produces the double logarithm
from the overlap of the collinear (the integration over $l^+$)
and soft (the integration over $l_T$) enhancements.

The existence of infrared divergences is a general feature of QCD corrections.
An amplitude is not a physical quantity, but a cross section is.
To examine whether the infrared divergences really call for attention,
we extend the calculation to the cross section of the process $e^-e^+\to X$,
the $e^-e^+$ annihilation into hadrons. A cross section is
computed as the square of an amplitude, whose Feynman diagrams
are composed of those for the amplitude connected by their
complex conjugate with a final-state cut between them. The cross section at
the Born level $e^-e^+\to \gamma^*\to q\bar q$ is written as
\begin{eqnarray}
\sigma^{(0)}=N_c\frac{4\pi \alpha^2}{3Q^2}\sum_f Q_f^2,\label{ee0}
\end{eqnarray}
where $N_c=3$ is the number of colors, $\alpha$ is the electromagnetic
coupling constant, $Q^2$ is the $e^-e^+$ invariant mass, and $Q_f$
is the quark charge in units of the electron charge. The virtual one-loop
corrections, including those to the gluon vertex in Eq.~(\ref{ver1}) and to
the quark self-energy, give in the dimensional regularization \cite{Sterman}
\begin{eqnarray}
\sigma^{(1)V}=-2N_c C_F\sum_fQ_f^2\frac{\alpha\alpha_s}{\pi}Q^2
\left(\frac{4\pi\mu^2}{Q^2}\right)^{2\epsilon}
\frac{1-\epsilon}{\Gamma(2-2\epsilon)}\left[\frac{1}{\epsilon^2}
+\frac{3}{2}\frac{1}{\epsilon}-\frac{\pi^2}{2}+4+O(\epsilon)\right],
\label{eev}
\end{eqnarray}
with the color factor $C_F=4/3$, the strong coupling constant $\alpha_s$,
the renormalization scale $\mu$, and the Gamma function $\Gamma$. The
double pole $1/\epsilon^2$ is a consequence of the overlap of the collinear
and soft divergences. The one-loop corrections from real gluons lead to
\cite{Sterman}
\begin{eqnarray}
\sigma^{(1)R}=2N_c C_F\sum_fQ_f^2\frac{\alpha\alpha_s}{\pi}Q^2
\left(\frac{4\pi\mu^2}{Q^2}\right)^{2\epsilon}
\frac{1-\epsilon}{\Gamma(2-2\epsilon)}\left[\frac{1}{\epsilon^2}
+\frac{3}{2}\frac{1}{\epsilon}-\frac{\pi^2}{2}+\frac{19}{4}+O(\epsilon)\right].
\label{eer}
\end{eqnarray}

\begin{figure}
\centering\includegraphics[width=.3\linewidth]{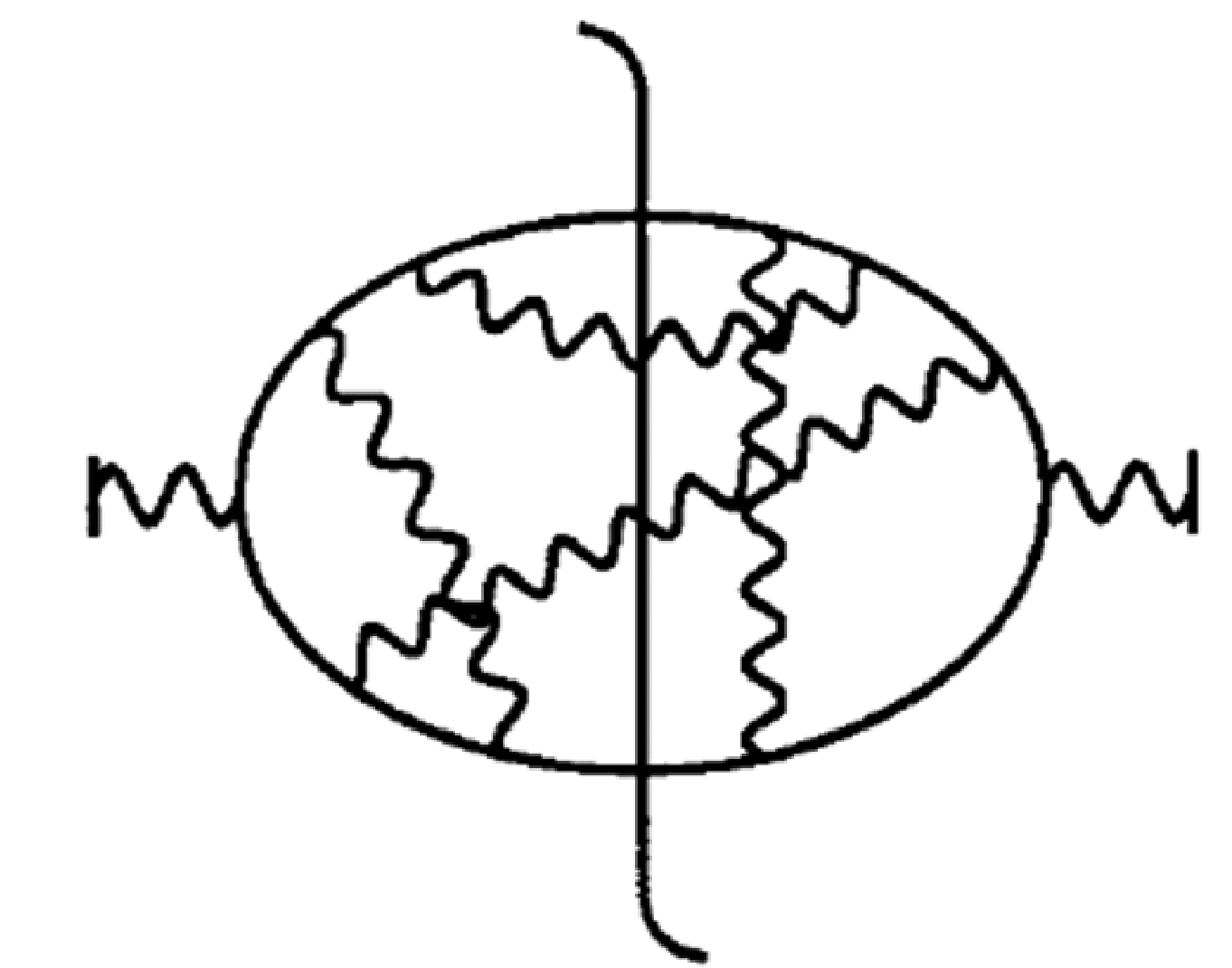}
\caption{Final-state cut on self-energy corrections to a virtual photon
propagator.}
\label{fact1}
\end{figure}

It is a crucial observation that the infrared divergences cancel in the
summation over the virtual and real corrections in Eqs.~(\ref{eev})
and (\ref{eer}), respectively: the double and single
poles have a minus sign in the former, but a plus sign in
the latter. It is easy to understand the infrared cancellation
by means of self-energy corrections to the propagator of a virtual photon.
Since a virtual photon does not involve a low characteristic scale,
the loop corrections must be infrared finite. As taking the final-state
cut shown in Fig.~\ref{fact1}, the imaginary piece of a particle
propagator is picked up, ${\rm Im}(1/(p^2+i\epsilon))\propto \delta(p^2)$,
which corresponds to the Feynman rule for an on-shell particle. Because
the self-energy corrections are infrared finite, their imaginary part,
i.e., the $e^-e^+\to X$ cross section, is certainly infrared finite.
The above observation has been formulated into the Kinoshita-Lee-Nauenberg
(KLN) theorem \cite{KLN}, which states that a cross section is infrared safe, as
integrating over all phase spaces of final states. Combining
Eqs.~(\ref{ee0}), (\ref{eev}), and (\ref{eer}), one derives the $e^-e^+\to X$ cross
section up to next-to-leading order (NLO)
\begin{eqnarray}
\sigma=N_c\frac{4\pi \alpha^2}{3Q^2}\sum_fQ_f^2
\left[1+\frac{3}{4}\frac{\alpha_s(Q)}{\pi}C_F\right],\label{ee1}
\end{eqnarray}
that has been used to determine the strong coupling constant
$\alpha_s(Q)$ at the scale $Q$.

\subsection{DIS and Collinear Factorization}

Though a naive perturbation theory applies to the $e^-e^+$ annihilation,
it fails for more complicated ones, such as the deeply inelastic scattering
(DIS) of a nucleon by a lepton, $\ell(k)N(p)\to\ell(k')+X$. Even as
the momentum transfer squared $-q^2=(k-k')^2\equiv Q^2$ is large,
the quark-level cross section for the DIS suffers infrared divergences at high orders,
which reflect the nonperturbative dynamics in the nucleon. A special treatment
of the infrared divergences is then required. It will be demonstrated that they can
be factorized out of the scattering process, and absorbed into a nucleon PDF.

Consider the two structure functions $F_{1,2}(x,Q^2)$ involved in the DIS,
where the Bjorken variable is defined as $x\equiv -q^2/(2p\cdot q)=Q^2/(2p\cdot q)$,
and take $F_2$ as an example. We shall not repeat loop integrations,
but quote the NLO corrections to the quark-level diagrams \cite{Sterman}:
\begin{eqnarray}
F_2^q(x,Q^2)&=&x\left\{\delta(1-x)+\frac{\alpha_s}{2\pi}
C_F\left[\frac{1+x^2}{1-x}\left(\ln\frac{1-x}{x}-\frac{3}{4}\right)
+\frac{1}{4}\left(9+5x\right)\right]_+\right.\nonumber\\
& &\left.+\frac{\alpha_s}{2\pi}
C_F\left(\frac{1+x^2}{1-x}\right)_+\left(4\pi\mu e^{-\gamma_E}\right)^\epsilon
\int_0^{Q^2}\frac{dk_T^2}{k_T^{2+2\epsilon}}+\cdots\right\},\label{dis1}
\end{eqnarray}
where the superscript $q$ denotes the initial-state quark, $\gamma_E$ is the
Euler constant, and the first term comes from the leading-order
(LO) contribution. The subscript $+$ represents the plus function, which
is understood as a distribution function via
\begin{eqnarray}
\int_0^1 dx\frac{f(x)}{(1-x)_+}\equiv \int_0^1 dx\frac{f(x)-f(1)}{1-x}.
\end{eqnarray}

The integration over $k_T^2$ generates an infrared divergence, that is regularized
in the dimensional regularization with $\epsilon<0$,
\begin{eqnarray}
\int_0^{Q^2}\frac{dk_T^2}{k_T^{2+2\epsilon}}=
\frac{1}{-\epsilon}\left(Q^2\right)^{-\epsilon}.
\end{eqnarray}
Hence, the infrared divergence does exist in the perturbative
evaluation of the DIS structure function, even after summing
over the virtual and real corrections. This divergence arises
from the collinear region with the loop momentum being parallel to
the nucleon momentum, since it can also be regularized by introducing
a mass to the initial-state quark. It is related to the confinement
mechanism, and corresponds to a long-distance phenomenon associated
with a group of collimated on-shell particles. The other terms in
Eq.~(\ref{dis1}) represent the hard NLO contribution to the structure
function. Comparing the results for the DIS and for the $e^-e^+$
annihilation, the former involves the integration over final-state
kinematics, but not over initial-state kinematics. This is the reason
why the KLN theorem does not apply to the infrared divergences
associated with the initial-state nucleon, and the above collinear divergence
exists. Note that the soft divergences cancel between virtual and real
diagrams due to the fact that a nucleon is color-singlet: a soft gluon
with a huge space-time distribution cannot resolve the color structure
of a nucleon, so it does not interact with it.

Besides, the collinear gluon emissions modify a quark momentum, such that
the initial-state quark can carry various momenta, as it
participates in hard scattering. It is then natural to absorb the collinear
divergences into a PDF for the nucleon, $\phi_{q/N}$, which describes the
probability for quark $q$ to carry certain amount of the nucleon momentum.
In other words, the quark-level collinear divergences are subtracted by those
in the PDF in perturbation theory, and the remaining infrared finite piece
contributes to the hard kernel $H$. We write the quark-level structure function
as the following expansion in the strong coupling constant,
\begin{eqnarray}
F_2^q(x,Q^2)=H^{(0)}\otimes\phi_{f/N}^{(0)}+\frac{\alpha_s}{2\pi}
H^{(1)}\otimes\phi_{q/N}^{(0)}+\frac{\alpha_s}{2\pi}
H^{(0)}\otimes\phi_{q/N}^{(1)}+\cdots,
\end{eqnarray}
where $H^{(i)}$ ($\phi_{q/N}^{(i)}$) is the hard kernel (PDF) of the
$i$-th order. The symbol $\otimes$
represents a convolution in the parton momentum fraction $\xi$:
\begin{eqnarray}
H\otimes\phi_{q/N}\equiv\int_x^1 \frac{d\xi}{\xi} H(x/\xi,Q,\mu)\phi_{q/N}(\xi,\mu).
\end{eqnarray}

We are ready to assign each term in Eq.~(\ref{dis1}) into either
$H^{(i)}$ or $\phi_{q/N}^{(i)}$. The first term $\delta(1-x)$ goes to
$H^{(0)}\otimes\phi_{q/N}^{(0)}$ with the definitions
\begin{eqnarray}
H^{(0)}(x/\xi,Q,\mu)=\delta(1-x/\xi),\;\;\;\;\phi_{q/N}^{(0)}(\xi,\mu)=\delta(1-\xi),
\label{h0}
\end{eqnarray}
which confirm $H^{(0)}\otimes\phi_{q/N}^{(0)}=\delta(1-x)$. The second term
in Eq.~(\ref{dis1}) is assigned to $H^{(1)}\otimes\phi_{q/N}^{(0)}$ and the third
term to $H^{(0)}\otimes\phi_{q/N}^{(1)}$ with
\begin{eqnarray}
H^{(1)}(x,Q,\mu)&=&P_{qq}^{(1)}(x)\ln\frac{Q^2}{\mu^2}+\cdots,\nonumber\\
\phi_{q/N}^{(1)}(\xi,\mu)&=&\left(4\pi\mu e^{-\gamma}\right)^\epsilon P_{qq}^{(1)}(\xi)
\int_0^{\mu^2}\frac{dk_T^2}{k_T^{2+2\epsilon}},\label{h1}
\end{eqnarray}
and the quark splitting function
\begin{eqnarray}
P_{qq}^{(1)}(x)=C_F\left(\frac{1+x^2}{1-x}\right)_+.\label{splitq}
\end{eqnarray}

The definition of the PDF in terms of a hadronic matrix element
is given by
\begin{eqnarray}
\phi_{q/N}(\xi,\mu)
&=&\int \frac{dy^-}{2\pi}\exp(-i\xi p^+y^-)\nonumber\\
& &\times\frac{1}{2}\sum_\sigma\langle N(p,\sigma)|\bar q(0,y^-,0_T)\frac{1}{2}\gamma^+
W(y^-,0)q(0,0,0_T)|N(p,\sigma)\rangle,\label{pdf1}
\end{eqnarray}
where $|N(p,\sigma)\rangle$ denotes the bound state of the nucleon
with momentum $p$ and spin $\sigma$, $y^-$ is the minus component
of the coordinate of the quark field after the final-state cut, the
first factor $1/2$ is attributed to the average over the nucleon spin,
and the matrix $\gamma^+/2$ is the spin projector for the nucleon. Here
$\mu$ is called the factorization scale, which is similar to a
renormalization scale, but introduced in perturbative computations for
an effective theory. The Wilson lines are defined by
$W(y^-,0)=W(0)W^\dagger(y^-)$ with
\begin{eqnarray}
W(y^-)= {\cal P} \exp\left[-ig \int_0^\infty dz
n_-\cdot A(y+z n_-)\right],\label{wilson}
\end{eqnarray}
where $\cal P$ represents a path-ordered exponential. The Wilson
line behaves like a scalar particle carrying a color source. The two quark
fields in Eq.~(\ref{pdf1}) are separated by a distance, so
the above Wilson links are demanded by the gauge invariance of the
nonlocal matrix element. Since Eq.~(\ref{pdf1})
depends only on the property of the nucleon, but not on the hard processes
it participates in, a PDF is universal (process-independent). This is the
most important observation, that warrants the predictive power of the
factorization theorem.

The Wilson line appears as a consequence of the eikonalization of the
final-state quark, to which the collinear gluons attach. The eikonalization
is illustrated below by considering the loop correction to the
virtual photon vertex. Assuming the initial-state quark
momentum $p=(p^+,0,\bf 0_T)$ and the final-state quark momentum
$p'=(0,p^{\prime -},\bf 0_T)$, we have the partial integrand
\begin{eqnarray}
\not\!p'\gamma^\nu\frac{\not\!p'+\not l}{(p'+l)^2}\gamma^\mu
\frac{\not\!p+\not l}{(p+l)^2}\gamma_\nu\approx
\not\!p'\gamma^-\frac{\not\!p'+\not l}{(p'+l)^2}\gamma^\mu
\frac{\not\!p+\not l}{(p+l)^2}\gamma^+\approx
\not\!p'\gamma^-\frac{\not\!p'}{2p'\cdot l}\gamma^\mu
\frac{\not\!p+\not l}{(p+l)^2}\gamma^+,
\end{eqnarray}
as the loop momentum $l$ is collinear to $p$, where $\not\!p'$ comes
from the Feynman rule for the final-state quark, $\gamma^\mu$ is the
photon vertex, and the subleading contribution from the transverse
components of $\gamma^\nu$ has been neglected. Applying the
identity $\gamma^-\not\!p'=2p^{\prime-}-\not\!p'\gamma^-$ and
$\not\!p'\not\!p'=p^{\prime 2}=0$ leads the above expression to
\begin{eqnarray}
\not\!p'\gamma^\mu
\frac{\not\!p+\not l}{(p+l)^2}\gamma^+\frac{p^{\prime-}}{p'\cdot l}
\approx\not\!p'\gamma^\mu
\frac{\not\!p+\not l}{(p+l)^2}\gamma^+\frac{n_-^{-}}{n_-\cdot l}
\approx\not\!p'\gamma^\mu
\frac{\not\!p+\not l}{(p+l)^2}\gamma_\nu\frac{n_-^\nu}{n_-\cdot l},
\label{eikonal}
\end{eqnarray}
where the dimensionless vector $n_-=(0,1,\bf 0_T)$ is parallel to
$p'$, and the subleading contribution from $\nu=T$ has been restored.
The factor $n_-^\nu$ and $1/n_-\cdot l$ are called the eikonal
vertex and the eikonal propagator, respectively.

It is then shown that the Feynman rule
$n_-^\nu/n_-\cdot l$ for the eikonalized final-state quark
is derived from the Wilson line in Eq.~(\ref{wilson}).
Consider the expansion of the path-order exponential in $W(0)$
up to order of $\alpha_s$, and Fourier transform
the gauge field into the momentum space,
\begin{eqnarray}
& &-ig \int_0^\infty dz n_-\cdot \int d^4l
\exp[iz (n_-\cdot l+i\epsilon)]\tilde A(l)\nonumber\\
&=&-ig\int d^4l\frac{\exp[iz (n_-\cdot l+i\epsilon)]}
{i(n_-\cdot l+i\epsilon)}\bigg |_{z=0}^{z=\infty}
n_-\cdot\tilde A(l)=\int d^4l\frac{gn_-^\nu}{n_-\cdot l+i\epsilon}\tilde A_\nu(l),
\end{eqnarray}
where the term $i\epsilon$ has been introduced to suppress the
contribution from $z=\infty$. The field $\tilde A(l)$ is
contracted with the gauge field from the initial-state quark
with interaction to form the gluon propagator $-i/(l^2+i\epsilon)$.
The expansion of the second piece $W(y^-)$ gives the Feynman
rules for the eikonal propagator appearing after the final-state cut.
In this case the additional exponential factor $\exp(i l\cdot y)$ is combined
with $\exp(-i\xi p^+y^-)$, implying that the valence quark
$q(0,y^-,0_T)$ after the final-state cut carries the momentum
$\xi p-l$. In summary, the first (second) piece of Wilson lines corresponds to
the configuration without (with) the loop momentum flowing through the hard
kernel. The above discussion verifies the Wilson lines in the PDF definition.

After detaching the collinear gluons from the final-state quark, the fermion flow
still connects the PDF and the hard kernel. To achieve the factorization
in the fermion flow, we insert the Fierz identity,
\begin{eqnarray}
I_{ij}I_{lk}&=& \frac{1}{4}I_{ik}I_{lj}
+ \frac{1}{4}(\gamma_\alpha)_{ik}(\gamma^\alpha)_{lj}
+ \frac{1}{4}(\gamma^5\gamma_\alpha)_{ik}(\gamma^\alpha\gamma^5)_{lj}\nonumber\\
& &+ \frac{1}{4}(\gamma^5)_{ik}(\gamma^5)_{lj}
+ \frac{1}{8}(\gamma^5\sigma_{\alpha\beta})_{ik}
(\sigma^{\alpha\beta}\gamma^5)_{lj},
\label{fierz0}
\end{eqnarray}
with $I$ being the identity matrix and
$\sigma_{\alpha\beta}\equiv i[\gamma_\alpha,\gamma_\beta]/2$.
At leading power, only the term $(\gamma_\alpha)_{ik}(\gamma^\alpha)_{lj}/4$
contributes, in which the structure $(\gamma^\alpha)_{lj}/2\approx
(\gamma^+)_{lj}/2$ goes to the definition of the PDF in Eq.~(\ref{pdf1}),
and $(\gamma_\alpha)_{ik}/2\approx(\gamma^-)_{ik}/2$ goes into the evaluation
of the hard kernel. The other terms in Eq.~(\ref{fierz0}) contribute at higher powers.
Similarly, we have to factorize the color flow between the PDF and the hard
kernel by inserting the identity
\begin{eqnarray}
I_{ij}I_{lk}=\frac{1}{N_c}I_{ik}I_{lj}+2(T^c)_{ik}(T^c)_{lj},\label{col}
\end{eqnarray}
where $I$ denotes the $3\times 3$ identity matrix, and $T^c$ is a color
matrix. The first term in the above expression contributes to the
present configuration, in which the valence quarks before and after the
final-state cut are in the color-singlet state. The structure $I_{lj}/N_c$
goes into the definition of the PDF, and $I_{ik}$ goes into the evaluation
of the hard kernel. The second term in Eq.~(\ref{col}) contributes to the
color-octet state of the valence quarks, together with which an additional
gluonic parton comes out of the nucleon and participates in the hard scattering.

The factorization formula for the nucleon DIS structure function is
written as
\begin{eqnarray}
F_2(x,Q^2)=\sum_f\int_x^1 \frac{d\xi}{\xi}H_f(x/\xi,Q,\mu)\phi_{f/N}(\xi,\mu),
\label{dis2}
\end{eqnarray}
with the subscript $f$ labeling the parton flavor, such as a valence quark,
a gluon, or a sea quark. The hard kernel $H_f$ is obtained following the
subtraction procedure for the collinear divergences, and its LO and NLO
expressions have been presented in Eqs.~(\ref{h0}) and (\ref{h1}), respectively.
The universal PDF $\phi_{f/N}$, describing the probability for
parton $f$ to carry the momentum fraction $\xi$ in the nucleon, takes a smooth
model function. It must be derived by nonperturbative methods, or extracted from data.

\subsection{Predictive Power}

The factorization theorem derived above is consistent with the well-known
parton model. The nucleon travels a long space-time, before it is hit by the
virtual photon. As $Q^2\gg 1$, the hard scattering occurs at point space-time.
Relatively speaking, the quark in the nucleon behaves like a
free particle before the hard scattering, and decouples from the
rest of the nucleon. Therefore, the cross section for the nucleon DIS reduces
to an incoherent sum over parton flavors under the collinear factorization.
That is, the approximation
\begin{eqnarray}
\left|\sum_i {\cal M}_{i/N}\right|^2\approx
\sum_i \left|{\cal M}_f\right|^2\phi_{f/N},
\end{eqnarray}
holds, where ${\cal M}_{i/N}$ represents the scattering amplitude for partonic
state $i$ of the nucleon $N$ (it could be a multi-parton state), and ${\cal M}_f$
represents the infrared finite scattering amplitude for parton $f$.

Comparing the factorization theorem with the operator product expansion (OPE),
the latter involves an expansion in short distance $y^\mu\sim 0$. A typical example
is the infrared safe $e^-e^+\to X$, whose cross section can be expressed as
a series $\sigma\approx \sum_i C_i(y)O_i(0)$. The Wilson coefficients $C_i$ and
the local effective operators $O_i$ appear in a product in the OPE.
A factorization formula involves an expansion on the light cone with small
$y^2\sim 0$, instead of $y^\mu\sim 0$. A typical example is the DIS structure
function, in which the existence of the collinear divergences implies that a
parton travels a finite longitudinal distance $y^-$. It is also the reason
why the hard kernel $H_f$ and the PDF $\phi_{f/N}$ appear in a convolution in
the momentum fraction.

\begin{figure}
\centering\includegraphics[width=0.4\linewidth]{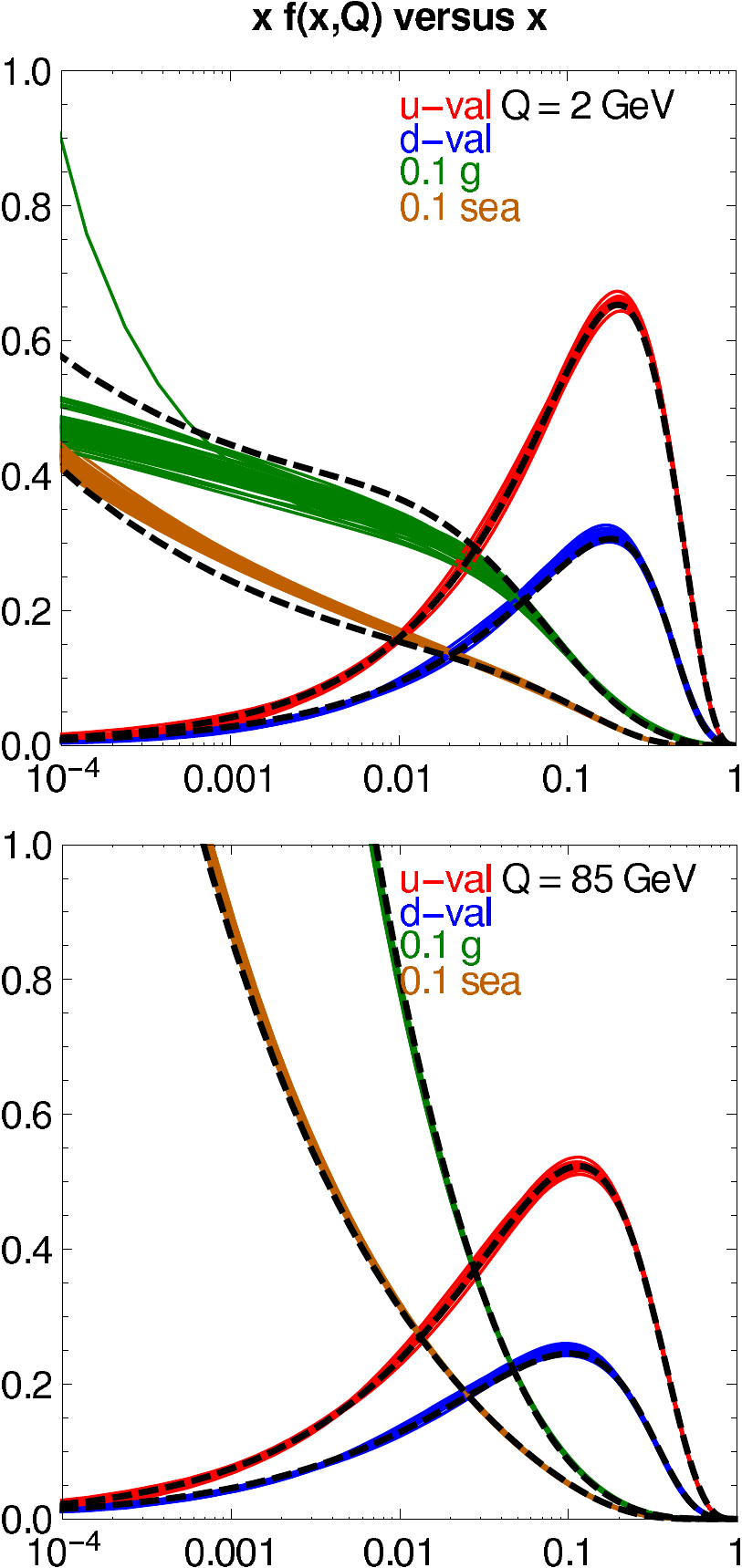}
\caption{CT10 NNLO (solid color) and NLO (dashed) parton distribution functions.
\label{CTEQ}}
\end{figure}

The factorization procedure introduces the factorization scale $\mu$
into the hard kernel $H_f$ and the PDF $\phi_{f/N}$, as indicated in
Eq.~(\ref{dis2}). Higher-order corrections produce the logarithms $\ln(Q/\mu)$
in $H_f$ and $\ln(\mu/Q_0)$ in $\phi_{f/N}$, which come from
the splitting of $\ln(Q/Q_0)$ in the structure function $F_2$,
$Q_0$ being a low scale characterizing $\phi_{f/N}$. One usually sets $\mu=Q$ to
eliminate the logarithm in $H_f$, such that the input $\phi_{f/N}(\xi,Q)$ for
arbitrary $Q$ is needed. The factorization scale does not exist in QCD diagrams,
but is introduced when a physical quantity like the structure function
is factorized. The independence of the factorization scale, $\mu dF_2/d\mu=0$,
leads to a set of renormalization-group (RG) equations
\begin{eqnarray}
& &\mu\frac{d}{d\mu}\phi_{f/N}(\xi,\mu)=\gamma_f \phi_{f/N}(\xi,\mu),\nonumber\\
& &\mu\frac{d}{d\mu}H_f(x/\xi,Q,\mu)=-\gamma_f H_f(x/\xi,Q,\mu),
\end{eqnarray}
where $\gamma_f$ denotes the anomalous dimension of the PDF.
A solution of the RG equations describes the evolution of the PDF
in $Q$
\begin{eqnarray}
\phi_{f/N}(\xi,Q)=\phi_{f/N}(\xi,Q_0)\exp\left[\int_{Q_0}^Q\frac{d\mu}{\mu}
\gamma_f(\alpha_s(\mu))\right],
\end{eqnarray}
as a result of the all-order summation of $\ln(Q/Q_0)$. Hence,
one just extracts the initial condition $\phi(\xi,Q_0)$ defined
at the initial scale $Q_0$ from data. The PDF at other higher scales
$Q$ is known through the evolution. That is, the inclusion of the RG evolution
increases the predictive power of the factorization theorem.

Fitting the factorization formulas for those processes, whose dynamics
is believed to be clear, such as Eq.~(\ref{dis2}) for DIS,
one has determined the PDFs for various partons in the proton.
The CTEQ-TEA CT10 models at the accuracy of NLO and next-to-next-to-leading order 
(NNLO) for hard kernels
are displayed in Figs.~\ref{CTEQ} \cite{CT10,Nadolsky12}. The increase of
the gluon and sea-quark PDFs with the decrease of the momentum fraction
$\xi$ is a consequence of more radiations in that region in order to
reach a lower $\xi$. The comparison of the PDFs at $Q=2$ GeV and $Q=85$
GeV indicates that the valence $u$-quark and $d$-quark PDFs become broader
with $Q$, while the gluon and sea-quark PDFs increase with $Q$.

Note that a choice of an infrared regulator is, like an ultraviolet regulator,
arbitrary; namely, we can associate an arbitrary finite piece with the infrared
pole $1/(-\epsilon)$ in $\phi^{(1)}_{f/N}$. Shifts of different finite pieces
between $\phi_{f/N}$ and $H_f$ correspond to different factorization schemes.
Hence, the extraction of a PDF depends not only on powers and orders, at which
QCD diagrams are computed, but on factorization schemes. Since perturbative
calculations are performed up to finite powers and orders, a factorization scheme
dependence is unavoidable. Nevertheless, the scheme dependence of pQCD predictions
would be minimized, if one sticks to the same factorization scheme.
Before adopting models for PDFs, it should be checked at which power and order,
at which initial scale, and in what scheme they are determined.

\begin{figure}
\centering\includegraphics[width=0.7\linewidth]{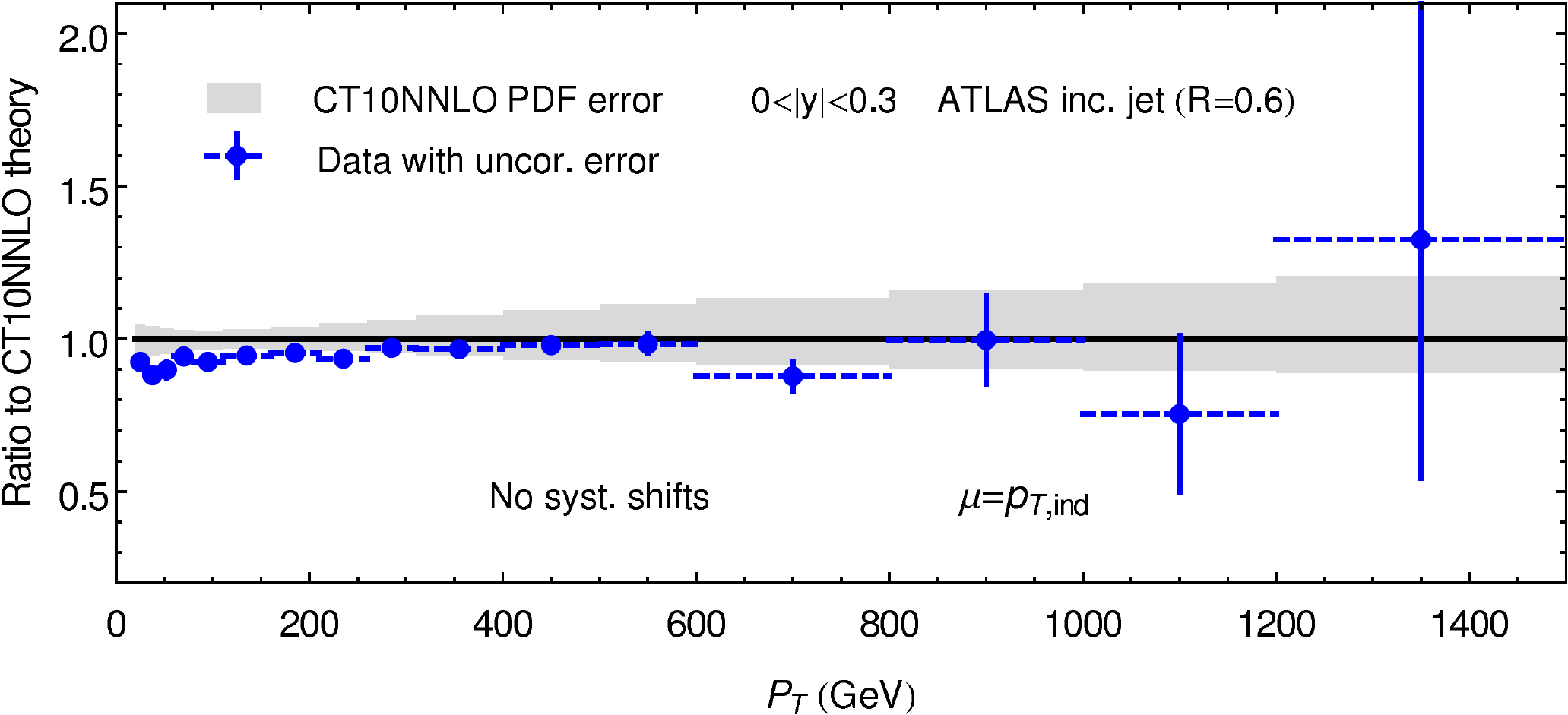}
\caption{Comparison of ATLAS data for inclusive jet $p_{T}$ distribution with
a theoretical prediction using CT10 NNLO. \label{jetpt}}
\end{figure}

At last, I explain how to apply the factorization theorem to make predictions
for QCD processes. A nucleon PDF $\phi_{f/N}$ is infrared divergent, if evaluated
in perturbation theory due to the confinement mechanism. The QCD
diagram for a DIS structure function involving quarks and gluons as the external
particles are also infrared divergent. It has been demonstrated that the infrared
divergences cancel between the QCD diagrams and the effective diagrams for $\phi_{f/N}$,
as taking their difference, which defines the hard kernel $H^{\rm DIS}$. One then
derives the factorization formula for other processes, such as
the Drell-Yan (DY) process $N(p_1)N(p_2)\to\ell^+\ell^-(q)+X$,
and computes the corresponding hard kernel $H^{\rm DY}$. The point is to verify that
the infrared divergences in the QCD diagrams for DY and in the effective diagrams
for the nucleon PDF cancel, and $H^{\rm DY}$ is infrared finite.
If it is the case, the universality of the nucleon PDF holds, and the factorization
theorem is applicable. If not, the factorization theorem fails. After verifying
the factorization theorem, one makes predictions for the DY cross section using the
formula $\sigma^{\rm DY}=\phi_{f_1/N}\otimes H^{\rm DY}\otimes \phi_{f_2/N}$.
As an example, the predictions for the inclusive jet $p_T$ distribution derived from
the factorization theorem \cite{Nadolsky12} are presented in Fig.~\ref{jetpt}.
The consistency between the predictions and the ATLAS data is obvious.

\subsection{$k_T$ Factorization}

The collinear factorization theorem introduced above has been intensively
investigated and widely applied to many QCD processes up to higher powers
and orders. The evolution of PDFs from low to high factorization scales is
governed by the DGLAP equation. The databases for PDFs have been constructed,
such as the CTEQ models. Other nonperturbative inputs like soft functions,
jet functions, and fragmentation functions have been all explored to some extent.
However, another more complicated framework, the $k_T$ factorization theorem
\cite{CCH,CE,LRS}, may be more appropriate in some kinematic regions or in
semi-inclusive processes. The collinear factorization applies, when the DIS
is measured at a finite Bjorken variable $x$. The cross section is written
as the convolution of a hard kernel with a PDF in a parton momentum fraction
$\xi$. As $x\to 0$, $\xi \ge x$ can reach a small value, at which the parton
transverse momentum $k_T$ is of the same order of magnitude as the longitudinal
momentum $\xi p$, and not negligible. Once $k_T$ is kept in a hard kernel,
a transverse-momentum-dependent (TMD) function $\Phi(\xi,k_T,\mu)$
is needed to describe the parton distribution not only in the momentum fraction
$\xi$, but also in the transverse momentum $k_T$. The DIS cross section
is then written, in the $k_T$ factorization theorem, as the convolution
\begin{eqnarray}
F_2(x,Q^2)=\sum_f\int_x^1 \frac{d\xi}{\xi}\int d^2 k_T
H_f(x/\xi,k_T, Q,\mu)\Phi_{f/N}(\xi,k_T,\mu).
\label{dis4}
\end{eqnarray}
The $k_T$ factorization theorem is also applicable to the analysis of
low $p_T$ spectra of final states, like direct photon and jet productions,
for which $k_T\sim p_T$ is not negligible.

A collinear gluon emission, modifying a parton longitudinal momentum,
generates a parton transverse momentum $k_T$ at the same time. The
factorization of a TMD from the DIS is similar to that of a PDF, which
relies on the eikonal approximation in the collinear region. This procedure
results in the eikonal propagator $n_-^\nu/n_{-}\cdot l$, represented by
the Wilson lines similar to that defined in Eq.~(\ref{wilson}). A naive
TMD definition as an extension of the PDF in Eq.~(\ref{pdf1}) is given by
\begin{eqnarray}
\Phi_{q/N}(\xi,k_T,\mu)&=&\int\frac{dy^-}{2\pi}\int\frac{d^2y_T}{(2\pi)^2}
e^{-i\xi p^+y^-+i{\bf k}_T\cdot {\bf y}_T}\nonumber\\
& &\times \frac{1}{2}\langle N(p,\sigma)|\bar q(0,y^-,y_T)\frac{1}{2}\gamma^+
W(y^-,y_T,0,0_T)q(0,0,0_T)|N(p,\sigma)\rangle,\label{deq}
\end{eqnarray}
with the Wilson links $W(y^-,y_T,0,0_T)=W(0,0_T)I_{0,y_T}W^\dagger(y^-,y_T)$.
Because the valence quark fields before and after the final-state cut
are separated by a transverse distance in this case, the vertical
links $I_{0,y_T}$ located at $y^-=\infty$ are demanded by the gauge
invariance of a TMD \cite{BJY}. More investigations on the vertical Wilson
links can be found in \cite{CS08}.

Though we do need the $k_T$ factorization theorem, many of its
aspects have not yet been completely understood. For example,
the naive definition in Eq.~(\ref{deq}) is actually ill-defined,
due to the existence of the light-cone singularity, that arises from a
loop momentum parallel to the Wilson line direction $n_-$. A plausible
modification is to rotate the Wilson line away from the light cone,
namely, to replace $n_-$ by a vector $n$ with $n^2\not=0$. This rotation
is allowed, since the collinear divergences are insensitive to the 
direction $n$ as illustrated in Eq.~(\ref{eikonal}) \cite{L5}: even 
when $n_-$ is rotated to $n$, only the minus component $n^-$ is relevant 
for the evaluation of the collinear divergences. A detailed
discussion on this subtle issue can be found in \cite{Co03}.
Besides, a parton is off-shell by $-k_T^2$, once $k_T$ is retained.
Then whether a hard kernel obtained in the $k_T$ factorization theorem
is gauge invariant becomes a concern \cite{LM09}. Dropping the $k_T$
dependence of the hard kernel in Eq.~(\ref{dis4}), the integration of
the TMD over $k_T$, $\int d^2 k_T\Phi_{f/N}(\xi,k_T)$, can be worked out.
How this integral is related to the PDF $\phi_{f/N}(\xi)$ in Eq.~(\ref{pdf1})
is worth of a thorough study.

\section{Evolution and resummation}

As stated in the previous section, radiative corrections in pQCD produce
large logarithms at each order of the coupling constant. Double logarithms
appear in processes involving two scales, such as $\ln^2(p^+b)$ with
$p^+$ being the large longitudinal momentum of a parton and $1/b$ being the 
small inverse impact parameter, where $b$ is conjugate to the parton 
transverse momentum $k_T$. In the region with large Bjorken
variable $x$, there exists $\ln^2(1/N)$ from the Mellin transformation of
$\ln(1-x)/(1-x)_+$, for which the two scales are the large $p^+$ and the
small infrared cutoff $(1-x)p^+$ for gluon emissions from a parton.
Single logarithms are generated in processes involving one scale, such
as $\ln p^+$ and $\ln(1/x)$, for which the relevant scales are the large
$p^+$ and the small $xp^+$, respectively. Various methods have been 
developed to organize these logarithmic corrections to a PDF or a TMD:
the $k_T$ resummation for $\ln^2(p^+b)$ \cite{CS,Collins:1984kg}, the
threshold resummation for $\ln^2(1/N)$ \cite{S,CT,KM}, the joint resummation
\cite{Li99,LSV00} that unifies the above two formalisms, the DGLAP equation
for $\ln p^+$ \cite{AP}, the BFKL equation for $\ln(1/x)$ \cite{BFKL}, and
the CCFM equation \cite{CCFM} that combines the above two evolution equations.
I will explain the basic ideas of all the single- and double-logarithmic
summations in the Collins-Soper-Sterman (CSS) resummation formalism
\cite{CS,Collins:1984kg}.

\subsection{Resummation Formalism}

Collinear and soft divergences may overlap to form double logarithms in extreme
kinematic regions, such as low $p_T$ and large $x$. The former includes low $p_T$
jet, photon, and $W$ boson productions, which all require real gluon emissions
with small $p_T$. The latter includes top pair production, DIS, DY production, 
and heavy meson decays $B\to X_u l\nu$ and $B\to X_s\gamma$ \cite{LY1,KS,L1} 
at the end points, for which parton momenta remain large, and radiations
are constrained in the soft region. Because of the limited phase space
for real gluon corrections, the infrared cancellation is not complete. The double
logarithms, appearing in products with the coupling constant $\alpha_s$, such as
$\alpha_s\ln^2(E/p_T)$ with the beam energy $E$ and $\alpha_s\ln(1-x)/(1-x)_+$,
deteriorate perturbative expansion. Double logarithms also occur in exclusive
processes, such as Landshoff scattering \cite{BS}, hadron form factors
\cite{LS}, Compton scattering \cite{CL93} and heavy-to-light transitions
$B\to\pi(\rho)$ \cite{LY2} and $B\to D^{(*)}$ \cite{L2} at maximal recoil.
In order to have a reliable pQCD analysis of these processes,
the important logarithms must be summed to all orders.

\begin{figure}
\begin{center}
\centering\includegraphics[width=0.6\linewidth]{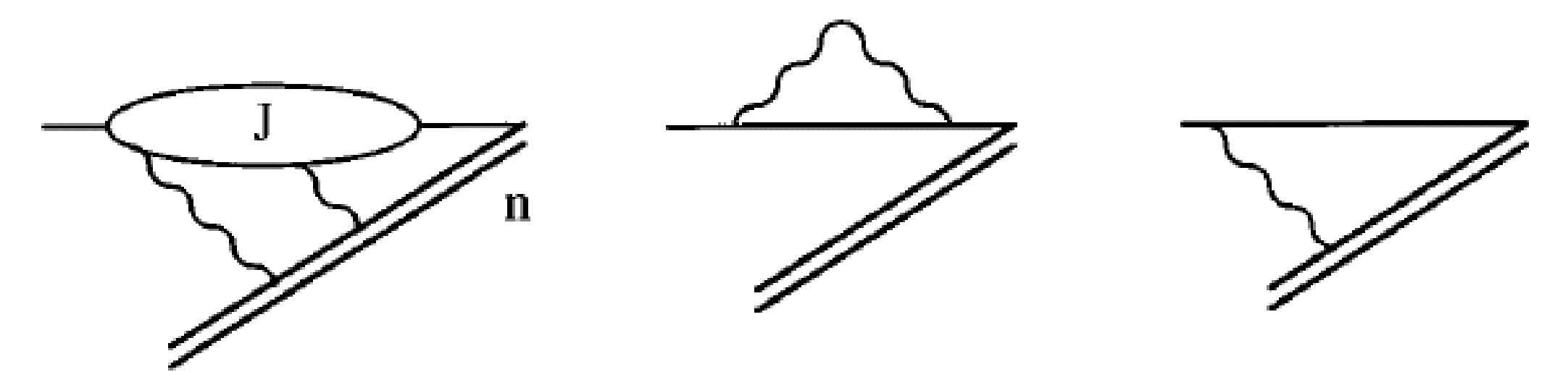}

(a)\hspace{2.5cm}(b)\hspace{2.5cm}(c)
\end{center}
\caption{(a) Jet subprocess defined in Eq.~(\ref{j}).
(b) and (c) LO diagrams of (a). \label{resum}}
\end{figure}

The resummation of large logarithms will be demonstrated in the covariant
gauge $\partial\cdot A=0$ \cite{L1}, in which the role of the Wilson line
direction $n$ and the key technique can be explained straightforwardly.
Take as an example a jet subprocess defined by the matrix element
\begin{equation}
J(p,n)u(p)=\langle 0|{\cal P}\exp\left[-ig\int_0^\infty dz n\cdot A(nz)\right]
q(0)|p\rangle\;,
\label{j}
\end{equation}
where $q$ is a light quark field with momentum $p$, and $u(p)$ is a spinor.
The abelian case of this subprocess has been discussed in \cite{C}. The
path-ordered exponential in Eq.~(\ref{j}) is the consequence of the
factorization of collinear gluons with momenta parallel to $p$
from a full process, as explained in the previous section. For convenience,
it is assumed that $p$ has a large light-cone component $p^+$, and all
its other components vanish. A general diagram of the jet function $J$ is
shown in Fig.~\ref{resum}(a), where the path-ordered exponential is
represented by a double line along the vector $n$. As explained before,
varying the direction $n$ does not change the collinear divergences 
collected by the Wilson line.

It is easy to see that $J$ contains double logarithms from the overlap of
collinear and soft divergences by calculating the LO diagrams in 
Fig.~\ref{resum}(b), the self-energy correction, and in Fig.~\ref{resum}(c), 
the vertex correction. In the covariant gauge both Figs.~\ref{resum}(b) and 
\ref{resum}(c) produce double logarithms. In the axial gauge $n\cdot A=0$ 
the path-ordered exponential reduces to an identity, and Fig.~\ref{resum}(c) 
does not exist. The essential step in the resummation technique is to derive a
differential equation $p^+dJ/dp^+=CJ$ \cite{L1,LY1,LY2}, where the
coefficient function $C$ contains only single logarithms, and can be
treated by RG methods. Since the path-ordered exponential is
scale-invariant in $n$, $J$ must depend on $p$ and $n$ through
the ratio $(p\cdot n)^2/n^2$. The differential operator $d/dp^+$
can then be replaced by $d/dn$ using a chain rule
\begin{equation}
p^+\frac{d}{dp^+}J=-\frac{n^2}{v\cdot n}v_\alpha\frac{d}{dn_\alpha}J,
\label{cr}
\end{equation}
with the vector $v=(1,0,{\bf 0}_T)$ being defined via $p=p^+v$.

\begin{figure}
\centering\includegraphics[width=0.5\linewidth]{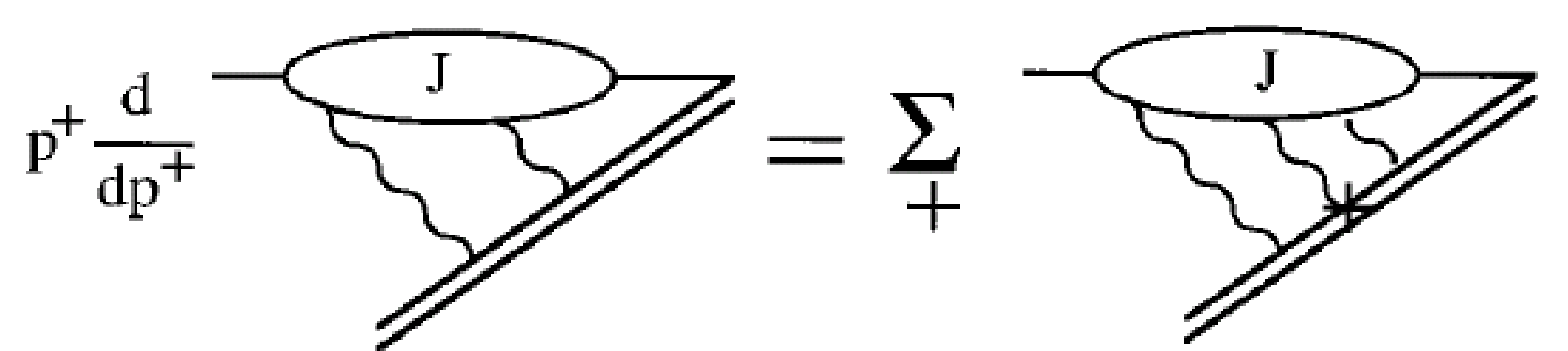}
\caption{Derivative $p^+dJ/dp^+$ in the covariant gauge. \label{resum2}}
\end{figure}

Equation (\ref{cr}) simplifies the analysis tremendously, because $n$
appears only in the Feynman rules for the Wilson line, while $p$ may flow
through the whole diagram in Fig.~\ref{resum}(a). The differentiation of each 
eikonal vertex and of the associated eikonal propagator with respect to 
$n_\alpha$,
\begin{eqnarray}
-\frac{n^2}{v\cdot n}v_\alpha\frac{d}{dn_\alpha}\frac{n_\mu}{n\cdot l}
=\frac{n^2}{v\cdot n}\left(\frac{v\cdot l}{n\cdot l}n_\mu-v_\mu\right)
\frac{1}{n\cdot l}\equiv\frac{{\hat n}_\mu}{n\cdot l},
\label{dp}
\end{eqnarray}
leads to the special vertex ${\hat n}_\mu$. The derivative $p^+dJ/dp^+$ is 
thus expressed as a summation over different attachments of ${\hat n}_\mu$, 
labeled by the symbol $+$ in Fig.~\ref{resum2}.
If the loop momentum $l$ is parallel to $p$, the factor $v\cdot l$
vanishes, and ${\hat n}_\mu$ is proportional to $v_\mu$. When this
${\hat n}_\mu$ is contracted with a vertex in $J$, in which all momenta
are mainly parallel to $p$, the contribution to
$p^+dJ/dp^+$ is suppressed. Therefore, the leading
regions of $l$ are soft and hard.

\begin{figure}
\centering\includegraphics[width=0.5\linewidth]{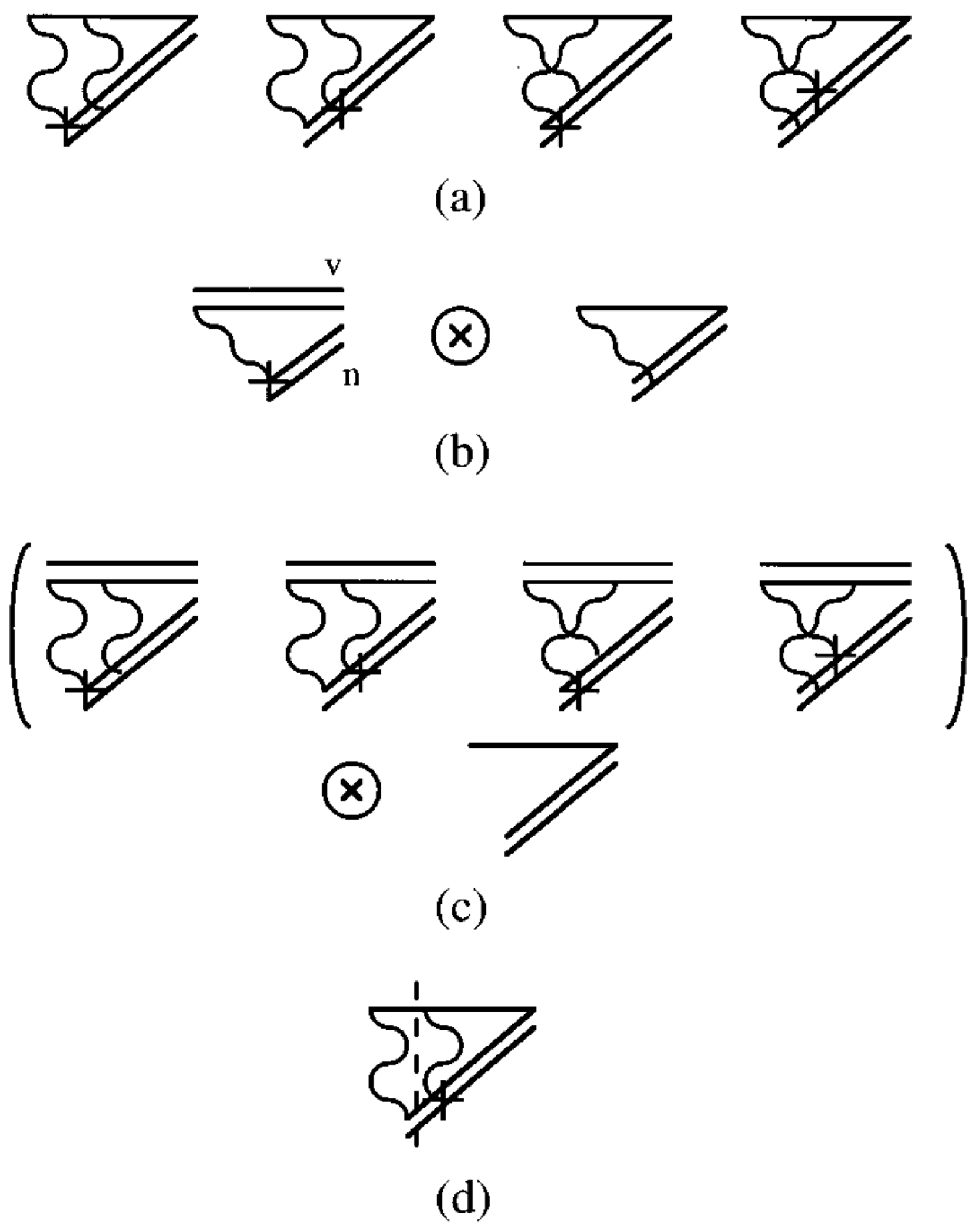}
\caption{(a) $O(\alpha_s^2)$ examples for the differentiated $J$.
(b) Factorization of $K$ at $O(\alpha_s)$.
(c) Factorization of $K$ at $O(\alpha_s^2)$.
(d) Factorization of $G$ at $O(\alpha_s)$. \label{resum3}}
\end{figure}

According to this observation, we investigate some two-loop examples
exhibited in Fig.~\ref{resum3}(a). If the loop momentum flowing through the 
special vertex is soft but another is not, only the first diagram is important,
giving a large single logarithm. In this soft region the subdiagram
containing the special vertex can be factorized using the eikonal approximation
as shown in Fig.~\ref{resum3}(b), where the symbol $\otimes$ represents a 
convoluting relation. The subdiagram is absorbed into a soft kernel $K$, and 
the remainder is identified as the original jet function $J$, both being 
$O(\alpha_s)$ contributions. If both the loop momenta are soft, the four 
diagrams in Fig.~\ref{resum3}(a) are equally important. The subdiagrams, 
factorized according to Fig.~\ref{resum3}(c), contribute to $K$ at 
$O(\alpha_s^2)$, and the remainder is the LO diagram of $J$. If the loop momentum 
flowing through the special vertex is hard and another is not, the second 
diagram in Fig.~\ref{resum3}(a) dominates. In this region the subdiagram 
containing the special vertex is factorized as shown in Fig.~\ref{resum3}(d). 
The right-hand side of the dashed line is absorbed into a hard kernel $G$ 
as an $O(\alpha_s)$ contribution, and the left-hand side is identified as the
$O(\alpha_s)$ diagram of $J$. If both the loop momenta are hard, all the 
diagrams in Fig.~\ref{resum3}(a) are absorbed into $G$, giving the 
$O(\alpha_s^2)$ contributions.

Extending the above reasoning to all orders, one derives the differential
equation
\begin{equation}
p^+\frac{d}{dp^+}J=\left[K(m/\mu,\alpha_s(\mu))+G(p^+\nu/\mu,
\alpha_s(\mu))\right]J,
\label{de1}
\end{equation}
where the coefficient function $C$ has been written as the sum of the
soft kernel $K$ and the hard kernel $G$. In the above expression $\mu$ is a
factorization scale, the gauge factor in $G$ is defined as
$\nu=\sqrt{(v\cdot n)^2/|n^2|}$, and a gluon mass $m$ has been
introduced to regularize the infrared divergence in $K$. It has been made
explicit that $K$ and $G$ depend on a single infrared scale $m$
and a single ultraviolet scale $p^+$, respectively.

The $O(\alpha_s)$ contribution to $K$ from Fig.~\ref{resum3}(b) is written as
\begin{eqnarray}
K&=&-ig^2 C_F\mu^\epsilon\int\frac{d^{4-\epsilon} l}
{(2\pi)^{4-\epsilon}}\frac{{\hat n}_\mu}{n\cdot l}
\frac{g^{\mu\nu}}{l^2-m^2}\frac{v_\nu}{v\cdot l}-\delta K,
\label{k1}
\end{eqnarray}
$\delta K$ being an additive counterterm. The $O(\alpha_s)$ contribution to
$G$ from Fig.~\ref{resum3}(d) is given by
\begin{eqnarray}
G=-ig^2 C_F\mu^\epsilon\int\frac{d^{4-\epsilon} l}
{(2\pi)^{4-\epsilon}}\frac{{\hat n}_\mu}{n\cdot l}\frac{g^{\mu\nu}}{l^2}
\left(\frac{\not p+\not l}{(p+l)^2}\gamma_\nu-\frac{v_\nu}{v\cdot l}\right)
-\delta G,
\label{g1}
\end{eqnarray}
where the second term in the parentheses acts as a soft subtraction to avoid 
double counting, and $\delta G$ is an additive counterterm. A straightforward 
evaluation shows that Eqs.~(\ref{k1}) and (\ref{g1}) contain only the single
logarithms $\ln(m/\mu)$ and $\ln(p^+\nu/\mu)$, respectively, as claimed
before. Organizing these single logarithms using RG methods, and then
solving Eq.~(\ref{de1}), one resums the double logarithms $\ln^2(p^+/m)$ in $J$.

To explain all the known resummations and evolution equations, we first 
construct a master equation for the TMD $\Phi(x,k_T)$, which is a differential
equation in the hadron momentum $p^+$. The dependence on the factorization
scale $\mu$ is implicit. If the parton is a quark, $\Phi$ is defined by 
Eq.~(\ref{deq}). If the parton is a gluon, the nonlocal operator in the 
hadronic matrix element of Eq.~(\ref{deq}) is replaced by $F^+_\mu(y^-,y_T)F^{\mu+}(0)$.
Similarly, $n$ is varied arbitrarily away from the light cone with $n^2\not= 0$. 
Then $\Phi$ depends on $p^+$ via the ratio $(p\cdot n)^2/n^2$, so
the chain rule in Eq.~(\ref{cr}) relating the derivative $d\Phi/dp^+$ to 
$d\Phi/dn_\alpha$ applies. Following the derivation in the previous subsection,
one obtains the master equation
\begin{eqnarray}
p^+\frac{d}{dp^+}\Phi(x,k_T)=2{\bar \Phi}(x,k_T),
\label{meq}
\end{eqnarray}
where $\bar \Phi$ contains the special vertex, and the coefficient 2 is 
attributed to the equality of $\bar\Phi$ with the special vertex on
either side of the final-state cut.

The function $\bar \Phi$ is factorized into the convolution of the
soft and hard kernels with $\Phi$:
\begin{eqnarray}
{\bar \Phi}(x,k_T)={\bar \Phi}_{s}(x,k_T)+
{\bar \Phi}_{h}(x,k_T),
\label{ssf}
\end{eqnarray}
with the soft contribution
\begin{eqnarray}
{\bar \Phi}_{s}&=&\left[-ig^2 C_F\mu^\epsilon\int\frac{d^{4-\epsilon} l}
{(2\pi)^{4-\epsilon}}\frac{{\hat n}\cdot v}{n\cdot ll^2 v\cdot l}
-\delta K\right]\Phi(x,k_T)\nonumber\\
& &-ig^2 C_F\mu^\epsilon\int\frac{d^{4-\epsilon} l}
{(2\pi)^{4-\epsilon}}\frac{{\hat n}\cdot v}{n\cdot l v\cdot l}2\pi i\delta(l^2)
\Phi(x+l^+/p^+,|{\bf k}_T+{\bf l}_T|),
\label{fsr}
\end{eqnarray}
where the first term is the same as in Eq.~(\ref{k1}), and
the second term proportional to $\delta(l^2)$ arises from the real soft
gluon emission. The hard contribution is given by 
${\bar\Phi}_h(x,k_T)=G(xp^+\nu/\mu,\alpha_s(\mu))\Phi(x,k_T)$, in which
the hard kernel $G$ is the same as in Eq.~(\ref{g1}).

\subsection{$k_T$ Resummation and BFKL Equation}

The TMD definition in Eq.~(\ref{deq}) contains three scales: 
$(1-x)p^+$, $xp^+$, and $k_T$. We first consider the soft approximation 
corresponding to the rapidity ordering of real gluon emissions in a 
ladder diagram. Assume that a parton carries the longitudinal momentum 
$xp^++l_2^++l_1^+$, which becomes $xp^++l_1^+$ after emitting a gluon 
of longitudinal momentum $l_2^+$ and transverse momentum $l_{2T}$,
and then becomes $xp^+$ after emitting a gluon of longitudinal
momentum $l_1^+$ and transverse momentum $l_{1T}$. In the kinematic
configuration with $l_2^+\gg l_1^+$ and $l_{2T}\sim l_{1T}$,
the original parton momentum is approximated by
$xp^++l_2^++l_1^+\approx xp^++l_2^+$. The loop integral
associated with the first gluon emission is then independent of $l_1^+$,
and can be worked out straightforwardly, giving a logarithm.
The loop integral associated with the second gluon emission, involving
only $l_1^+$, also gives a logarithm. Therefore, a ladder diagram with 
$N$ rung gluons generates the logarithmic correction $(\alpha_s L)^N$ 
under the above rapidity ordering, where $L$ denotes the large 
logarithm. Following the rapidity ordering, we adopt
the approximation for the real gluon emission in Eq.~(\ref{fsr})
\begin{equation}
\Phi(x+l^+/p^+,|{\bf k_T+\bf l_T}|)\approx
\Phi(x,|{\bf k_T+\bf l_T}|),
\label{nl}
\end{equation}
where the $l^+$ dependence has been neglected. The transverse momenta
$l_T$, being of the same order as $k_T$ in this kinematic configuration,
is kept. The variable $l^+$ in $K$ is then integrated up to infinity,  
such that the scale $(1-x)p^+$ disappears.

Equation~(\ref{fsr}) is Fourier transformed into the impact parameter $b$
space to decouple the $l_T$ integration. Hence, in the intermediate $x$ region
$\Phi$ involves two scales, the large $xp^+$ that characterizes the hard
kernel $G$ and the small $1/b$ that characterizes the
soft kernel $K$. The master equation (\ref{meq}) becomes
\begin{eqnarray}
p^+\frac{d}{dp^+}\Phi(x,b)=2\left[K(1/(b\mu),\alpha_s(\mu))+
G(xp^+\nu/\mu,\alpha_s(\mu))\right]\Phi(x,b),
\label{dph}
\end{eqnarray}
whose solution with $\nu=1$ leads to the $k_T$ resummation
\begin{eqnarray}
\Phi(x,b)=\Delta_k(x,b)\Phi_i(x),
\label{sph}
\end{eqnarray}
with the Sudakov exponential
\begin{eqnarray}
\Delta_k(x,b)=\exp\left[-2\int_{1/b}^{xp^+}\frac{d p}{p}
\int_{1/b}^{p}\frac{d\mu}{\mu}\gamma_{K}(\alpha_s(\mu))\right],
\label{fb}
\end{eqnarray}
and the initial condition $\Phi_i$ of the Sudakov evolution.
The anomalous dimension of $K$, $\lambda_K=\mu d\delta K/d\mu$,
is given, up to two loops, by \cite{KT82}
\begin{eqnarray}
\gamma_K=\frac{\alpha_s}{\pi}C_F+\left(\frac{\alpha_s}{\pi}
\right)^2C_F\left[{C}_A\left(\frac{67}{36}
-\frac{\pi^{2}}{12}\right)-\frac{5}{18}n_{f}\right]\;,
\label{lk}
\end{eqnarray}
with $n_{f}$ being the number of quark flavors and $C_A=3$ being a color factor.

\begin{figure}
\centering\includegraphics[width=.4\linewidth]{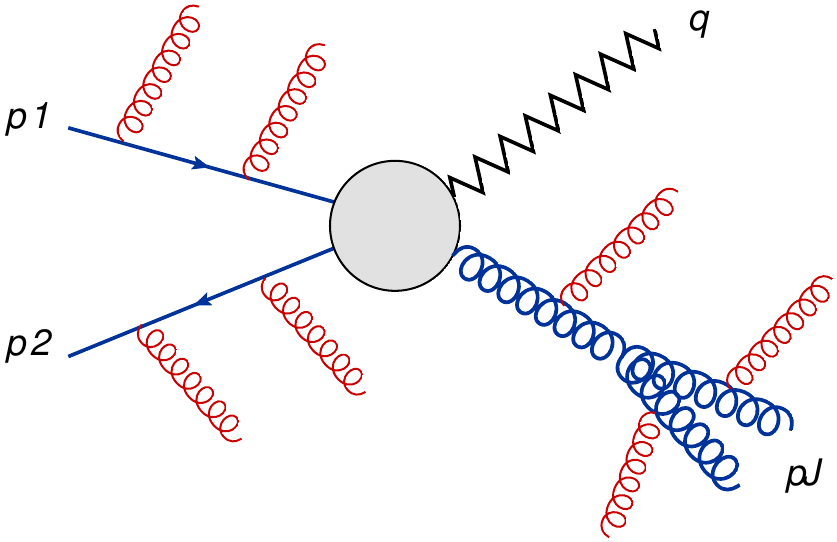}
\caption{Scattering amplitude for direct photon production.}
\label{factorization}
\end{figure}

\begin{figure}
\centering\includegraphics[angle=-90,width=.7\linewidth]{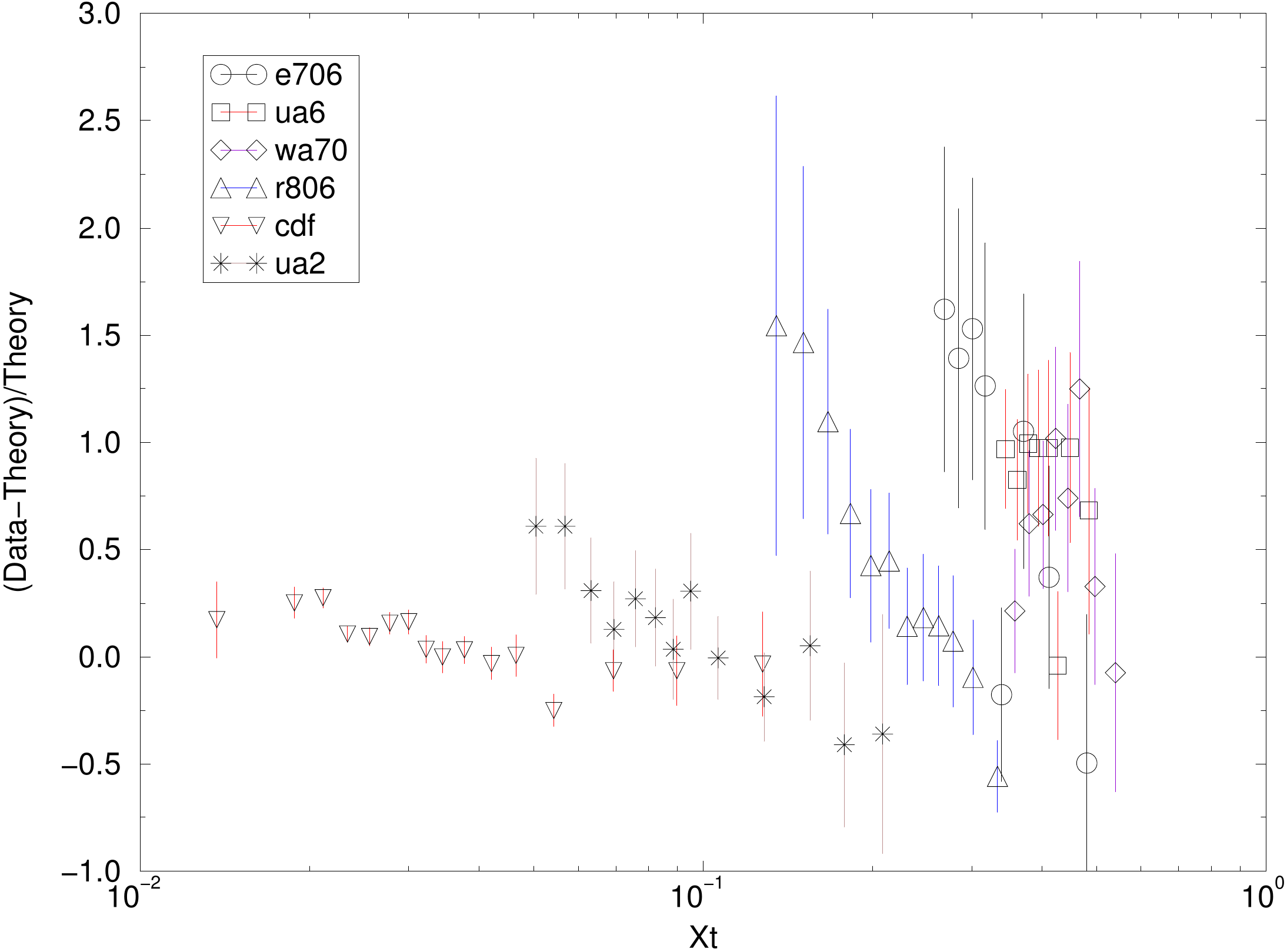}
\centering\includegraphics[angle=-90,width=.7\linewidth]{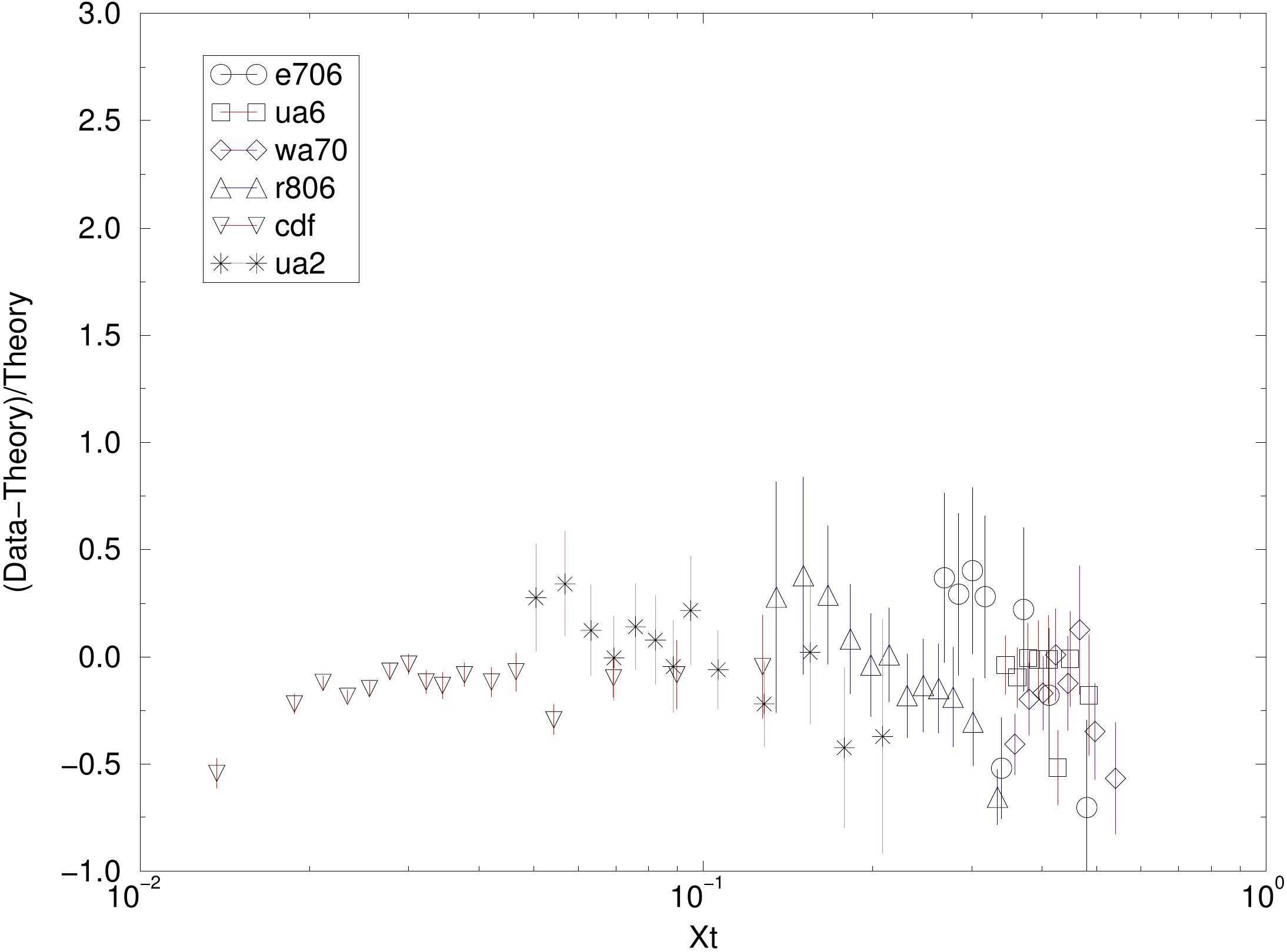}
\caption{Low $p_T$ direct photon spectra before (upper) and after (lower) 
including the $k_T$ resummation.}
\label{direct}
\end{figure}

\begin{figure}
\begin{center}
\centering\includegraphics[width=.6\linewidth]{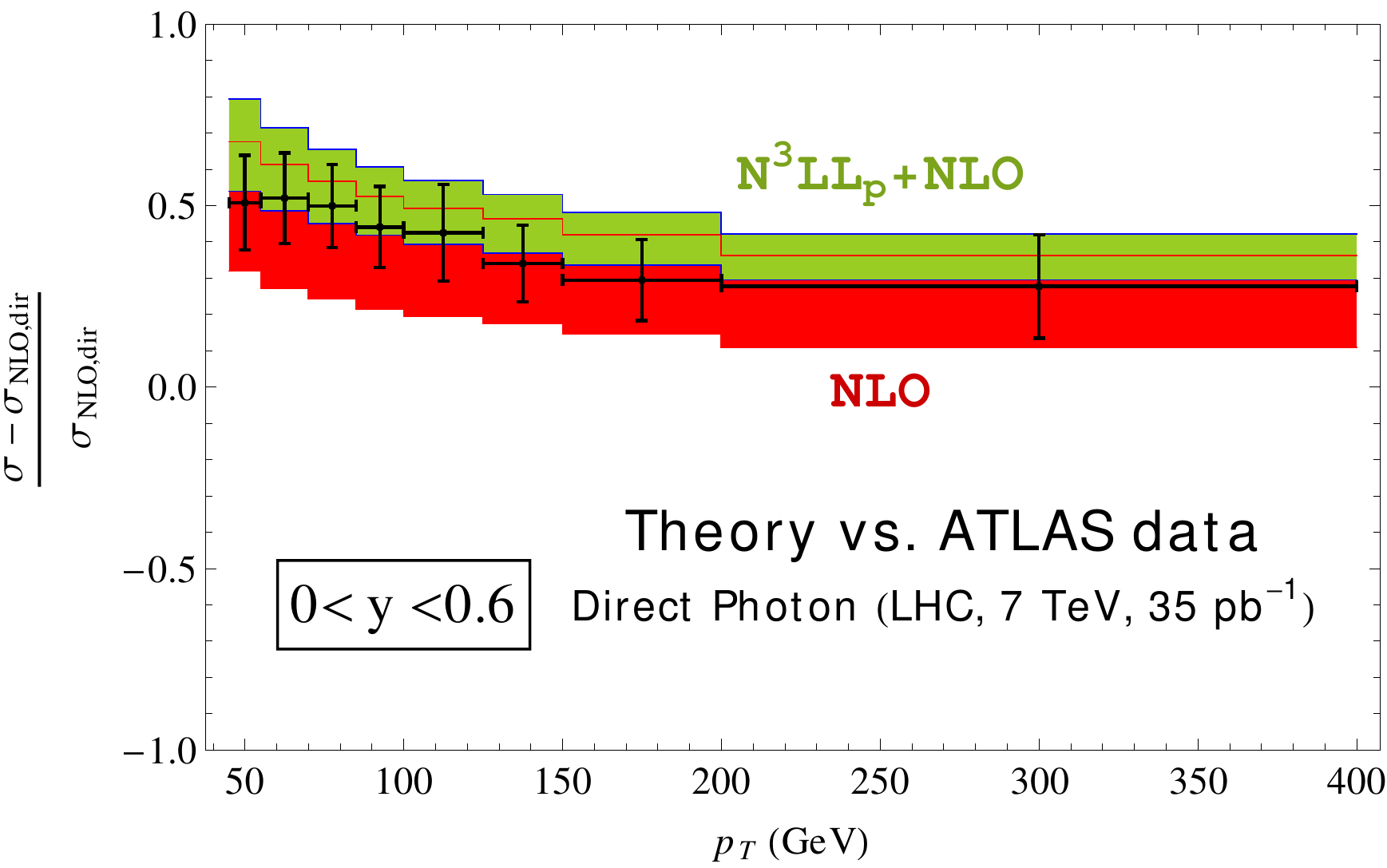}
\end{center}
\caption{High $p_T$ direct photon spectrum under the threshold resummation.}
\label{DPAtlas}
\end{figure}

The $k_T$ resummation effect on the low $p_T$ spectra of the direct photon 
production depicted in Fig.~(\ref{factorization})has been analyzed \cite{LL98}. 
The initial-state and final-state radiations are constrained
in the low $p_T$ region, where the $k_T$ resummation is necessary for
improving the perturbation theory. Figure~\ref{direct} shows the
deviation (Data -Theory)/Theory of the NLO pQCD predictions, obtained using
the CTEQ4M PDFs \cite{cteq4}, from the experimental data as
a function of $x_t=2p_T/\sqrt{s}$, $\sqrt{s}$ being the center-of-mass energy.
The deviation is huge as expected, especially at low $x_t$ of each
set of the data. After including the $k_T$ resummation
effect \cite{LL98}, it is clear that a significant improvement on the
agreement between theoretical predictions and the data is achieved.
As to the intermediate- and high-$p_T$ regions of the direct photon production,
NLO pQCD works reasonably well in accommodating the data as indicated in 
Fig.~\ref{DPAtlas}. The threshold resummation effect, which will be introduced 
in the next subsection, is more relevant in these regions: it slightly improves 
the consistency between predictions and the data \cite{Becher12}.

In the small $x$ region with $xp^+\sim k_T$, or $xp^+\sim 1/b$ in the
$b$ space, the two-scale case reduces to the single-scale one. In this region 
contributions from gluonic partons dominate, so $\Phi$ represents the gluon 
TMD below. The source of double logarithms, i.e., the integral containing 
the anomalous dimension $\gamma_K$, is less important. Because only the soft 
scale exists, one drops the hard kernel $G$, and keeps the soft kernel with an
ultraviolet cutoff. The right-hand side of Eq.~(\ref{meq}) becomes
\begin{eqnarray}
{\bar \Phi}(x,k_T)&=&
-ig^2N_c\int\frac{d^{4}l}{(2\pi)^4}\frac{{\hat n}\cdot v}{n\cdot l v\cdot l}
\left[\frac{\theta(k_T^2-l_T^2)}{l^2}\Phi(x,k_T)\right.
\nonumber \\
& &\left.+2\pi i\delta(l^2)\phi(x,|{\bf k}_T+{\bf l}_T|)\right],
\label{kf1}
\end{eqnarray}
where the color factor $C_F$ has been replaced by $N_c$ for the gluon TMD. 
The $\theta$ function introduces the ultraviolet cutoff on $l_T$ mentioned 
above. To make variation in $x$ via variation in $p^+$, a fixed
parton momentum is assumed. Under this assumption, the momentum fraction $x$ is
proportional to $1/p^+$, and one has $p^+d\Phi/dp^+=-xd\Phi/dx\Phi$ \cite{L0}.
Performing the integrations over $l^+$ and $l^-$ in Eq.~(\ref{kf1}), the master
equation (\ref{meq}) reduces to the BFKL equation \cite{KMS},
\begin{eqnarray}
\frac{d\phi(x,k_T)}{d\ln(1/x)}=
{\bar \alpha}_s\int\frac{d^{2}l_T}{\pi l_T^2}
\left[\phi(x,|{\bf k}_T+{\bf l}_T|)
-\theta(k_T^2-l_T^2)\phi(x,k_T)\right],
\label{bfkl}
\end{eqnarray}
with the coupling constant ${\bar \alpha}_s=N_c\alpha_s/\pi$.

\begin{figure}
\centering\includegraphics[width=.7\linewidth]{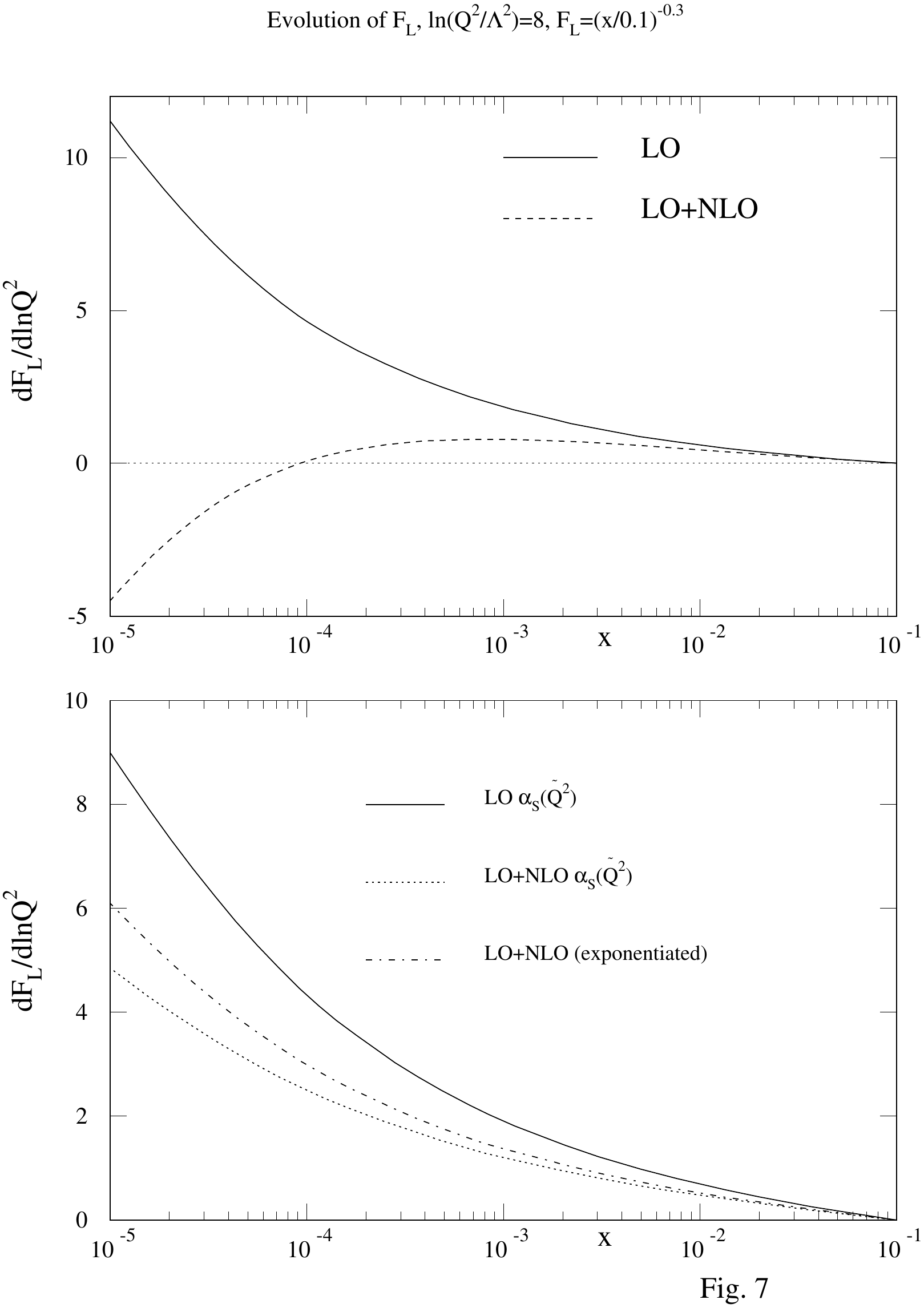}
\caption{Effects from LO and NLO BFKL equations.}
\label{bfklfig}
\end{figure}

A remarkable prediction of the above LO BFKL equation is that a 
high-energy cross section increases with the center-of-mass energy,
\begin{eqnarray}
\sigma\approx\frac{1}{t}
\left(\frac{s}{t}\right)^{\omega_P-1},\label{llb}
\end{eqnarray}
with the momentum transfer squared $t$. It turns out that Eq.~(\ref{llb}), 
with the Pomeron intercept $\omega_P-1=4{\bar \alpha}_s\ln 2$, violates the 
Froissart (unitarity) bound $\sigma< {\rm const.}\times \ln^2 $ \cite{F61}. 
The unsatisfactory prediction of the LO BFKL equation called for
the NLO corrections \cite{BFKLNLO}, which were, however, found to be dramatic
as indicated by the $x$ dependence of the derivative of the structure function
$dF_L/d\ln Q^2$ in Fig.~\ref{bfklfig} \cite{Thorne99}: the NLO effect is 
nearly as large as the LO result for $x \sim 0.001$, and becomes dominant at 
lower $x$. It even turns $dF_L/d\ln Q^2$ negative below $x \sim 0.0001$ in the 
upper of Fig.~\ref{bfklfig}. That is, the perturbative solution is not at all 
stable. Choosing a running coupling constant \cite{Thorne99}, the NLO effect 
is not overwhelming, but still significant as exhibited in the lower
of Fig.~\ref{bfklfig}.

\subsection{Threshold Resummation and DGLAP Equation}

We then consider the soft approximation corresponding to
the $k_T$ ordering of real gluon emissions in a ladder diagram.
Assume that a parton without the transverse momentum,
carries $-{\bf l}_{1T}$ after emitting a gluon of longitudinal
momentum $l_1^+$ and transverse momentum ${\bf l}_{1T}$,
and then carries $-{\bf l}_{1T}-{\bf l}_{2T}$ after emitting a gluon of
longitudinal momentum $l_2^+$ and transverse momentum ${\bf l}_{2T}$.
In the kinematic configuration with $l_{2T}\gg l_{1T}$
and $l_2^+\sim l_1^+$, the final parton momentum can be approximated by
$-{\bf l}_{2T}-{\bf l}_{1T}\approx -{\bf l}_{2T}$, such that the loop
integral associated with the first gluon emission involves only $l_{1T}$,
and can be worked out straightforwardly, giving a logarithm.
The loop integral associated with the second gluon emission involves
only $l_{2T}$, and also gives a logarithm. Hence, a ladder diagram
with $N$ rung gluons generates the logarithmic correction $(\alpha_s L)^N$
under the above $k_T$ ordering. In this case $\Phi$ is independent of $l_T$, 
and we have the approximation for the real gluon emission in Eq.~(\ref{fsr})
\begin{equation}
\Phi(x+l^+/p^+,|{\bf k_T+\bf l_T}|)\approx
\Phi(x+l^+/p^+,k_T),
\label{nk}
\end{equation}
in which $x$ and $l^+/p^+$ are of the same order.
The dependence on $k_T$ can then be integrated out from both sides of
the master equation (\ref{meq}), and the TMD $\Phi$ reduces to the PDF $\phi$.
The scale $k_T$ disappears, and the scale $(1-x)p^+$ is retained.

The Mellin transformation is employed to bring ${\bar\phi}_{s}$ from the 
momentum fraction $x$ space to the moment $N$ space,
\begin{eqnarray}
{\bar\phi}_{s}(N)&=&\int_0^1 dxx^{N-1}{\bar\phi}_{s}(x),
\end{eqnarray}
under which the $l^+$ integration decouples.
In the large $x$ region $\phi$ involves two scales, the large
$xp^+\sim p^+$ from the hard kernel $G$ and the small
$(1-x)p^+\sim p^+/N$ from the soft kernel $K$.
To sum $\ln(1/N)$, we rewrite the derivative $p^+d\phi/dp^+$ as
\begin{equation}
p^+\frac{d\phi}{dp^+}=
\frac{p^+}{N}\frac{d\phi}{d (p^+/N)}.
\end{equation}
The solution of the master equation (\ref{meq}) then gives the threshold resummation,
\begin{eqnarray}
\phi(N)=\Delta_t(N)\phi_i
\label{pht}
\end{eqnarray}
with the exponential
\begin{eqnarray}
\Delta_t(N)=\exp\left[-2\int_{p^+/N}^{p^+}\frac{d p}{p}
\int_{p^+}^{p}\frac{d\mu}{\mu}
\gamma_{K}(\alpha_s(\mu))\right],
\label{fbt}
\end{eqnarray}
or its equivalent expression
\begin{eqnarray}
\Delta_t(N)=\exp\left[\int_{0}^{1}dz\frac{1-z^{N-1}}{1-z}
\int_{(1-z)^2}^{1}\frac{d\lambda}{\lambda}
\gamma_{K}(\alpha_s(\sqrt{\lambda}p^+))\right].
\end{eqnarray}

\begin{figure}
\centering\includegraphics[width=.6\linewidth]{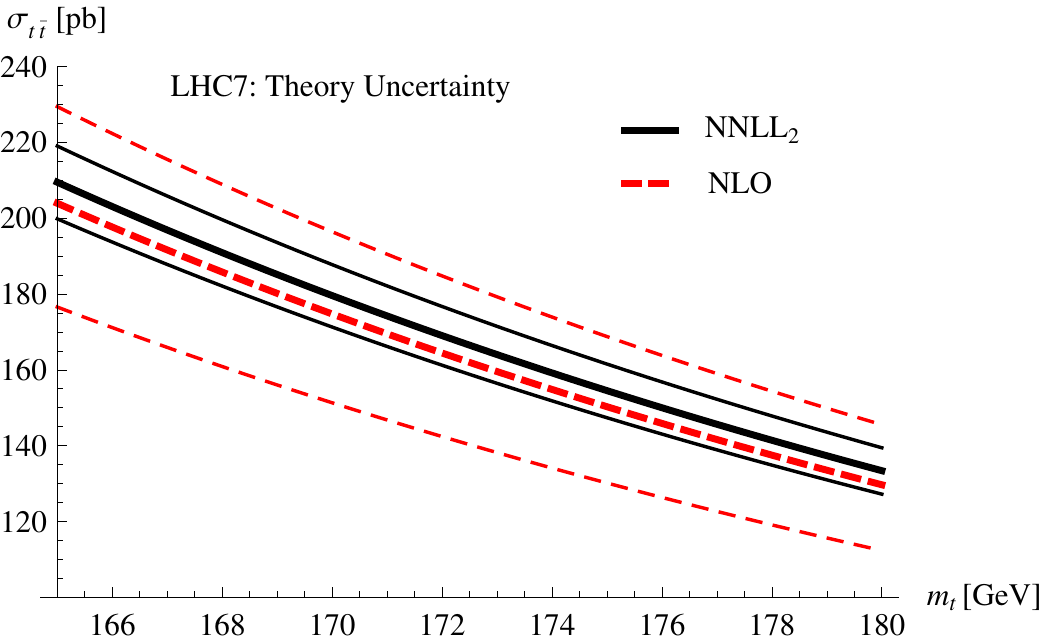}
\caption{Dependence of the total cross section for the top-pair production
on the top mass at the LHC with $\sqrt{s} =$ 7 TeV.}
\label{top1}
\end{figure}

\begin{figure}
\centering\includegraphics[width=.6\linewidth]{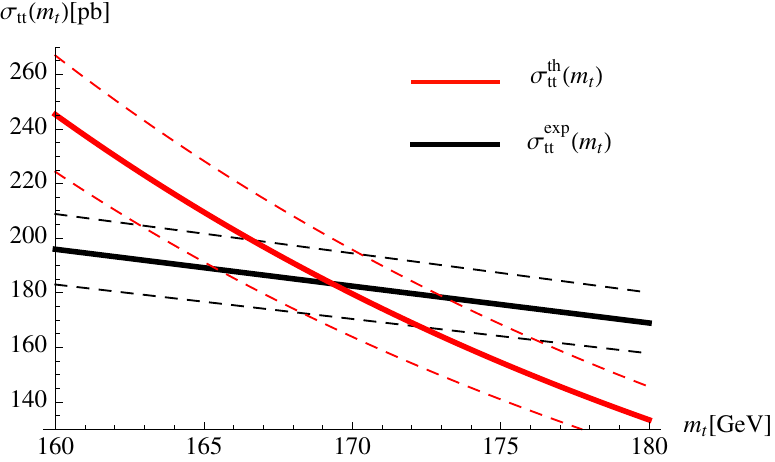}
\caption{Mass dependence of the theoretical cross section with the threshold
resummation effect (red) and of the measured cross section (black).}
\label{top2}
\end{figure}

An application of the threshold resummation is found in the analysis of the
top-quark pair production, which was performed at the 
next-to-next-to-leading-logarithmic (NNLL)
accuracy \cite{Beneke11}. It has been observed that
the threshold resummation effect enhances the NLO total cross section by
few percents as shown in Fig.~\ref{top1}, where the bands sandwiched by the thinner
lines denote the theory uncertainty. The above formalism can be used to
determine of the top quark mass as indicated in Fig.~\ref{top2}, where
the solid lines represent the central values, and the total uncertainties
of the theoretical and experimental results \cite{ATLAS11}
are given by the external dashed lines.

In the intermediate $x$ region the two-scale case reduces to the
single-scale one because of $xp^+\sim (1-x)p^+$, and the source of double
logarithms is less important. Without the Mellin transformation,
the sum in Eq.~(\ref{ssf}), with the approximation
in Eq.~(\ref{nk}) being inserted, leads to the DGLAP equation \cite{L0},
\begin{eqnarray}
p^+\frac{d}{dp^+}\phi(x)
=\int_x^1 \frac{d\xi}{\xi}P(x/\xi)\phi(\xi)\;,
\label{con}
\end{eqnarray}
with the kernel
\begin{eqnarray}
P(z)=\frac{\alpha_s(p^+)}{\pi}C_F\frac{2}{(1-z)_+}\;,
\label{kgir}
\end{eqnarray}
where the variable change $\xi=x+l^+/p^+$ has been made. The argument
of $\alpha_s$, i.e, the factorization scale $\mu$,
has been set to the scale $xp^+\sim (1-x)p^+\sim O(p^+)$. Note that the 
kernel $P$ differs from the splitting function $P_{qq}$ in Eq.~(\ref{splitq})
by the term $(z^2-1)/(1-z)_+$, which is finite in the $z\to 1$ limit.
The reason is that the real gluon emission was evaluated under the soft
approximation as deriving $P$, while it was calculated exactly as deriving
$P_{qq}$.

\begin{figure}
\centering\includegraphics[width=.3\linewidth]{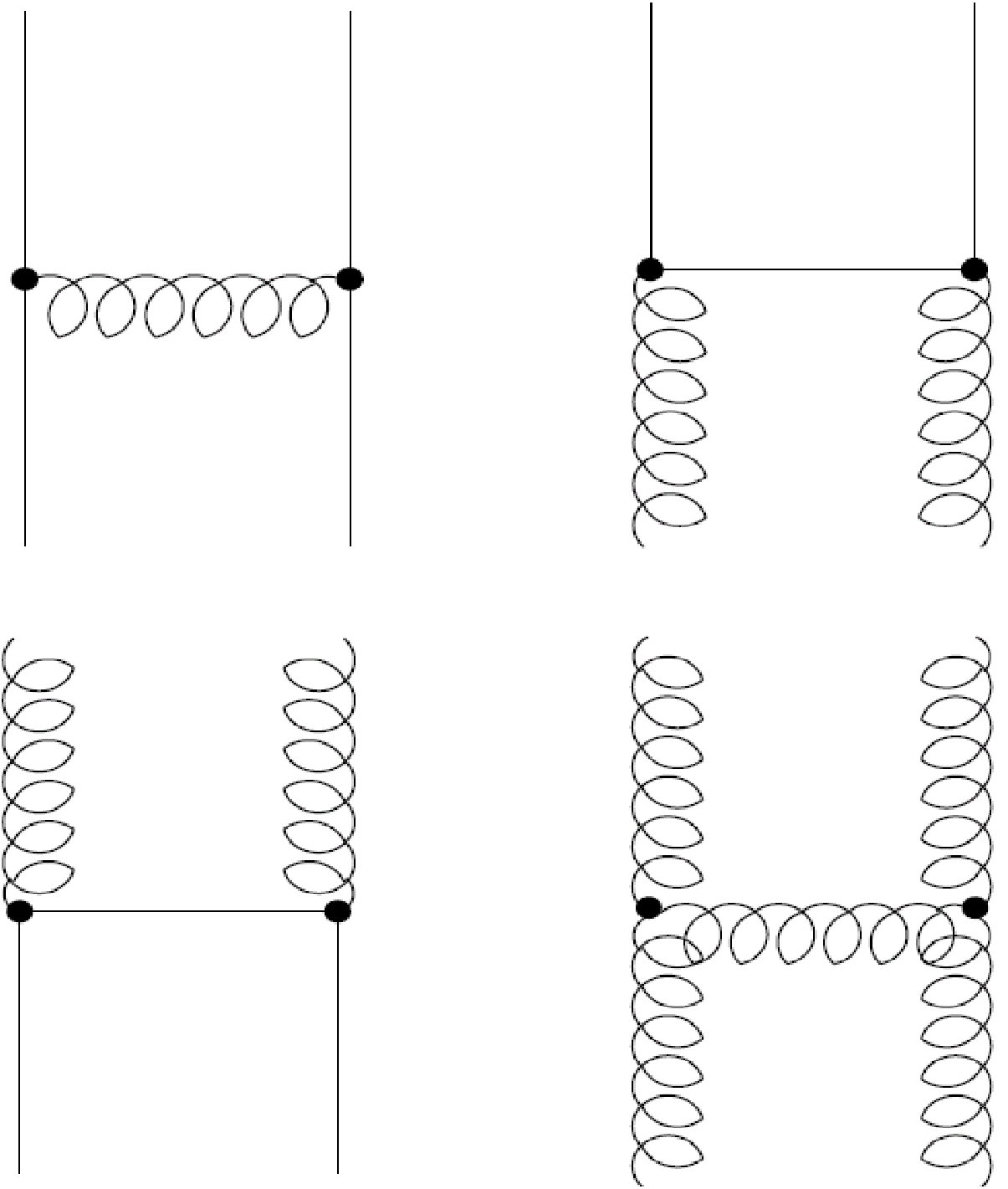}
\caption{Diagrams for the DGLAP splitting functions.}
\label{dglap1}
\end{figure}

\begin{figure}
\centering\includegraphics[width=.7\linewidth]{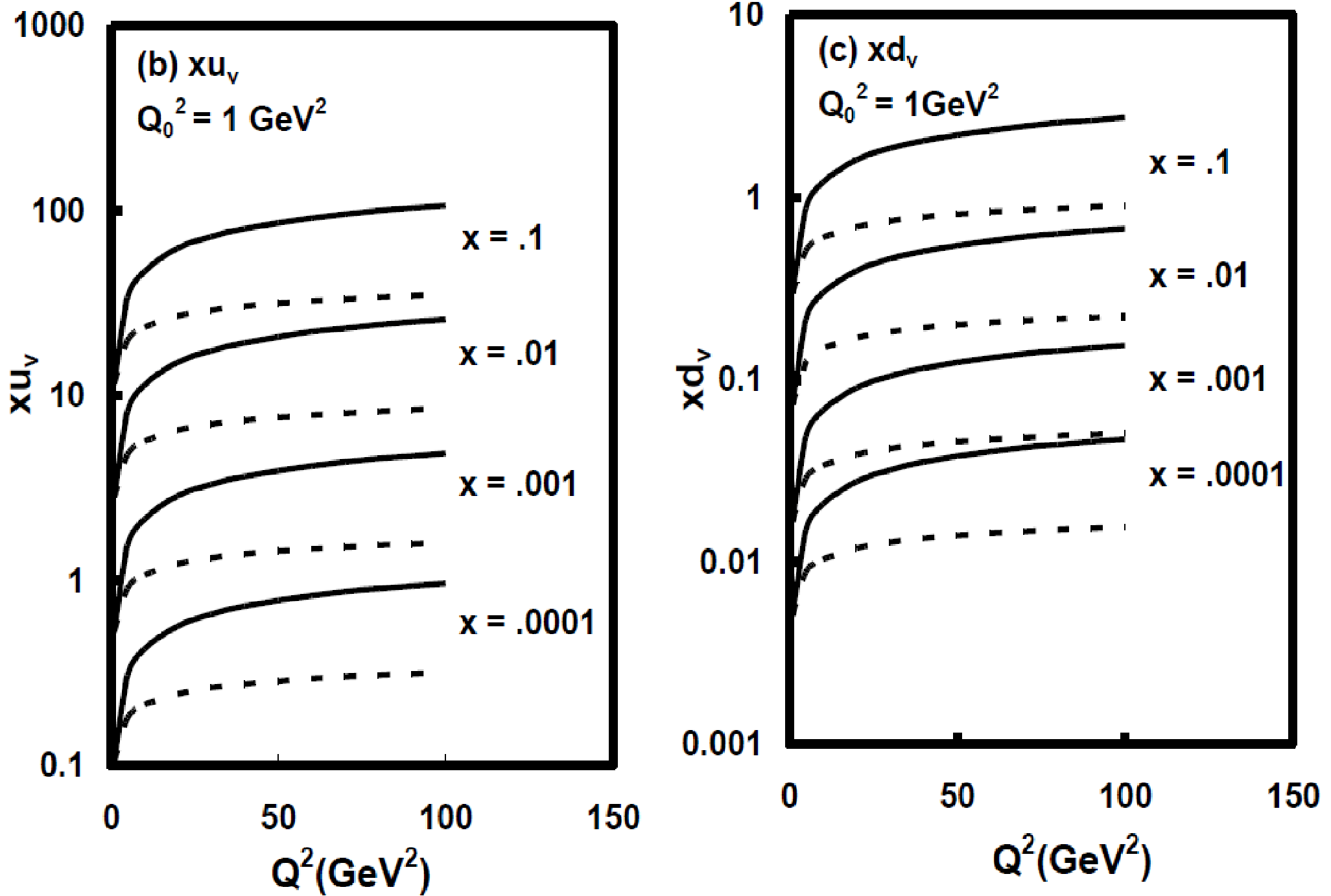}
\caption{$Q^2$ evolutions of the valence quark PDFs for some parameter values
in the DGLAP solutions (solid and dashed lines). 
}
\label{dglap3}
\end{figure}

\begin{figure}
\begin{center}
\includegraphics[width=.4\linewidth]{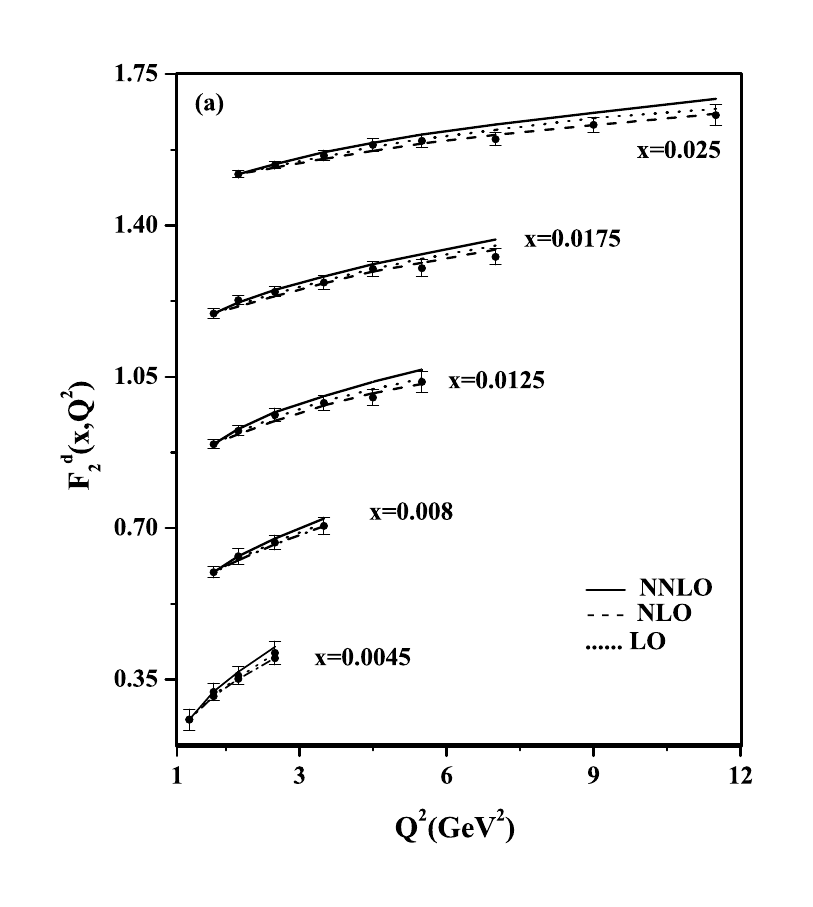}
\includegraphics[width=.4\linewidth]{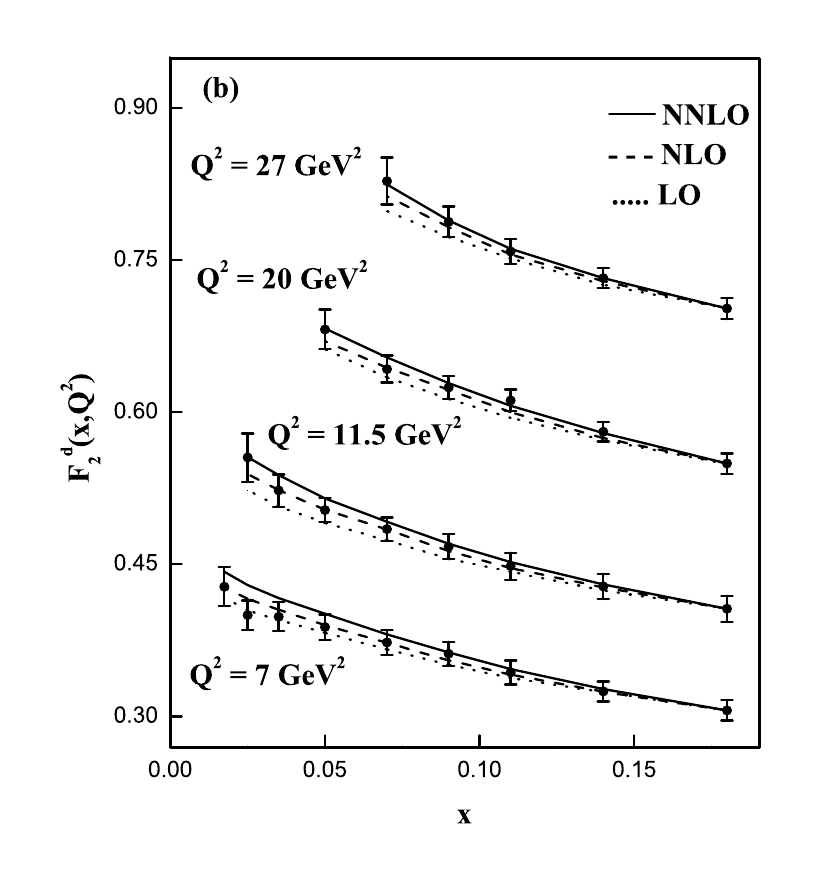}
\end{center}
\caption{Predictions from the DGLAP equation and the NMC data for
the deuteron structure function. For clarity, data are scaled up by +0.2i (in
Fig.(a)) and +i (in Fig.(b)) (with i = 0,1,2,3) starting from the bottom of
the graphs.}
\label{dglap2}
\end{figure}

Gluon emissions in Fig.~\ref{dglap1} cause the mixing between the quark and
gluon PDFs, giving the complete set of DGLAP equations with four splitting
functions
\begin{equation}\label{qs}
   \frac{\partial}{\partial \ln Q^2}\left( \begin{array}{c}
    \phi_q \\ \phi_g
   \end{array} \right)
   = \left( \begin{array}{cc}
    P_{qq} & P_{qg} \\ P_{gq} & P_{gg}
   \end{array} \right)\otimes
   \left( \begin{array}{c}
    \phi_q \\ \phi_g
   \end{array} \right).
\end{equation}
The evolution of the $u$-quark and $d$-quark PDFs in $Q^2$ predicted
by the LO DGLAP equation \cite{RS12} is shown in Fig.~\ref{dglap3},
where the inputs at the initial scale $Q_0=1$ GeV were taken from
MRST2001 \cite{MRST01}. It is observed that the valence quark PDFs increase
with $Q^2$ at small $x$, namely, they become broader with $Q^2$, a feature
consistent with what was stated in the previous section.
The predictions for the deuteron structure function derived from the LO,
NLO, and NNLO DGLAP equations are displayed in Fig.~\ref{dglap2} \cite{Devee12},
which agree with the NMC data \cite{NMC97}.

\subsection{Joint Resummation and CCFM Equation}

At last, a unified resummation formalism for large and intermediate
$x$ and a unified evolution equation for intermediate and small $x$ can be derived
by retaining the $l^+$ and $l_T$ dependencies of $\Phi$ in Eq.~(\ref{fsr}),
which corresponds to the so-called angular ordering.
In this case both the Fourier and Mellin transformations are applied
to Eq.~(\ref{fsr}), leading to
\begin{eqnarray}
{\bar\Phi}_s(N,b)=K(p^+/(N\mu),1/(b\mu),\alpha_s(\mu))\Phi(N,b)\;,
\end{eqnarray}
with the soft kernel \cite{Li99}
\begin{eqnarray}
K&=&-ig^2C_F\mu^\epsilon\int_0^1dz
\int\frac{d^{4-\epsilon}l}{(2\pi)^{4-\epsilon}}
\frac{{\hat n}\cdot v}{n\cdot l v\cdot l}
\left[\frac{\delta(1-z)}{l^2}\right.
\nonumber\\
& &\left.+2\pi i\delta(l^2)\delta\left(1-z-\frac{l^+}{p^+}\right)
z^{N-1}e^{i{\bf l}_T\cdot{\bf b}}\right]-\delta K,
\nonumber\\
&=&\frac{\alpha_s(\mu)}{\pi}C_F\left[\ln\frac{1}{b\mu}
-K_0\left(\frac{2\nu p^+b}{N}\right)\right],
\label{uk}
\end{eqnarray}
$K_0$ being the modified Bessel function. As $p^+b\gg N$, we have $K_0\to 0$, 
and the soft scale inferred by the above expression approaches $1/b$ for the 
$k_T$ resummation. As $N\gg p^+b$, we have $K_0\approx -\ln(\nu p^+b/N)$, 
and the soft scale approaches $p^+/N$ for the threshold resummation.

\begin{figure}
\centering\includegraphics[width=.5\linewidth]{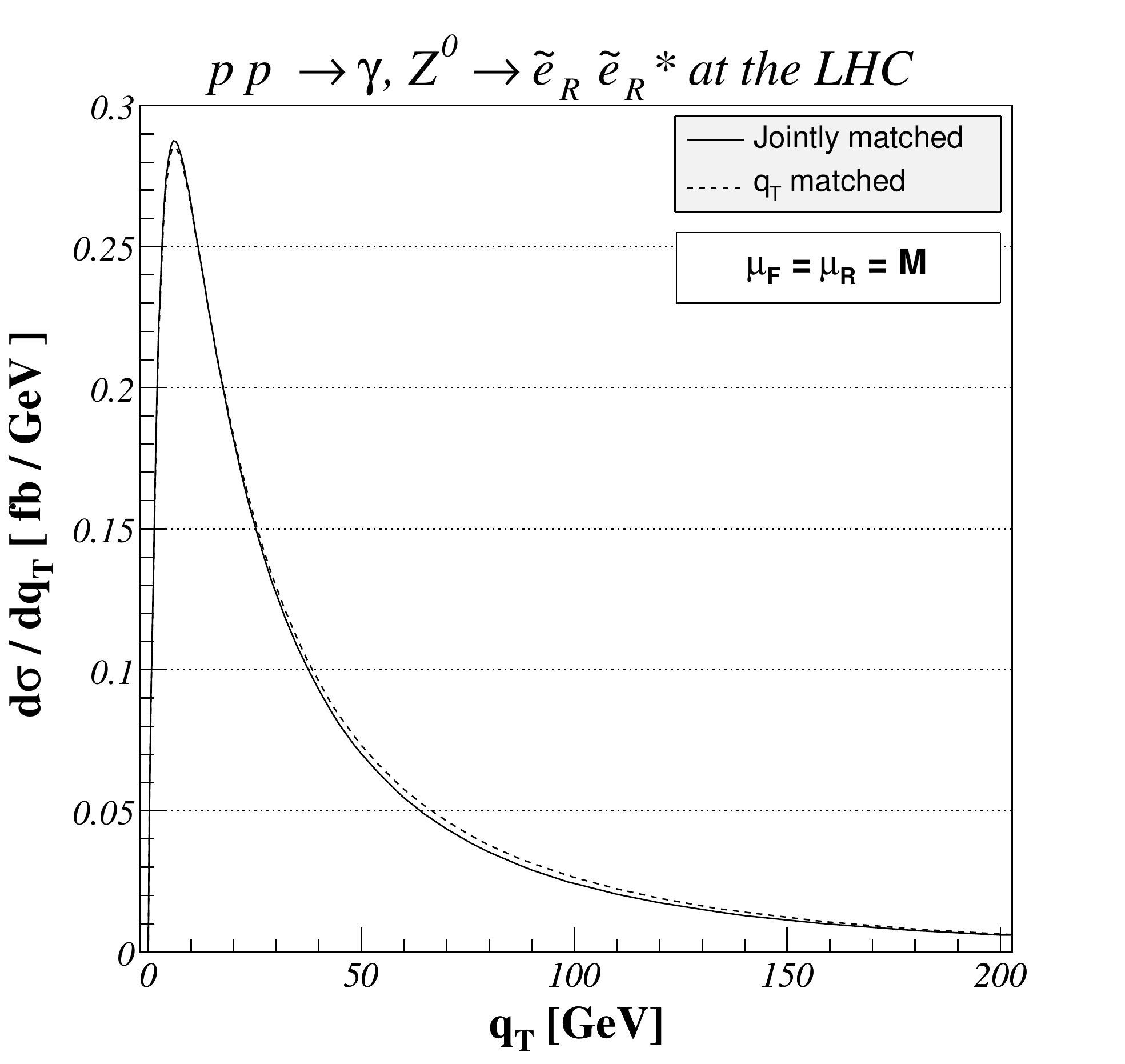}
\caption{Transverse-momentum distribution of selectron pairs at the LHC
in the framework of joint (full) and $k_T$ (dotted) resummations.
}
\label{joint1}
\end{figure}

Following the procedures similar to Eqs.~(\ref{dph})-(\ref{fb}), 
we derive the joint resummation
\begin{equation}
\Phi(N,b)=\Delta_u(N,b)\Phi_i,
\end{equation}
with the exponential
\begin{equation}
\Delta_u(N,b)=
\exp\left[-2\int_{p^+\chi^{-1}(N,b)}^{p^+}\frac{d p}{p}
\int_{p^+\chi^{-1}(1,b)}^{p}\frac{d\mu}{\mu}
\gamma_{K}(\alpha_s(\mu))\right].
\label{fb3}
\end{equation}
The dimensionless function \cite{LSV00}
\begin{eqnarray}
\chi(N,b)=\left(N+\frac{p^+b}{2}\right)e^{\gamma_E},
\end{eqnarray}
is motivated by the limits discussed above. It is apparent that Eq.~(\ref{fb3}) 
reduces to Eq.~(\ref{fb}) and Eq.~(\ref{fbt}) in the $b\to\infty$ and 
$N\to \infty$ limits, respectively. The effect from the joint resummation on 
the $q_T$ spectra of selectron pairs produced at the LHC with $\sqrt{S}=14$ TeV 
has been investigated in \cite{Bozzi07}.
It is seen in Fig.~\ref{joint1} that the joint and $k_T$ resumations exhibit a 
similar behavior in the small-$q_T$ region as expected, but the jointly-resummed 
cross section is about 5\%-10\% lower than the $k_T$-resummed cross section in 
the range 50 GeV $< q_T <$ 100 GeV.

In the intermediate and small $x$ regions, it is not necessary to resum the
double logarithms $\ln^2(1/N)$. After extracting the $k_T$ resummation, the
remaining single-logarithmic summation corresponds to a unification of the
DGLAP and BFKL equations, since both the $l^+$ and $l_T$ dependencies have
been retained. The function $\Phi(x+l^+/p^+,b)$ in Eq.~(\ref{fsr}) is reexpressed,
after the Fourier transformation, as
\begin{eqnarray}
& &\Phi(x+l^+/p^+,b)=\theta((1-x)p^+-l^+)\Phi(x,b)
\nonumber\\
& &\hspace{1.5cm} +[\Phi(x+l^+/p^+,b)-\theta((1-x)p^+-l^+)
\Phi(x,b)].
\label{fre}
\end{eqnarray}
The contribution from the first term is combined with the first term in
Eq.~(\ref{fsr}), giving the soft kernel $K$ for the $k_T$ resummation.
The second term in Eq.~(\ref{fre}) contributes
\begin{eqnarray}
-iN_cg^2\int\frac{d^4l}{(2\pi)^4}
\frac{{\hat n}\cdot v}{n\cdot l v\cdot l}
2\pi i\delta(l^2)e^{i{\bf l}_T\cdot {\bf b}}
[\Phi(x+l^+/p^+,b)-\theta((1-x)p^+-l^+)\Phi(x,b)],
\label{fs2}
\end{eqnarray}
which will generate the splitting function below. The color factor has
been replaced by $N_c$, since the gluon TMD is considered here.

The master equation (\ref{meq}) then becomes
\begin{eqnarray}
p^+\frac{d}{dp^+}\Phi(x,b)&=&-2\left[\int_{1/b}^{xp^+}
\frac{d\mu}{\mu}\gamma_K(\alpha_s(\mu))
-{\bar \alpha}_s(xp^+)\ln(p^+b)\right]\Phi(x,b)
\nonumber\\
& &+2{\bar\alpha}_s(xp^+)\int_x^1 dz P_{gg}(z)\Phi(x/z,b),
\label{ue2}
\end{eqnarray}
with the splitting function
\begin{equation}
P_{gg}=\left[\frac{1}{(1-z)_+}+\frac{1}{z}-2+z(1-z)\right],
\label{pgg}
\end{equation}
obtained from Eq.~(\ref{fs2}). The term $-2+z(1-z)$ finite as
$z\to 0$ and $z\to 1$ has been added. The exponential $\Delta$ is extracted 
from the $k_T$ resummation,
\begin{eqnarray}
\Delta(x,b,Q_0)
=\exp\left(-2\int_{xQ_0}^{xp^+}\frac{dp}{p}
\left[\int_{1/b}^{p}
\frac{d\mu}{\mu}\gamma_K(\alpha_s(\mu))
-{\bar \alpha}_s(p)\ln\frac{p b}{x}\right]\right),
\label{del}
\end{eqnarray}
$Q_0$ being an arbitrary low energy scale. It is trivial to justify by
substitution that the solution is given by
\begin{eqnarray}
\Phi(x,b)&=&\Delta(x,b,Q_0)\Phi_i
\nonumber\\
& &+2\int_x^1 dz\int_{Q_0}^{p^+}\frac{d\mu}{\mu}
{\bar\alpha}_s(x\mu)\Delta_k(x,b)P_{gg}(z)\Phi(x/z,b),
\label{nunif}
\end{eqnarray}
which can be regarded as a modified version of the CCFM equation \cite{CCFM}.

\section{PQCD for jet physics}

Jets, abundantly produced at colliders \cite{SW77}, carry information of 
hard scattering and parent particles, which is crucial for particle 
identification and new physics search. Study of jet physics is usually done 
using event generators, which, however, suffer ambiguity from parameter 
tuning. Hence, we are motivated to establish an alternative approach free of 
the ambiguity. I will demonstrate that jet dynamics can be explored and jet 
properties can be predicted in the pQCD resummation formalism.

\begin{figure}
\centering\includegraphics[width=.4\linewidth]{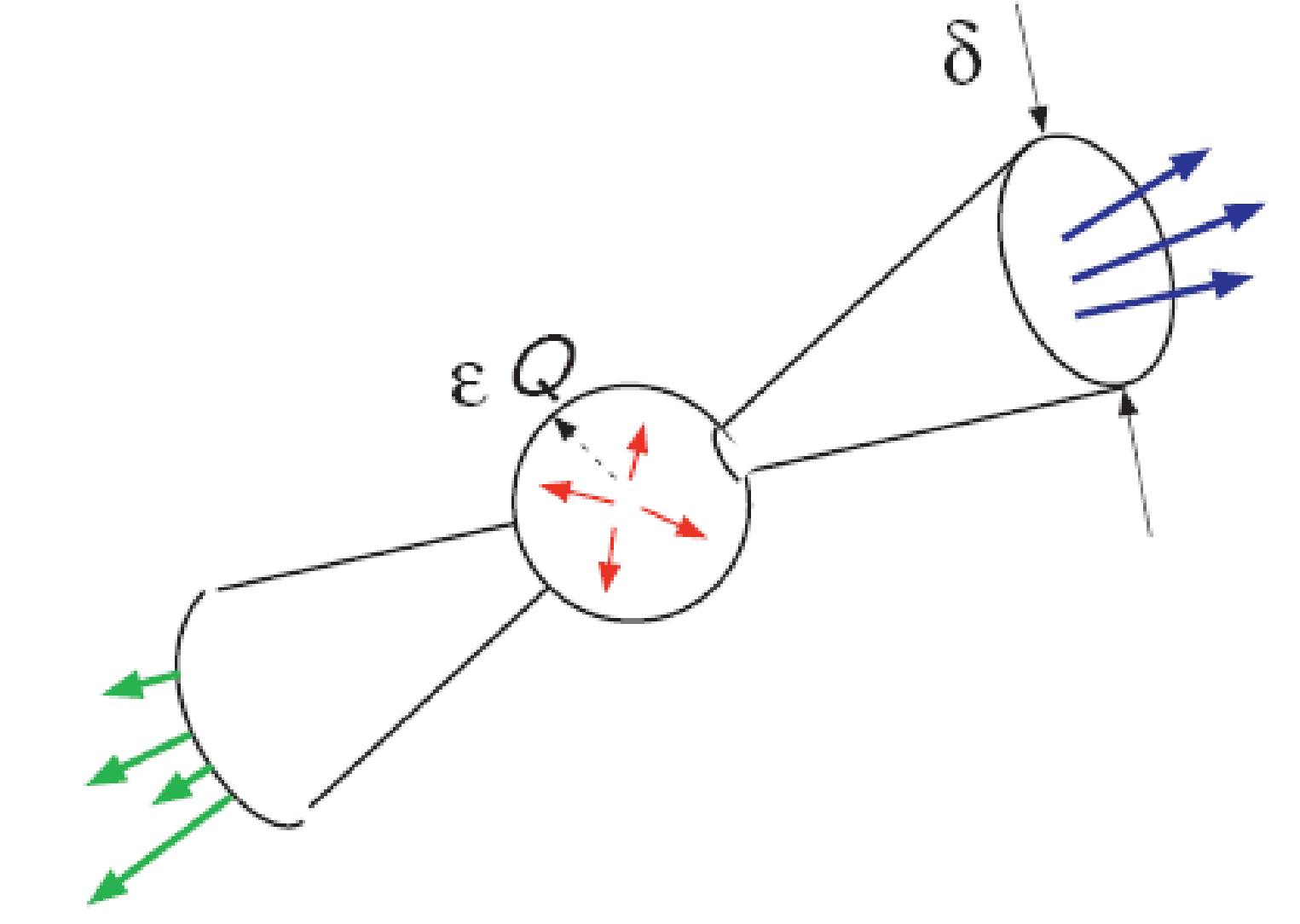}
\caption{Dijet final state in $e^-e^+$ annihilation.}
\label{jetf1}
\end{figure}

We start from the dijet production in the $e^-e^+$ annihilation, which is part of
its total cross section. The physical dijet final state, described in Fig.~\ref{jetf1},
contains two jet cones of half angle $\delta$ and isotropic soft gluons 
within the energy resolution $\epsilon Q$, $Q$ being the $e^-e^+$ invariant mass. 
The Born cross section is the same as the total one in Eq.~(\ref{ee0}). With the 
constrained phase space for real gluons, the infrared cancellation is not complete, 
and logarithmic enhancement appears. The explicit NLO calculations imply that the 
isotropic soft gluons give a contribution proportional to $2\ln^2(2\epsilon Q/\mu)-\pi^2/6$, 
the collinear gluons in the cones with energy higher than the resolution give 
$-3\ln(Q\delta/\mu)-2\ln^2(2\epsilon)-4\ln(Q\delta/\mu)\ln(2\epsilon)+17/4-\pi^2/3$, 
and the virtual corrections contribute $-2\ln^2(Q/\mu)+3\ln(Q/\mu)-7/4+\pi^2/6$. The total 
NLO corrections indicate that the dijet cross section is infrared finite, but 
logarithmically enhanced:
\begin{eqnarray}
3\ln\delta+4\ln\delta\ln(2\epsilon)+\frac{\pi^2}{3}-\frac{5}{2},
\end{eqnarray}
where the double logarithm $\ln\delta\ln(2\epsilon)$ is attributed to the
overlap of the collinear and soft logarithms.

\subsection{Jet in Experiments}

To describe the kinematics for jets, we define the pseudorapidity
$\eta=\ln[\cot(\theta/2)]$, which is related to the polar angle
$\theta$ with respect to the beam direction, and the azimuthal angle
$\phi$. That is, $\theta=0$, $90^\circ$, and $180^\circ$ correspond
to $\eta=+\infty$, 0 and $-\infty$, respectively. Comparison of theoretical 
and experimental descriptions for jet observables is nontrivial. One needs 
jet algorithms that map experimental measurements with theoretical 
calculations as close as possible. The infrared safety \cite{SW77} is an 
important guideline for setting up a jet algorithm. There are two major 
classes of jet algorithms in the literature: cone algorithms and sequential 
algorithms. The former is a geometrical method, which stamps out jets on
the $\eta$-$\phi$ plane as with a cookie cutter. The latter combines 
particle four-momenta one by one following given kinematic criteria.

I take the seeded cone algorithm as an example to explain the operation 
in the first class of jet algorithms, which aims at finding stable cones 
via an iterative procedure. Start from a seed particle $i$, and consider 
a set of particles $j$ with separations smaller than jet cone of radius $R$,
\begin{eqnarray}
\Delta R_{ij}^2\equiv (\eta_i-\eta_j)^2+(\phi_i-\phi_j)^2< R^2.
\end{eqnarray}
Calculate the new cone center $J$ by summing all particle four-momenta
in the cone. A stable cone is composed of a set of
particles $i$ satisfying $\Delta R_{iJ}<R$. If the cone is stable, the
procedure stops. Otherwise, take $J$ as a new seed, and repeat the above
procedure.

However, the seeded cone algorithm suffers the problem of infrared
divergences. Such a geometrical algorithm does not differentiate infrared
gluons from energetic gluons, so final outcomes depend on soft radiation and
collinear splitting. This problem can be illustrated by considering a system of
two particles 1 and 2, separated by $R_{12}$ with $R<R_{12}<2R$.
Each of particles 1 and 2, taken as a seed, forms a stable jet. One then adds
a soft gluon to this system. It is obvious that a virtual soft gluon exchanged
between jets 1 and 2 does not change the outcome; namely, a virtual soft gluon
contributes to the dijet cross section. On the contrary, adding a real soft
seed between jets 1 and 2 will merge the two jets because of $R<R_{12}<2R$.
Therefore, a real soft gluon contributes to the single jet cross section.
As a result, the soft divergences do not cancel between the virtual and real
corrections. One may speculate that starting from the hardest
particle may avoid the difficulty caused by the soft seed. It turns out that
the collinear splitting would change the outcome. Including a more energetic
particle into the above system, which is emitted between particles 1 and 2.
Taking this central particle as the seed, one constructs a single stable jet
formed by the three particles. A self-energy correction to
the central particle does not change this final state, and contributes to the
single jet cross section. However, the splitting of the central particle
may produce two particles, which are less energetic than particles 1 and 2.
Then one has to take particle 1 or 2 as the seed, and ends up with two stable
jets. That is, the collinear splitting contributes to the dijet cross section,
and there is no cancellation between virtual and real corrections. It is 
concluded that a seeded cone algorithm is not infrared safe.

Next I introduce sequential algorithms by taking the $k_T$ algorithm as
an example. For any pair of particles $i$ and $j$, find the minimum of the
following three distances
\begin{eqnarray}
d_{ij}=\min(k_{Ti}^2,k_{Tj}^2)\frac{\Delta R_{ij}^2}{R^2},\;\;\;\;
d_{iB}=k_{Ti}^2,\;\;\;\;d_{jB}=k_{Tj}^2,
\end{eqnarray}
with $k_T$ being is a jet transverse momentum.
If the minimum is $d_{iB}$ or $d_{jB}$, $i$ or $j$ is a jet, and removed
from the list of particles. Otherwise, $i$ and $j$ are merged into a new jet.
Repeat the above procedure until no particles are left. The other
sequential algorithms include the Cambridge/Aachen and anti-$k_T$ ones
with the definitions of the distances
\begin{eqnarray}
& &d_{ij}=\frac{\Delta R_{ij}^2}{R^2},\;\;\;\;
d_{iB}=1,\;\;\;\;d_{jB}=1,\nonumber\\
& &d_{ij}=\min(k_{Ti}^{-2},k_{Tj}^{-2})\frac{\Delta R_{ij}^2}{R^2},\;\;\;\;
d_{iB}=k_{Ti}^{-2},\;\;\;\;d_{jB}=k_{Tj}^{-2},
\end{eqnarray}
respectively. The grouping starts from soft (energetic) particles and usually
leads to an irregular (round) jet shape in the $k_T$ (anti-$k_T$) algorithm.
Note that a sequential algorithm differentiates infrared gluons from energetic
ones: adding a soft real gluon does not modify a cone center, so
it does not change the outcome.

\subsection{Jets in Theory}

As outlined in the Introduction,
we intend to establish a theoretical framework for jet study, following
the idea of the factorization theorem for the DIS in Sec.~2. At NLO, a jet is
produced in DIS, as the gluon emitted by the initial-state or final-state
quark is collimated to the final-state quark. The restricted phase space
of the final-state quark and the gluon in a small angular separation
renders an incomplete cancellation between the virtual and real
corrections. Hence, jet production is expected to be enhanced by collinear
dynamics. Similarly, the initial-state quark propagator can be eikonalized in
this collinear region, such that collinear gluons are detached
from the initial-state quark and absorbed into a jet function. To all orders, 
the collinear gluons are collected by the Wilson link with the path-ordered 
exponential
\begin{eqnarray}
W={\cal P}\exp\left[-ig \int_0^\infty dz n\cdot
A(zn)\right],\label{wil}
\end{eqnarray}
with an arbitrary vector $n$. The collinear gluon emitted by the final-state
quark can be factorized into the jet function straightforwardly by
applying the Fierz transformation. A more sophisticated factorization
formula for the jet production in the DIS is then written as a convolution
of a hard kernel $H$ with a PDF and a jet function $J$.
$H$ denotes the contribution with the collinear pieces for the
initial and final states being subtracted. This factorization formalism
is the basis for the application of pQCD to jet physics.

The light-quark and gluon jet functions are defined by \cite{Almeida:2008tp}
\begin{eqnarray}
J_q(M_J^2,P_T,\nu^2,R,\mu^2)&=&\frac{(2\pi)^3}
{2\sqrt{2}(P_J^0)^2N_c}\sum_{N_J}Tr\left\{\not\xi\langle
0|q(0)W^{(\bar q)\dagger}|N_J\rangle\langle N_J|W^{(\bar q)}
\bar q(0)|0\rangle\right\}\nonumber\\
& &\times\delta(M_J^2-\hat M_J^2(N_J,R))\delta^{(2)}(\hat e-\hat
e(N_J))\delta(P_J^0-\omega(N_J)),
\nonumber \\
J_g(M_J^2,P_T,\nu^2,R,\mu^2)&=&\frac{(2\pi)^3}
{2(P_J^0)^3N_c}\sum_{N_J}\langle
0|\xi_\sigma F^{\sigma\nu}(0)W^{(g)\dagger}
|N_J\rangle\langle N_J|W^{(g)}
F_\nu^\rho(0)\xi_\rho|0\rangle\nonumber\\
& &\times\delta(M_J^2-\hat M_J^2(N_J,R))\delta^{(2)}(\hat e-\hat
e(N_J))\delta(P_J^0-\omega(N_J)),\label{jet1}
\end{eqnarray}
where $|N_J\rangle$ denotes the final state with $N_J$ particles
within the cone of size $R$ centered in the direction of the unit
vector $\hat e$, $\hat M_J(N_J,R)$ ($\omega(N_J)$) is the invariant mass 
(total energy) of all $N_J$ particles, and $\mu$ is the factorization scale. 
The above jet functions absorb the collinear divergences from all-order
radiations associated with the energetic light jet of momentum 
$P_J^\mu=P_J^0 v^\mu$, in which $P_J^0$ is the jet energy, and
the vector $v$ is given by $v^\mu=(1,\beta,0,0)$ with
$\beta=\sqrt{1-(M_J/P_J^0)^2}$. $\xi^\mu=(1,-1,0,0)$ is a vector on the 
light cone. The coefficients in Eq.~(\ref{jet1}) have been chosen, such 
that the LO jet functions are equal to $\delta(M_J^2)$ in a perturbative 
expansion.

Underlying events include everything but hard scattering, such as
initial-state radiation, final-state radiation, and multiple parton
interaction (MPI). The Wilson lines in Eq.~(\ref{jet1}) have collected
gluons radiated from both initial states and other final states in a
scattering process, and collimated to the light-particle jets. Gluon 
exchanges between the quark fields $q$ (or the gluon fields $F^{\sigma\nu}$
and $F_\nu^\rho$) correspond to the final-state radiations. Both the
initial-state and final-state radiations are leading-power effects
in the factorization theorem, and have been included in the jet
function definition. A chance of involving more partons in hard scattering
is low, so the contribution from MPI is regarded as being subleading-power.
This contribution should be excluded from data, but it is certainly 
difficult to achieve in experiments. Nevertheless, it still makes sense
to compare predictions for jet observables based on Eq.~(\ref{jet1})
at the current leading-power accuracy with experimental data. At last,
pile-up events must be removed in experiments \cite{SS13}, since they 
cannot be handled theoretically so far.

\begin{figure}
\centering\includegraphics[width=.6\linewidth]{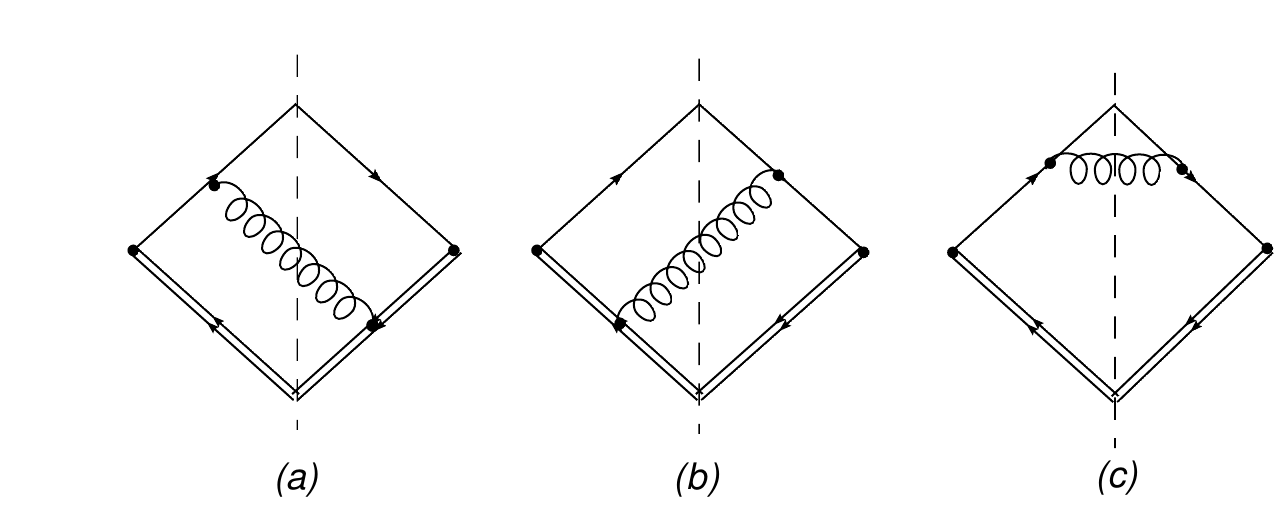}
\caption{Some NLO real corrections to the quark jet function.}
\label{quark}
\end{figure}

\begin{figure}
\centering\includegraphics[width=.5\linewidth]{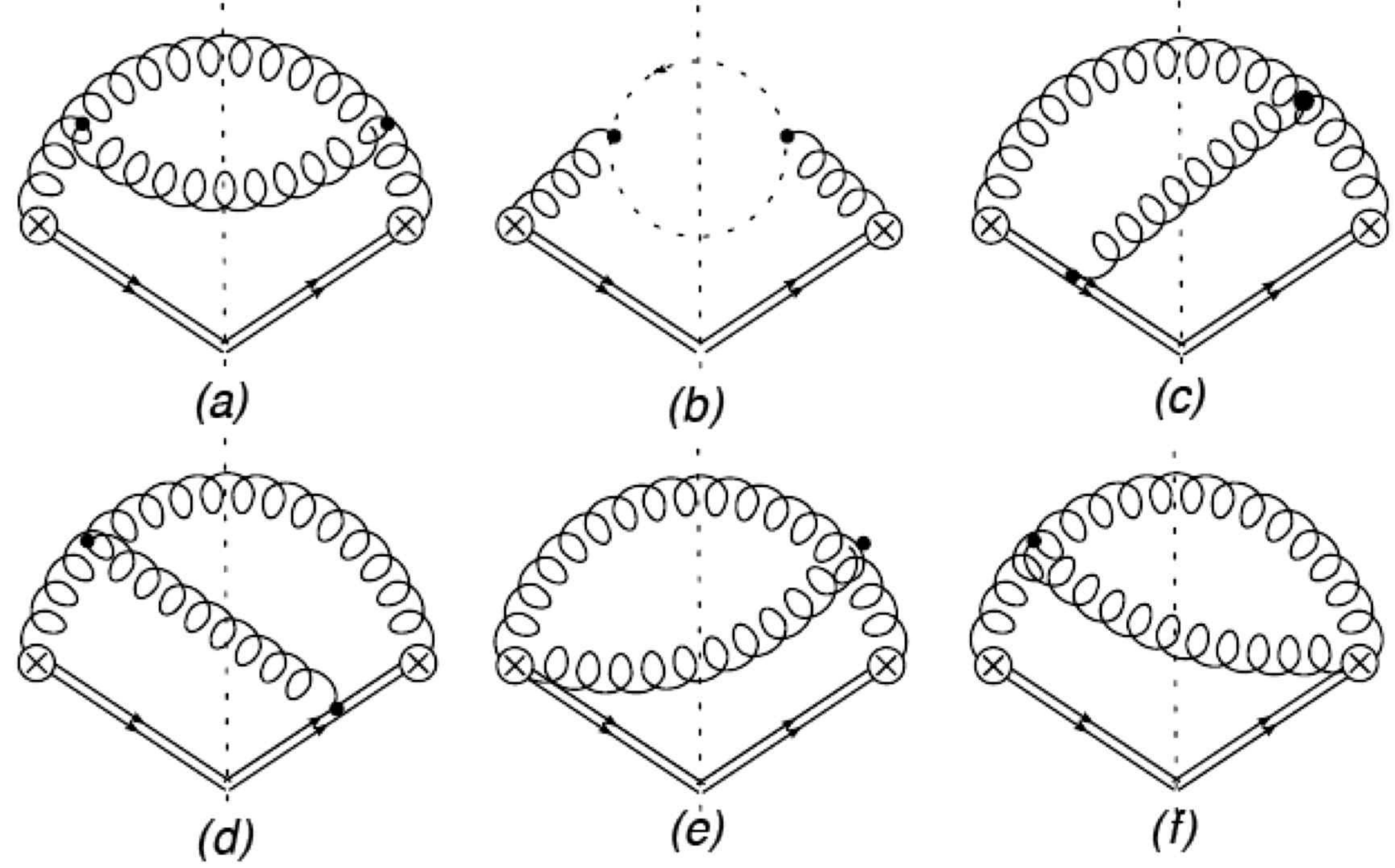}
\caption{Some NLO real corrections to the gluon jet function, where the
dashed line represents a ghost field.}
\label{gluon}
\end{figure}

\begin{figure}
\centering\includegraphics[width=.5\linewidth]{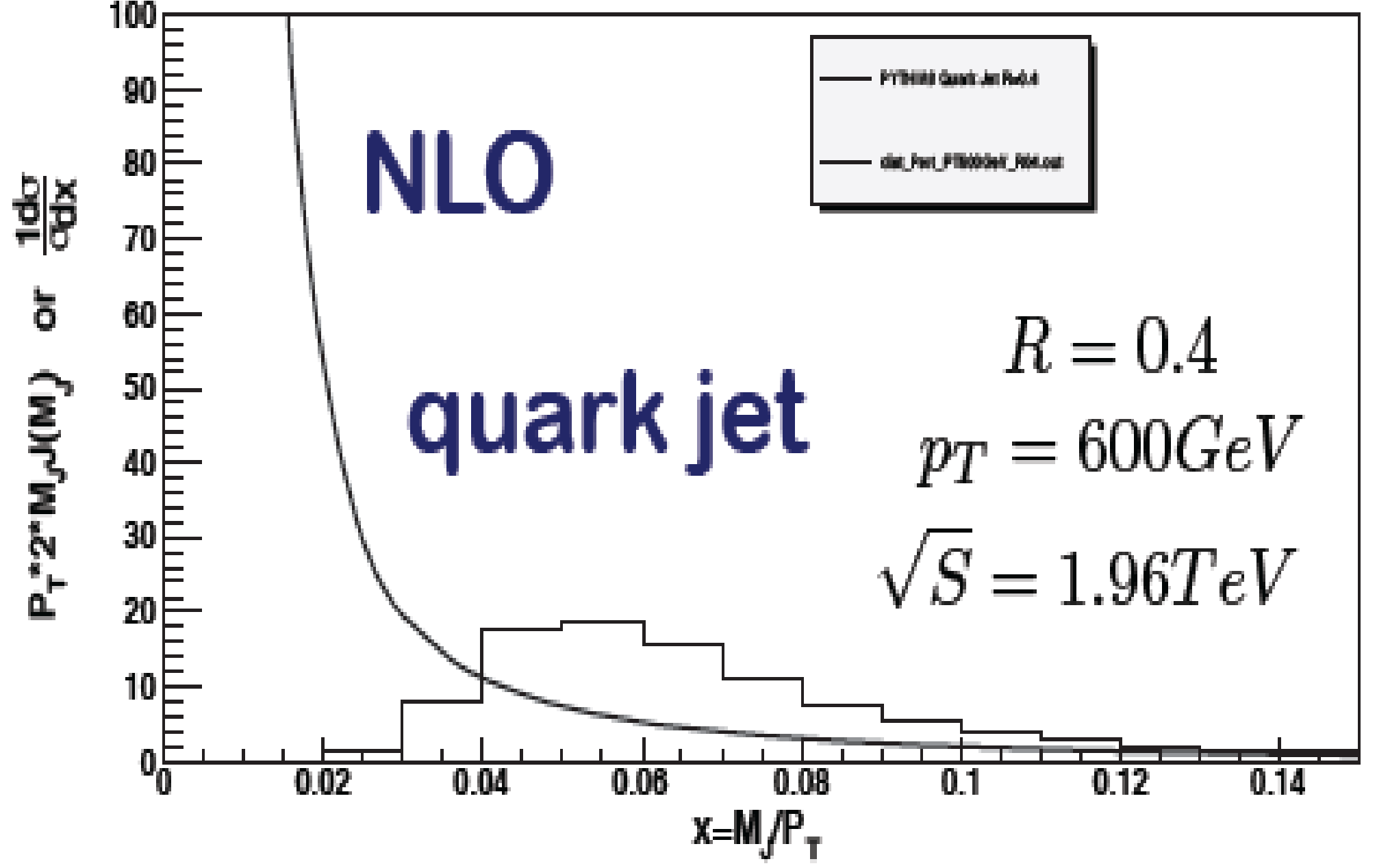}
\caption{Jet mass distribution at NLO.}
\label{jet4}
\end{figure}

The NLO diagrams for the light-quark and gluon jet functions are displayed in 
Figs.~\ref{quark} and \ref{gluon}, respectively. Evaluating the jet functions 
up to NLO, a divergence is observed at small jet invariant mass $M_J$ as 
shown in Fig.~\ref{jet4}, that implies the nonperturbtive nature of the 
jet functions. The total NLO corrections in Mellin space indicate the 
existence of double logarithms, which hint the implementation of the 
resummation technique. Both the angular and energy resolutions are related to 
the jet mass: when $M_J$ is not zero, particles in a jet cannot be completely
collimated, and the jet must have finite minimal energy. This accounts for
the source of the double logarithms. Recall that low $p_T$ spectra of
direct photons, dominated by soft and collinear radiations, are treated
by the $k_T$ resummation. The jet invariant mass is attributed to soft and
collinear radiations, so the mass distribution can also be derived in the
resummation formalism.

\begin{figure}[!htb]
\centering\includegraphics[width=0.7\textwidth]{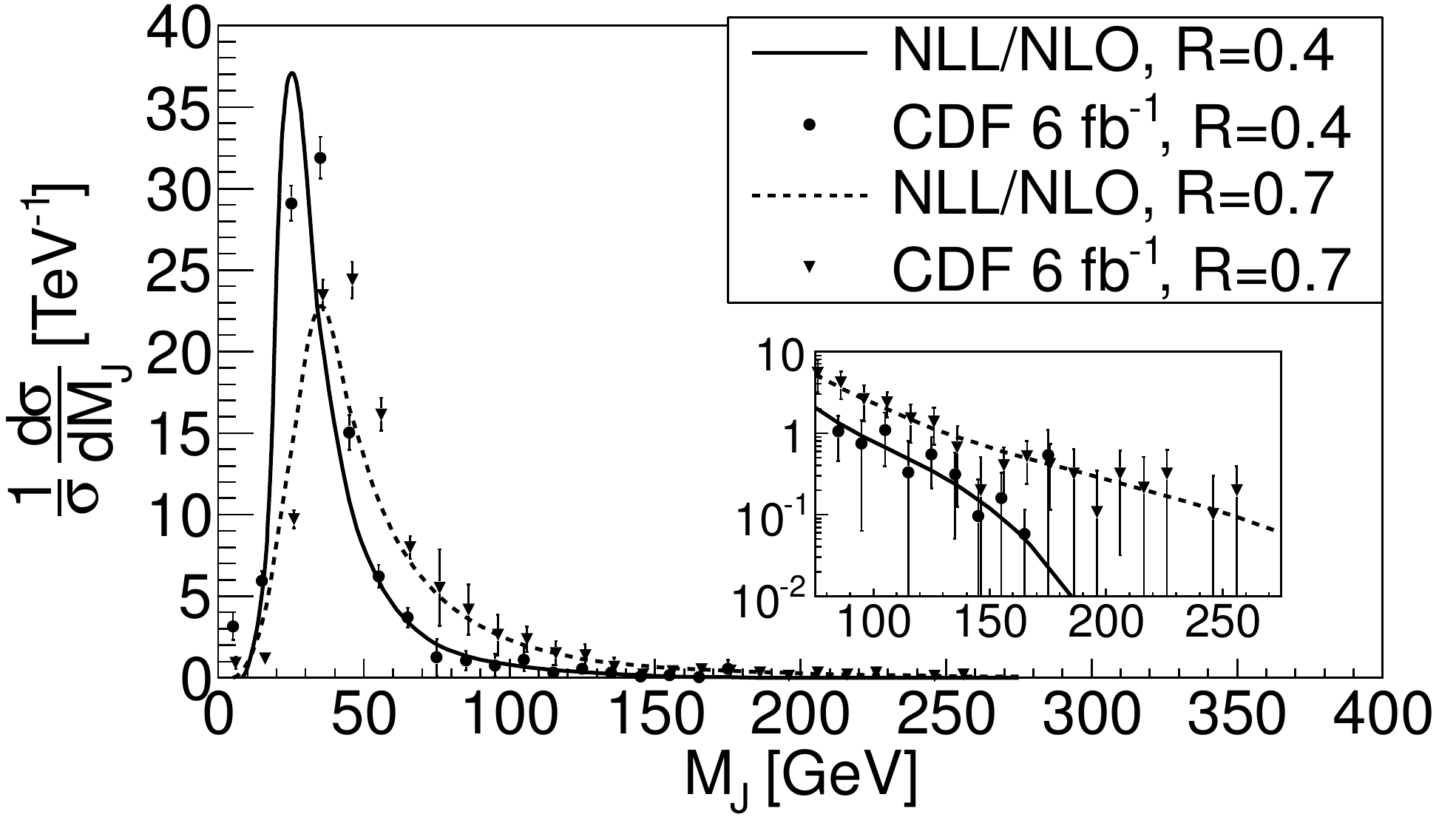}
\caption{Comparison of resummation predictions for the jet mass
distributions to Tevatron CDF data with the kinematic cuts $P_T>400$
GeV and $0.1<|Y|<0.7$ at $R=0.4$ and $R=0.7$. The inset shows the
detailed comparison in large jet mass region.} \label{CONVP1}
\end{figure}

Varying the Wilson line direction $n$, we derive the differential equation for the 
light-quark jet function \cite{Li:2011hy}
\begin{eqnarray}
-\frac{n^2}{v\cdot n}v_{\alpha}\frac{d}{dn_\alpha}{
J}_q(M_J^2,P_T,\nu^2,R,\mu^2)=2(K+G)\otimes {J}_q(M_J^2,P_T,\nu^2,R,\mu^2). \label{cr2}
\end{eqnarray}
The above equation implies that
the soft gluons in $K$ are associated with the jet function $J$, a feature
consistent with the anti-$k_T$ algorithm. The solution to Eq.~(\ref{cr2})
resums the double logarithms in the jet function. One then convolutes the
light-quark and gluon jet functions with the constituent cross sections of 
LO partonic dijet processes at the Tevatron and the PDF CTEQ6L \cite{Pumplin:2002vw}.
The resummation predictions for the jet mass distributions at $R=0.4$ and
$R=0.7$ are compared to the Tevatron CDF data \cite{Aaltonen:2011pg} in
Fig.~\ref{CONVP1} \cite{LLY12} with the kinematic cuts $P_T>400$ GeV and the
rapidity interval $0.1<|Y|<0.7$. The abbreviation NLL refers to the accuracy of 
the resummation, and NLO to the accuracy of the initial condition of the jet 
function solved from Eq.~(\ref{cr2}). The consistency of the resummation
results with the CDF data is satisfactory.

\subsection{Jet Substructure}

It is known that a top quark produced almost at rest at the Tevatron
can be identified by measuring isolated jets from its decay.
However, this strategy does not work for identifying a
highly-boosted top quark produced at the LHC. It has been observed 
that an ordinary high-energy QCD jet
\cite{Skiba:2007fw,Holdom:2007ap} can have an invariant mass close
to the top quark mass. A highly-boosted top quark, producing only 
a single jet, is then difficult to be distinguished from a QCD jet. 
This difficulty also appears in the identification of a 
highly-boosted new-physics resonance decaying into standard-model 
particles, or Higgs boson decaying into a bottom-quark pair. Hence,
additional information needs to be extracted from jet internal
structures in order to improve the jet identification at the LHC.
The quantity, called planar flow \cite{Almeida:2008yp}, has been
proposed for this purpose, which utilizes the geometrical shape of a
jet: a QCD jet with large invariant mass mainly involves one-to-two
splitting, so it leaves a linear energy deposition in a detector. A
top-quark jet, proceeding with a weak decay, mainly involves
one-to-three splitting, so it leaves a planar energy deposition.
Measuring this additional information, it has been shown with event
generators that the top-quark identification can be improved to some
extent. Investigations on various observables associated with jet
substructures are usually done using event generators. For a review on recent
theoretical progress and the latest experimental results in jet
substructures, see \cite{Altheimer:2012mn}.

Here I focus on a jet substructure, called the energy profile,
and explain how to calculate it in the resummation formalism \cite{Li:2011hy}.
This quantity describes the energy fraction accumulated
in the cone of size $r$ within a jet cone $R$, i.e., $r<R$. Its
explicit definition is given by \cite{Acosta:2005ix}
\begin{equation}
\Psi(r)=\frac{1}{N_{J}}\sum_{J}
\frac{\sum_{r_i<r,i\in {J}}P_{Ti}}{\sum_{r_i<R, i\in {J}}P_{Ti}},\label{pro}
\end{equation}
with the normalization $\Psi(R)=1$, where $P_{Ti}$ is the transverse
momentum carried by particle $i$ in the jet $J$, and $r_i<r$
$(r_i<R)$ means the flow of particle $i$ into the jet cone $r$
$(R)$. Different types of jets are expected to exhibit different
energy profiles. For example, a light-quark jet is narrower than a
gluon jet; that is, energy is accumulated faster with $r$ in a
light-quark jet than in a gluon jet. A heavy-particle jet certainly
has a distinct energy profile, which can be used for its identification.
The importance of higher-order corrections and
their resummation for studying a jet energy profile have been first
emphasized in \cite{Seymour:1997kj}. Another approach based on the
soft-collinear effective theory and its application to jet
production at an electron-positron collider can be found in Refs.
\cite{Ellis:2010rwa,Kelley:2011tj,Kelley:2011aa}.

We first define the jet energy
functions $J^E_f(M_J^2,P_T,\nu^2,R,r)$ with $f=q(g)$ denoting the
light-quark (gluon), which describe the energy accumulation
within the cone of size $r<R$. The definition is chosen, such that
$J^{E(0)}_f=P_T\delta(M_J^2)$ at LO. The Feynman rules for $J^E_f$
are similar to those for the jet functions $J_f$ at each order of
$\alpha_s$, except that a sum of the step functions
$\sum_ik_i^0\Theta(r-\theta_i)$  is inserted, where $k_i^0$
($\theta_i$) is the energy (the angle with respect to the jet axis)
of particle $i$. For example, the jet energy
functions $J^E_f$ are expressed, at NLO, as
\begin{eqnarray}
J_q^{E(1)}(M_J^2,P_T,\nu^2,R,r,\mu^2)&=&\frac{(2\pi)^3}
{2\sqrt{2}(P_J^0)^2N_c}\sum_{\sigma,\lambda}
\int\frac{d^3p}{(2\pi)^3 2p^0}\frac{d^3k}{(2\pi)^3 2k^0}\nonumber\\
& &\times [p^0\Theta(r-\theta_p)+k^0\Theta(r-\theta_k)]
\nonumber \\
&& \times{\rm Tr}\left\{\not\xi\langle 0|q(0)W^{(\bar
q)\dagger}|p,\sigma;k,\lambda\rangle \langle
k,\lambda;p,\sigma|W^{(\bar q)} \bar
q(0)|0\rangle\right\}\nonumber\\
& &\times\delta(M_J^2-(p+k)^2)\delta^{(2)}(\hat e-\hat
e_{\bf{p}+\bf{k}})\delta(P_J^0-p^0-k^0), \nonumber \\
J_g^{E(1)}(M_J^2,P_T,\nu^2,R,r,\mu^2)&=&\frac{(2\pi)^3}
{2(P_J^0)^3N_c}\sum_{\sigma,\lambda} \int\frac{d^3p}{(2\pi)^3
2p^0}\frac{d^3k}{(2\pi)^3 2k^0}\nonumber\\
& &\times [p^0\Theta(r-\theta_p)+k^0\Theta(r-\theta_k)]
\nonumber \\
&& \times \langle 0|\xi_\sigma
F^{\sigma\nu}(0)W^{(g)\dagger}
|p,\sigma;k,\lambda\rangle\langle
k,\lambda;p,\sigma|W^{(g)}
F_\nu^\rho(0)\xi_\rho|0\rangle\nonumber\\
& &\times\delta(M_J^2-(p+k)^2)\delta^{(2)}(\hat e-\hat
e_{\bf{p}+\bf{k}})\delta(P_J^0-p^0-k^0),
\label{jetENLO1}
\end{eqnarray}
where the expansion of the Wilson links in $\alpha_s$ is understood.
The factorization scale is set to $\mu=P_T$ to remove the associated
logarithms, so its dependence will be suppressed below.

\begin{figure}[!htb]
\centering\includegraphics[width=0.5\textwidth]{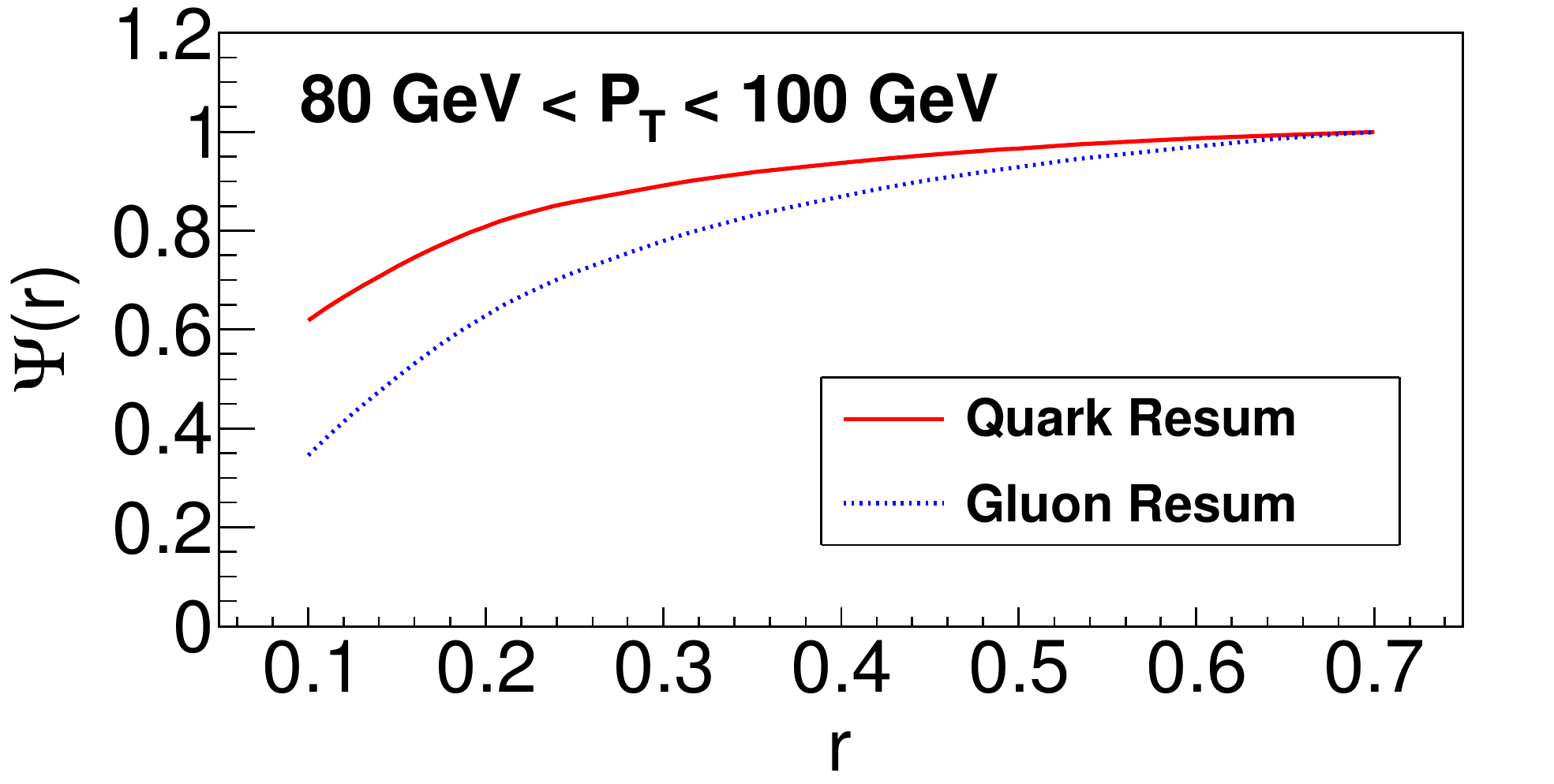}
\caption{Resummation predictions for the energy profiles of the
light-quark (solid curve) and gluon (dotted curve) jets with
$\sqrt{s}=7$ TeV and 80 GeV $< P_T<$ 100 GeV.} \label{COMP}
\end{figure}

The Mellin-transformed jet energy function ${\bar J}_q^E$ obeys a
similar differential equation \cite{Li:2011hy}
\begin{eqnarray}
-\frac{n^2}{v\cdot n}v_{\alpha}\frac{d}{dn_\alpha}{\bar
J}_q^E(N=1,P_T,\nu^2,R,r)=2({\bar K}+ G){\bar
J}_q^E(N=1,P_T,\nu^2,R,r), \label{er2}
\end{eqnarray}
which can be solved simply. Inserting the solutions to Eq.~(\ref{er2})
into Eq.~(\ref{pro}), the jet energy profile is derived. Note that a
jet energy profile with $N=1$ is not sensitive to
the nonperturbative contribution, so the predictions are free of the
nonperturbative parameter dependence, in contrast to the case of
the jet invariant mass distribution. It has been found
that the light-quark jet has a narrower energy profile than the
gluon jet, as exhibited in Fig.~\ref{COMP} for $\sqrt{s}=7$ TeV and
the interval $80$ GeV $< P_T<$ $100$ GeV of the jet transverse
momentum. The broader distribution of the gluon jet results from
stronger radiations caused by the larger color factor $C_A=3$,
compared to $C_F=4/3$ for a light-quark jet.

\begin{figure}[!htb]
\includegraphics[width=0.32\textwidth]{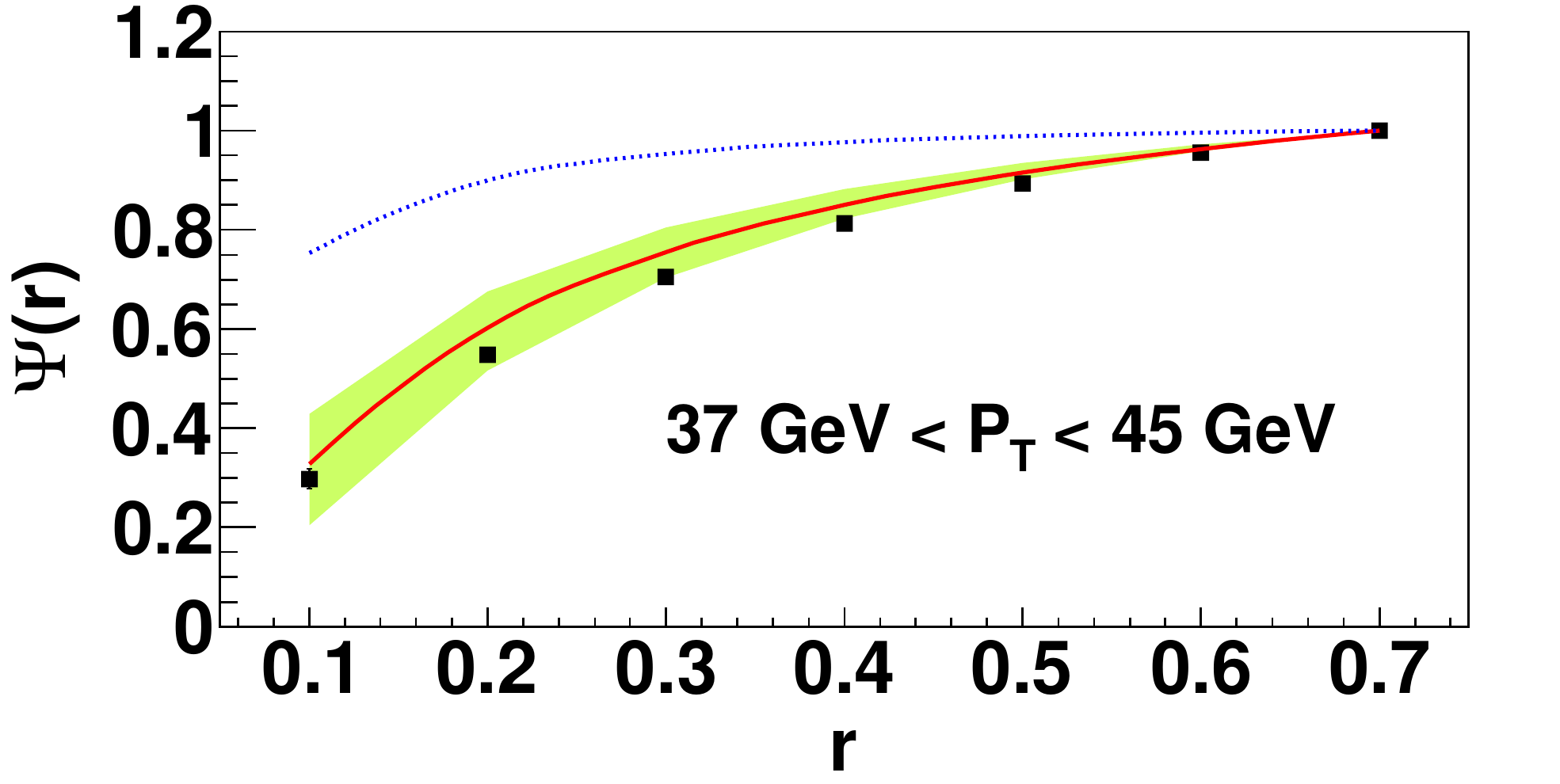}
\includegraphics[width=0.32\textwidth]{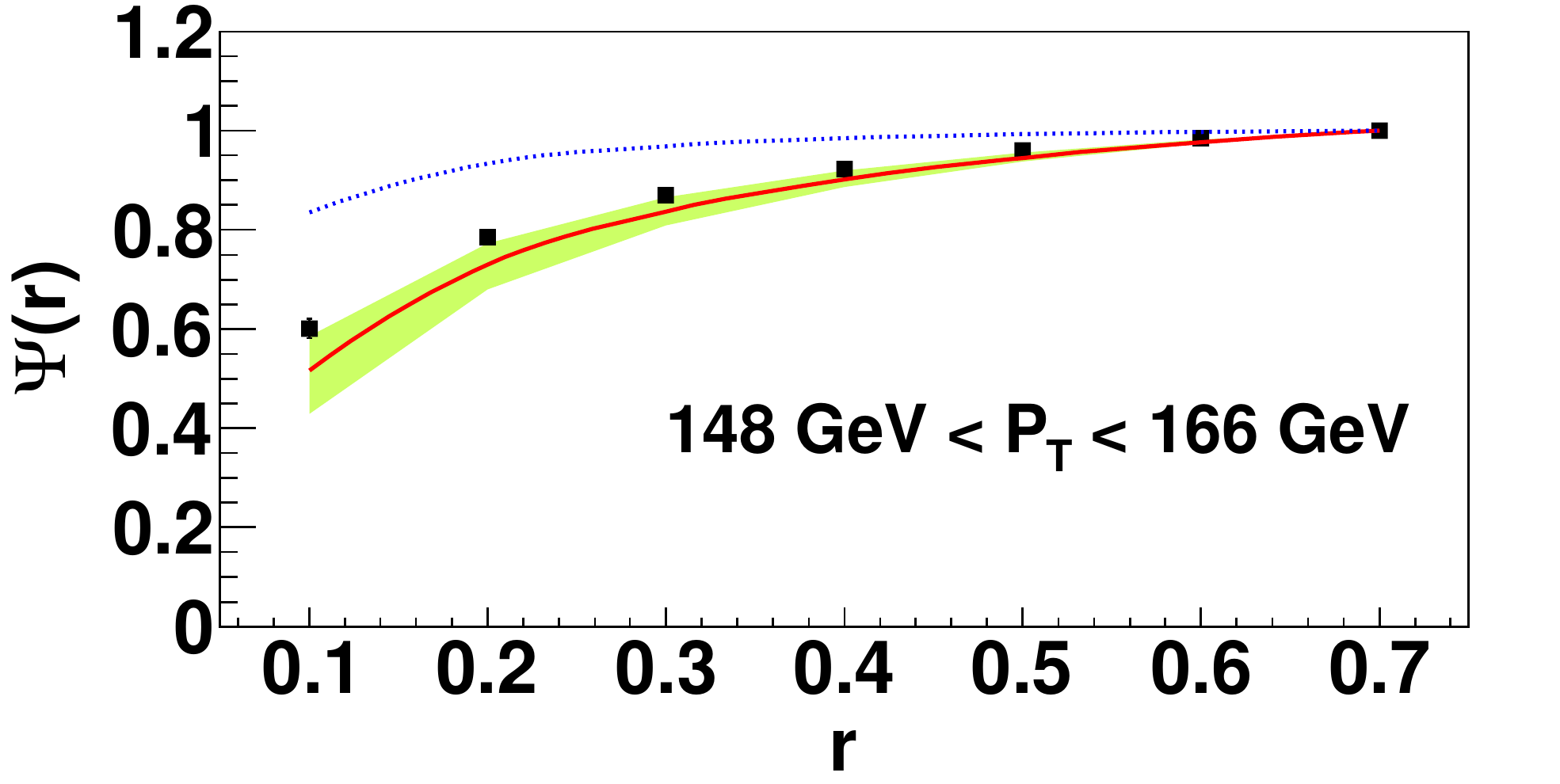}
\includegraphics[width=0.32\textwidth]{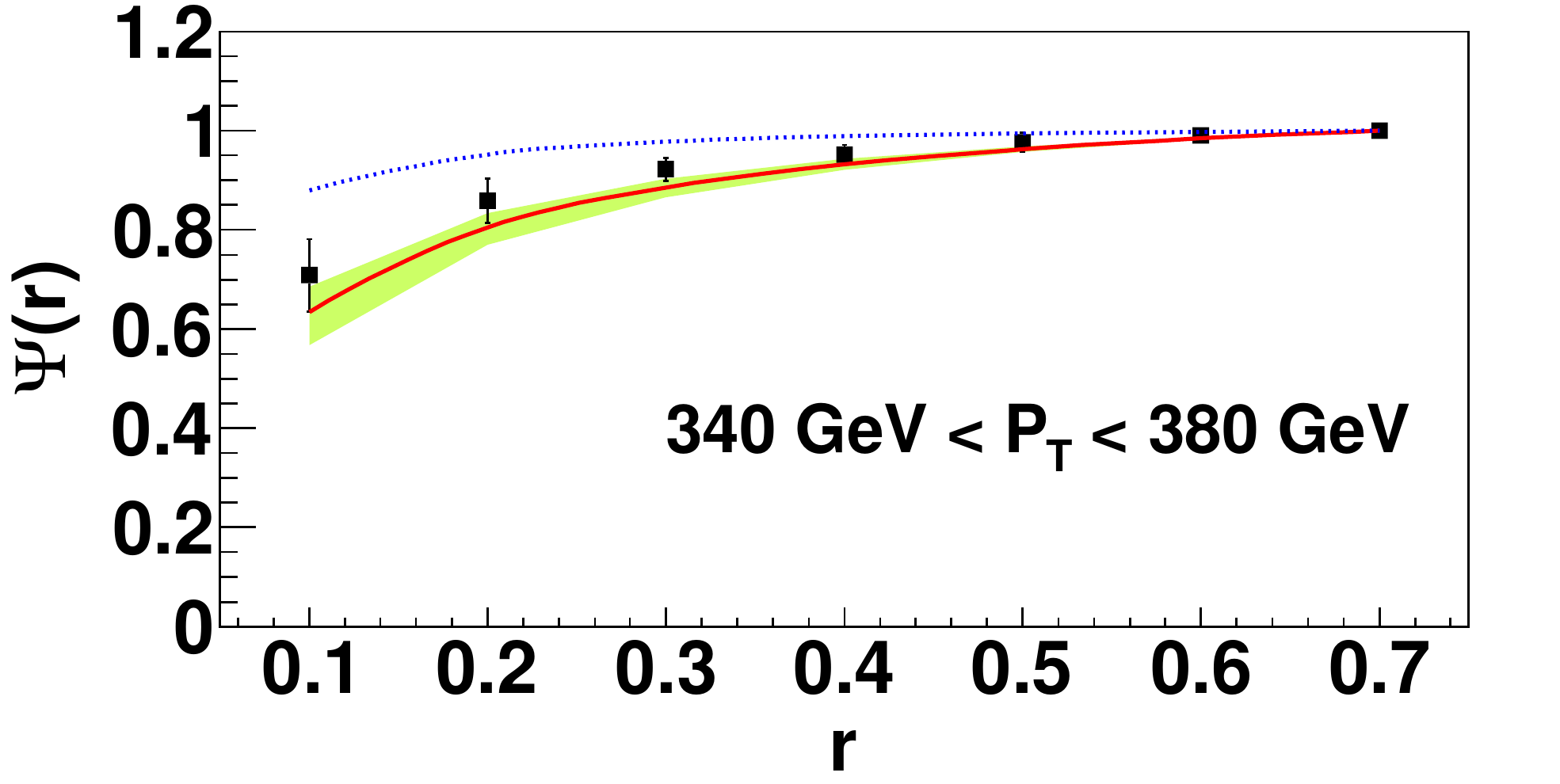}
\caption{Comparison of resummation predictions for the jet energy
profiles with $R=0.7$ to Tevatron CDF data in various $P_T$ intervals. The NLO
predictions denoted by the dotted curves are also displayed.}
\label{CDFJE}
\end{figure}

\begin{figure}[!htb]
\includegraphics[width=0.32\textwidth]{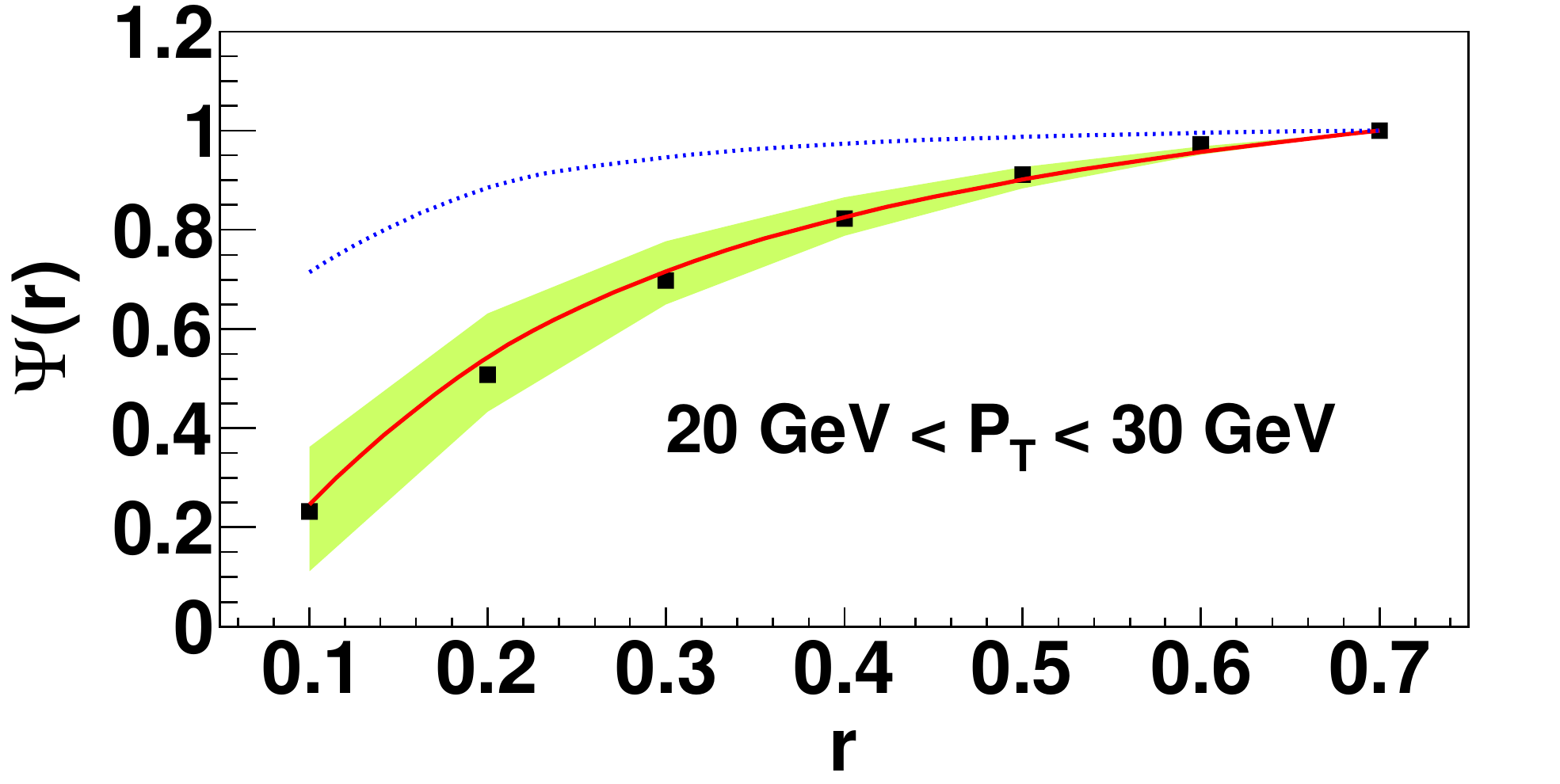}
\includegraphics[width=0.32\textwidth]{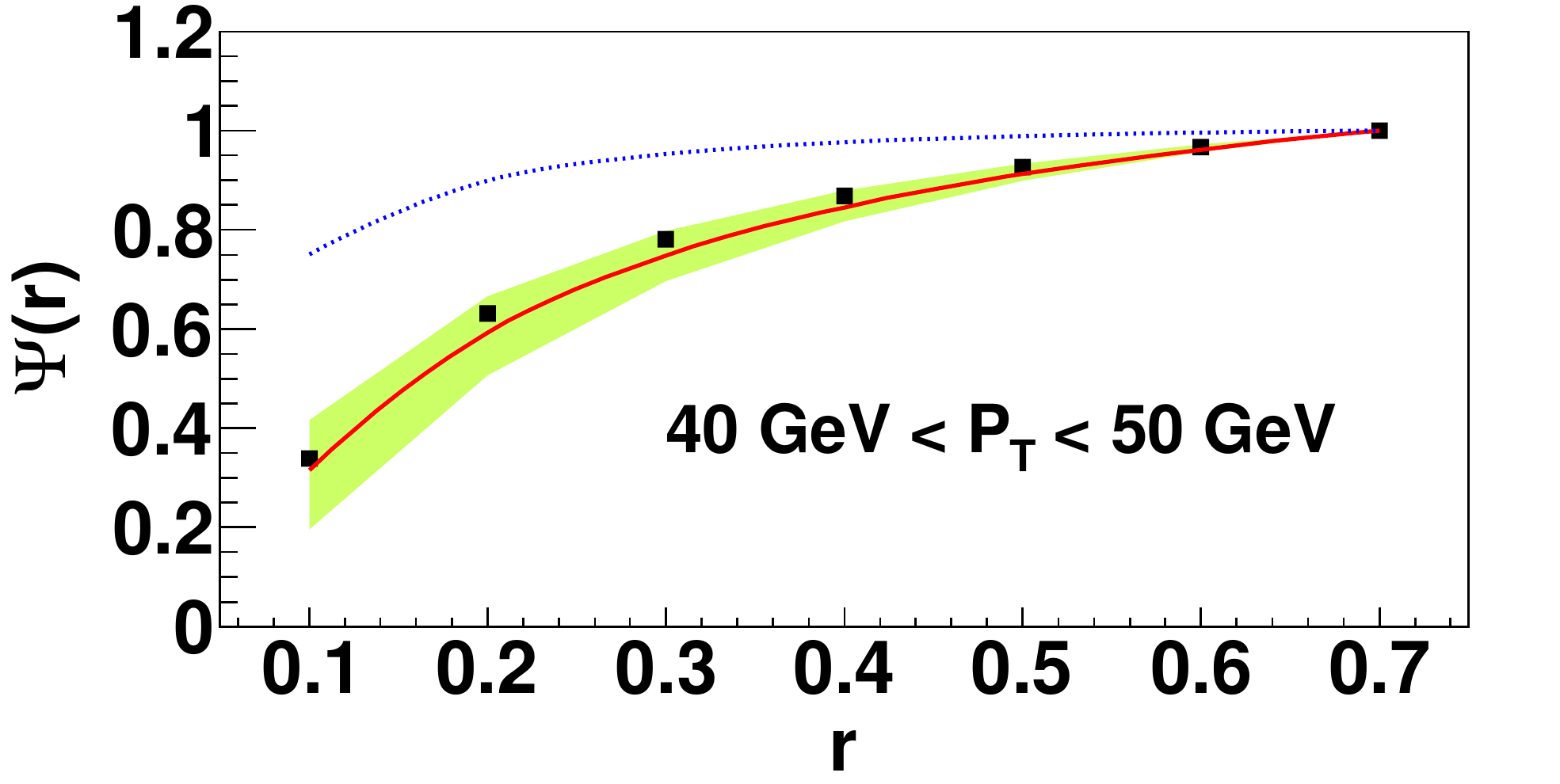}
\includegraphics[width=0.32\textwidth]{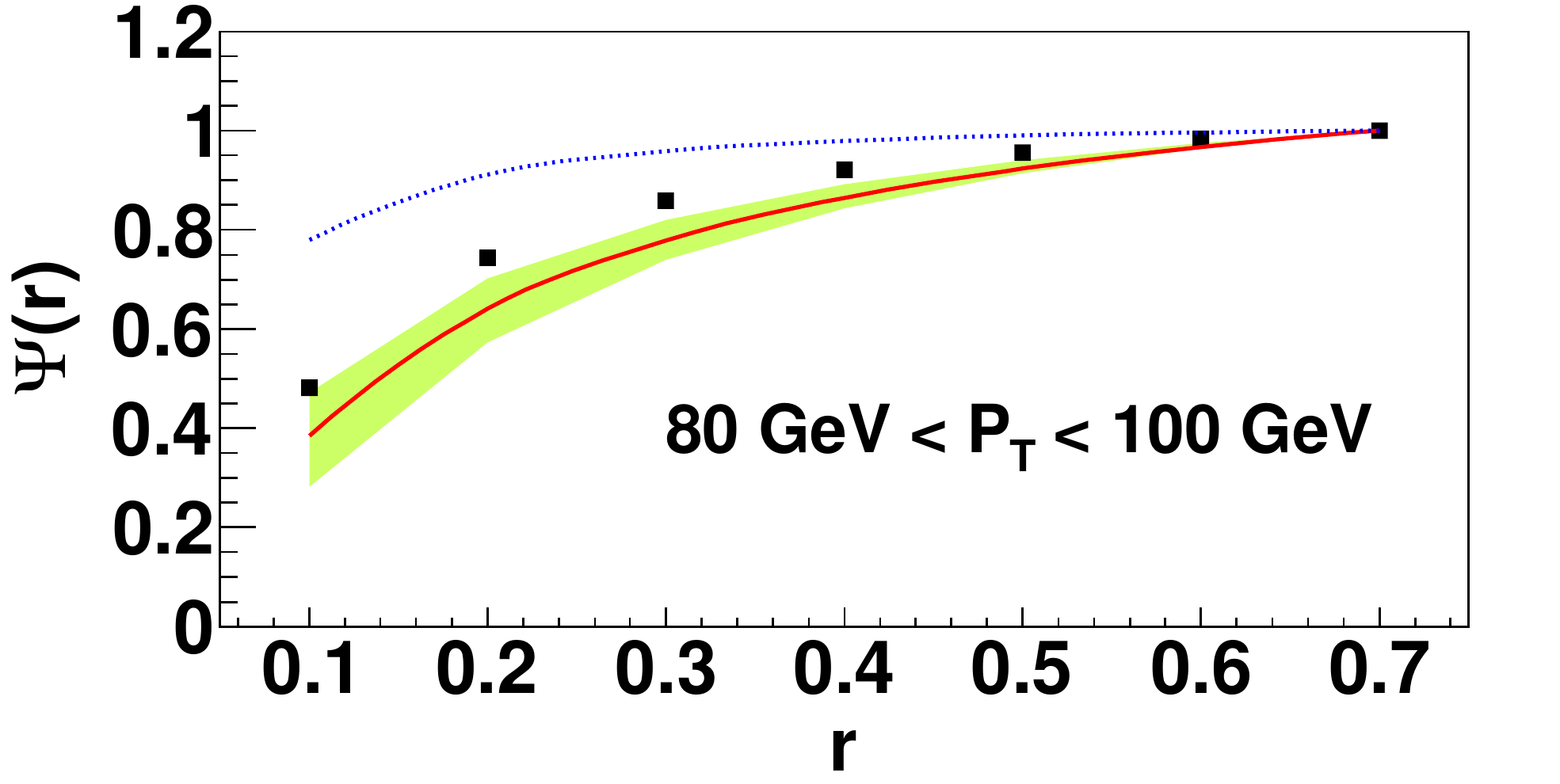}
\caption{Resummation predictions for the
jet energy profiles with $R=0.7$ compared to LHC CMS data
in various $P_T$ intervals. The NLO
predictions denoted by the dotted curves are also displayed.}
\label{CMSJE}
\end{figure}

One then convolutes the light-quark and gluon jet energy functions
with the constituent cross sections of the LO partonic subprocess
and CTEQ6L PDFs \cite{Pumplin:2002vw} at certain collider energy.
The predictions are directly compared with
the Tevatron CDF data \cite{Acosta:2005ix} as shown in Fig.~\ref{CDFJE}.
It is evident that the resummation predictions agree well
with the data in all $P_T$ intervals. The NLO predictions derived from 
$\bar J_f^{E(1)}(1,P_T,\nu_{\rm fi}^2, R, r)$ are also displayed for
comparison, which obviously overshoot the data. The resummation
predictions for the jet energy profiles are compared with the LHC
CMS data at 7 TeV \cite{CMSJE} from the anti-$k_T$ jet algorithm
\cite{Cacciari:2008gp} in Fig.~\ref{CMSJE}, which are also
consistent with the data in various $P_T$ intervals. Since one can
separate the contributions from the light-quark jet and the gluon
jet, the comparison with the CDF and CMS data implies that
high-energy (low-energy) jets are mainly composed of the light-quark
(gluon) jets. Hence, a precise measurement of the jet energy profile 
as a function of jet transverse momentum can be used to experimentally 
discriminate the production mechanism of jets in association with other 
particles, such as electroweak gauge bosons, top quarks and Higgs bosons.

\section{Hadronic heavy-quark decays}

Hadronic decays of heavy-quark bound states, such as $B$, $B_s$,
and $\Lambda_b$, are one of the focuses of LHCb physics, whose precision
measurement may reveal new physics in the flavor sector.
They are difficult to analyze theoretically because of
complicated QCD dynamics and multiple characteristic scales they
involve: the $W$ boson mass $m_W$, the $b$ quark mass $m_b$, and the
QCD scale $\Lambda_{\rm QCD}$. The standard procedure is first to
integrate out the scale $m_W$, such that QCD dynamics is organized
into an effective weak Hamiltonian \cite{REVIEW}. For the $B\to
D\pi$ decays, the effective Hamiltonian is written as
\begin{eqnarray}
{\cal H}_{\rm eff} = {G_F\over\sqrt{2}}\, V_{cb}V_{ud}^*
\Big[C_1(\mu)O_1(\mu)+C_2(\mu)O_2(\mu)\Big],
\end{eqnarray}
where $G_F$ is the Fermi coupling constant, $V_{cb}V_{ud}^*$ is the
product of the Cabibbo-Kobayashi-Maskawa matrix elements, $\mu$ is
the renormalization scale, $C_{1,2}$ are the Wilson coefficients,
and the four-fermion operators are defined by
\begin{eqnarray}
O_1 = (\bar db)_{V-A}(\bar cu)_{V-A}\;,\qquad\qquad O_2= (\bar
cb)_{V-A}(\bar du)_{V-A}.
\end{eqnarray}

For exclusive processes,
such as hadron form factors, the collinear factorization was developed in
\cite{BL,ER,CZS,CZ}. The range of a parton momentum fraction $x$,
contrary to that in the inclusive case, is not experimentally controllable,
and must be integrated over between 0 and 1. Hence, the end-point region
with a small $x$ is not avoidable. If there is no end-point singularity
developed in a hard kernel, the collinear factorization works. If such a
singularity occurs, indicating the breakdown of the collinear factorization,
the $k_T$ factorization should be employed, because the parton transverse
momentum $k_T$ is not negligible. To derive $B\to D\pi$ decay amplitudes,
one evaluates the hadronic matrix elements $\langle D\pi|O_i(\mu)|B\rangle$.
Different theoretical approaches have been developed for this purpose,
which include the factorization assumption, the QCD-improved
factorization, the perturbative QCD, the soft-collinear effective
theory, the light-cone QCD sum rules, and the quark-diagram
parametrization. In this section I briefly introduce the basic ideas of
the first three approaches \cite{L03}.

\subsection{Factorization Assumption}

Intuitively, decay products from a heavy $b$ quark move fast without
further interaction between them. This naive picture is supported by
the color-transparency argument \cite{transparency}: the Lorentz
contraction renders energetic final states emitted from the weak
vertex have small longitudinal color dipoles, which cannot be
resolved by soft gluons. Therefore, the hadronic matrix element
$\langle O(\mu)\rangle$ is factorized into a product of two matrix
elements of single currents, governed by decay constants and form
factors, without soft gluon exchanges between them. This
factorization assumption (FA) \cite{BSW} was first proved in the
framework of large energy effective theory \cite{leet}, and
justified in the large $N_c$ limit \cite{largeN}.
For the $B\to D\pi$ decays, the color-allowed
(color-suppressed) amplitude, involving the $B\to D$ ($B\to\pi$)
transition form factor, is proportional to the Wilson coefficient
$a_1=C_2+C_1/N_c$ ($a_2=C_1+C_2/N_c$).

In spite of its simplicity, the FA encounters three principal
difficulties. First, a hadronic matrix element under the FA is
independent of the renormalization scale $\mu$, as the vector or
axial-vector current is partially conserved. Consequently, the
amplitude $C(\mu)\langle O\rangle_{\rm fact}$ is not truly physical
as the scale dependence of the Wilson coefficient does not get
compensation from the matrix element. This problem may not be
serious for color-allowed modes, because the parameter $a_1$ is
roughly independent of $\mu$. It is then not a surprise that the
simple FA gives predictions in relatively good agreement with data
of these modes. However, the parameter $a_2$ depends strongly on the
renormalization scale and on the renormalization scheme, because of
the similar magnitude and different sign of the $C_1(\mu)$ and
$C_2(\mu)/N_c$ terms (calculated in the NDR scheme and for
$\Lambda_{\overline{MS}}^{(5)} = 225$ GeV, the Wilson coefficients
have the values $C_1(m_B) = -0.185$ and $C_2(m_B) = 1.082$
\cite{REVIEW}, $m_B$ being the $B$ meson mass). This may be the
reason why the FA fails to accommodate data of color-suppressed modes.
It also means that $a_2$ is more sensitive to subleading
contributions.

The second difficulty is related to the first one: nonfactorizable
effects have been neglected in the FA. This neglect may be justified
for color-allowed modes due to the large and roughly
$\mu$-independent value of $a_1$, but not for color-suppressed
modes, such as $B\to J/\psi K^{(*)}$. The $J/\psi$ meson emitted
from the weak vertex is not energetic, and the color-transparency
argument does not apply. To circumvent this difficulty,
nonfactorizable contributions were parameterized into the parameters
$\chi_i$ \cite{Cheng94,Soares},
\begin{eqnarray}
a_1^{\rm eff}& =& C_2(\mu) + C_1(\mu) \left[{1\over N_c}
+\chi_1(\mu)\right],
\nonumber\\
a_2^{\rm eff}& =& C_1(\mu) + C_2(\mu)\left[{1\over N_c} +
\chi_2(\mu)\right].
\end{eqnarray}
The $\mu$ dependence of the Wilson coefficients is assumed to be
exactly compensated by that of $\chi_i(\mu)$ \cite{NRSX}. It is
obvious that the introduction of $\chi_i$ does not really resolve
the scale problem in the FA.

Third, strong phases are essential for predicting CP asymmetries in
exclusive $B$ meson decays. These phases, arising from the
Bander-Silverman-Soni (BSS) mechanism \cite{BSS}, are ambiguous in
the FA: the charm quark loop contributes an imaginary piece
proportional to
\begin{equation}
\int du u(1-u)\theta(q^2 u(1-u)-m_c^2), \label{stp}
\end{equation}
where $q^2$ is the invariant mass of the gluon attaching to the charm
loop. Since $q^2$ is not precisely defined in the FA, one cannot
obtain definite information of strong phases from Eq.~(\ref{stp}).
Moreover, it is legitimate to question whether the BSS mechanism is
an important source of strong phases in $B$ meson decays. Viewing
the above difficulties, the FA is not a complete model, and it is
necessary to go beyond the FA by developing reliable and systematic
theoretical approaches.

\subsection{QCD-improved Factorization}

\begin{figure}
\begin{center}
\centering\includegraphics[width=.6\linewidth]{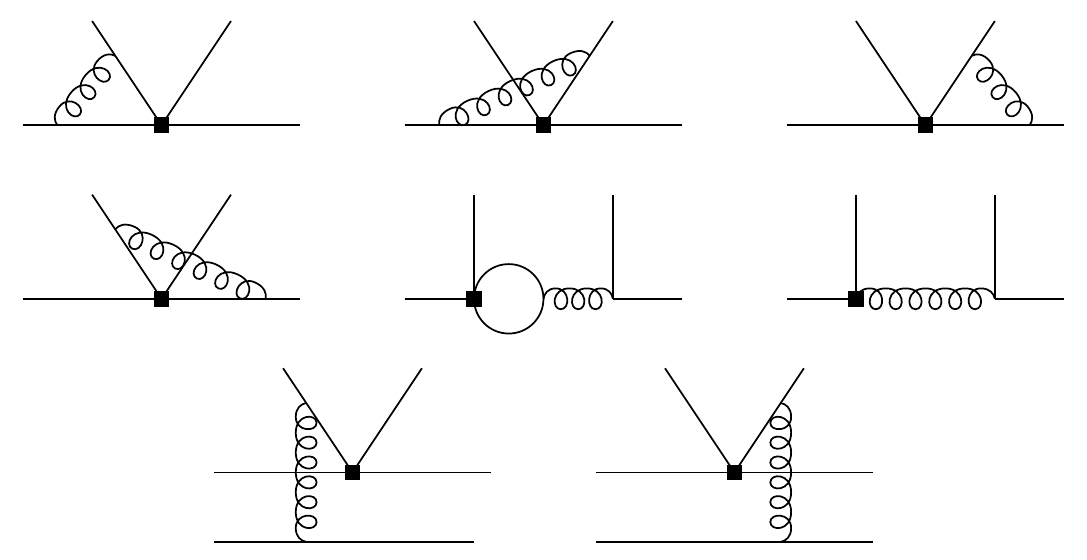}
\end{center}
\caption{$O(\alpha_s)$ corrections to the FA in the QCDF approach.}
\label{fig1}
\end{figure}

The color-transparency argument allows the addition of hard gluons
between the energetic mesons emitted from the weak vertex and the
$B$ meson transition form factors. These hard gluon exchanges lead
to higher-order corrections in the coupling constant $\alpha_s$ to
the FA. By means of Feynman diagrams, they appear as the vertex
corrections in the first two rows of Fig.~\ref{fig1} \cite{BBNS}. It
has been shown that soft divergences cancel among them, when
computed in the collinear factorization theorem. These $O(\alpha_s)$
corrections weaken the $\mu$ dependence in the Wilson coefficients, and
generate strong phases. Besides, hard gluons can also be added to
form the spectator diagrams in the last row of Fig.~\ref{fig1}.
Feynman rules of these two diagrams differ by a minus sign in the
soft region resulting from the involved quark and anti-quark
propagators. Including the above nonfactorizable corrections to the
FA leads to the QCD-improved factorization (QCDF) approach
\cite{BBNS}. The gluon invariant mass $q^2$ in the BSS mechanism can
be unambiguously defined and related to parton momentum fractions in
QCDF. Hence, the theoretical difficulties in the FA are resolved.
This is a breakthrough towards a rigorous framework for two-body
hadronic $B$ meson decays in the heavy quark limit.

\begin{figure}
\begin{center}
\includegraphics[width=.9\textwidth]{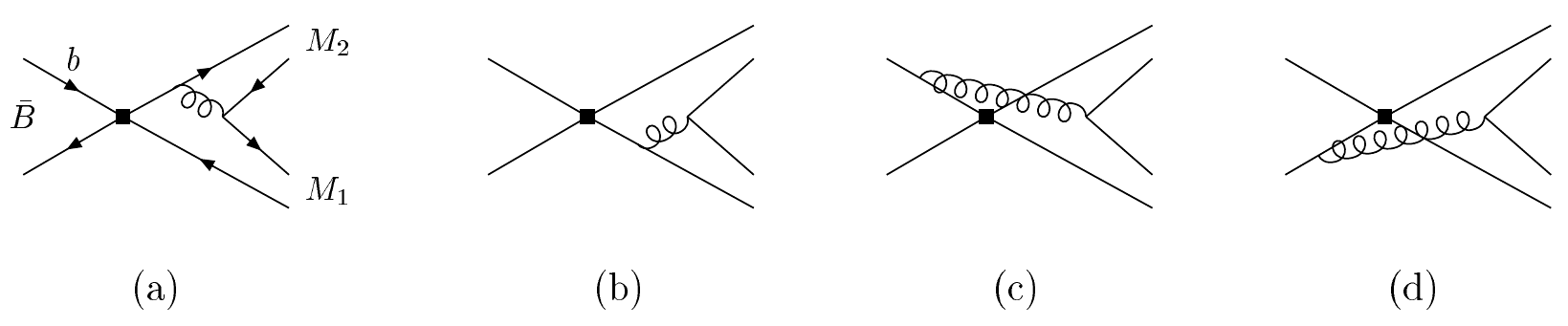}
\end{center}
\caption{Annihilation contributions.} \label{fig2}
\end{figure}

Corrections in higher powers of $1/m_b$ to the FA can also be
included into QCDF, such as those from the annihilation topology in
Fig.~\ref{fig2}, and from twist-3 contributions to the spectator
amplitudes. However, it has been found that endpoint singularities
exist in these high-power contributions, which arise from the
divergent integral $\int_0^1 dx/x$, $x$ being a momentum fraction.
These singularities have the same origin as those in the collinear
collinear factorization formulas for $B$ meson transition form factors \cite{SHB}.
Because of the endpoint singularities, the annihilation and twist-3
spectator contributions must be parameterized as \cite{BBNS}
\begin{eqnarray}
\ln\frac{m_B}{\Lambda_h}\left(1+\rho_Ae^{i\delta_A}\right)\;,\;\;\;\;
\ln\frac{m_B}{\Lambda_h}\left(1+\rho_He^{i\delta_H}\right)\;,
\label{rhoa}
\end{eqnarray}
respectively, with the hadronic scale $\Lambda_h$. A QCDF formula
then contains the arbitrary parameters $\rho_{A,H}$ and
$\delta_{A,H}$. Setting these parameters to zero, one obtains
predictions in the ``default'' scenario, and the variation of the
arbitrary parameters gives theoretical uncertainties. If tuning
these parameters to fit data, one obtains results in the scenarios
``S'', ``S2'',...\cite{BN03}.

\subsection{Perturbative QCD}

\begin{figure}
\begin{center}
\includegraphics[width=.6\textwidth]{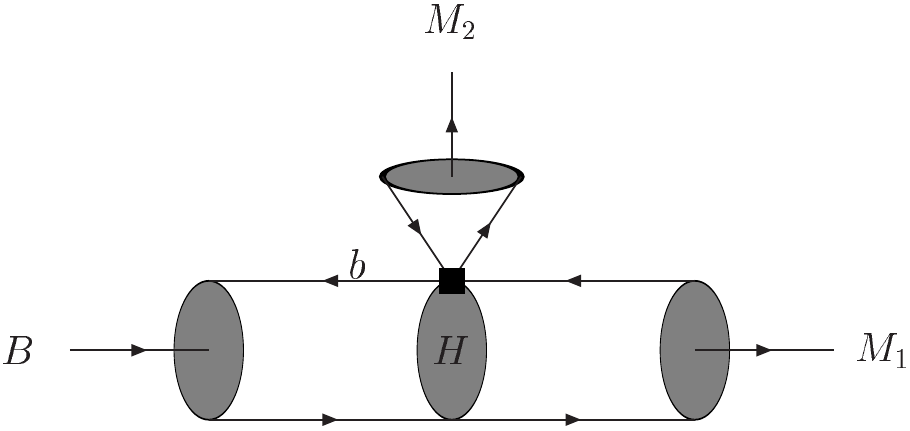}
\end{center}
\caption{Perturbative QCD factorization.} \label{fig3}
\end{figure}

The endpoint singularities signal the breakdown of the collinear
factorization for two-body hadronic $B$ meson decays. Motivated by
the removal of these singularities, the perturbative QCD (PQCD) approach
based on the $k_T$ factorization theorem was developed
\cite{LY1,CL,KLS,LUY}. A parton transverse momentum $k_T$ can be
generated by gluon radiations, before hard scattering occurs. The
endpoint singularities from the small $x$ region simply indicate that
$k_T$ is not negligible. Taking into
account $k_T$, a particle propagator does not diverge as $x\to 0$.
The $B$ meson transition form factors, and the spectator and
annihilation contributions are then all calculable in the framework
of the $k_T$ factorization theorem. It has been shown that a $B\to M_1M_2$
decay amplitude is factorized into the convolution of the six-quark
hard kernel, the jet function and the Sudakov factor with the
bound-state wave functions as shown in Fig.~\ref{fig3},
\begin{eqnarray}
A(B\to M_1M_2)=\phi_B\otimes H\otimes J\otimes S
\otimes\phi_{M_1}\otimes \phi_{M_2}. \label{six}
\end{eqnarray}
The jet function $J$ comes from the threshold resummation, which
exhibits suppression in the small $x$ region \cite{THRE}. The
Sudakov factor $S$ comes from the $k_T$ resummation, which exhibits
suppression in the small $k_T$ region \cite{BS,LS}.
These resummation effects guarantee the removal of the
endpoint singularities. $J$ ($S$), organizing double logarithms in
the hard kernel (meson wave functions), is hidden in $H$ (the three
meson states) in Fig.~\ref{fig3}. The arbitrary parameters
introduced in QCDF are not necessary, and PQCD involves
only universal and controllable inputs.

The theoretical difficulties in the FA are also resolved in PQCD but
in a different manner. The FA limit of the PQCD approach at large
$m_b$, which is not as obvious as in QCDF, has been examined
\cite{THRE}. It was found that the factorizable emission amplitude
decreases like $m_b^{-3/2}$, if the $B$ meson decay constant $f_B$
scales like $f_B\propto m_b^{-1/2}$. This power-law behavior is
consistent with that obtained in \cite{BBNS,Chernyak:1990ag}. The
higher-order corrections to the FA have been included in PQCD, which
moderate the dependence on the renormalization scale $\mu$. The
ratio of the spectator contribution over the factorizable emission
contribution decreases with $m_b$ in PQCD, showing a behavior close
to that in QCDF. The gluon invariant mass $q^2$ in the BSS mechanism
is unambiguously defined and related to parton momentum fractions. The
penguin annihilation amplitude is almost imaginary in PQCD
\cite{KLS}, whose mechanism is similar to the BSS one \cite{BSS}: in
the annihilation topology, the loop is formed by the internal
particles in the LO hard kernel and by infinitely many Sudakov
gluons exchanged between two partons in a light meson. A sizable
strong phase is generated, when the internal particles go on mass
shell. In terms of the principal-value prescription for the internal
particle propagator, the strong phase is given by \cite{KLS}
\begin{eqnarray}
\frac{1}{xm_B^2-k_T^2+i\epsilon}=\frac{P}{xm_B^2-k_T^2}
-i\pi\delta(xm_B^2-k_T^2).
\end{eqnarray}

\subsection{Soft-Collinear Effective Theory}

\begin{figure}
\begin{center}
\includegraphics[width=.3\textwidth]{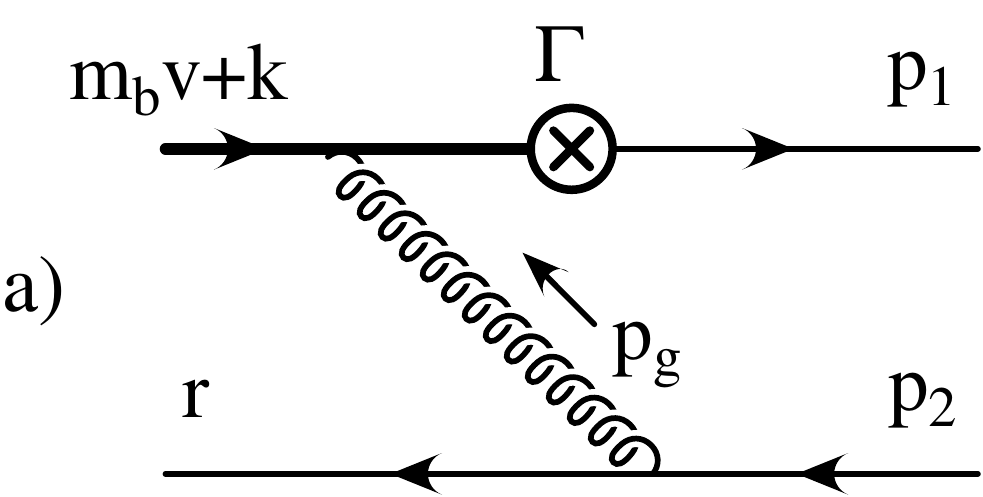}\hspace{1.0cm}
\includegraphics[width=.3\textwidth]{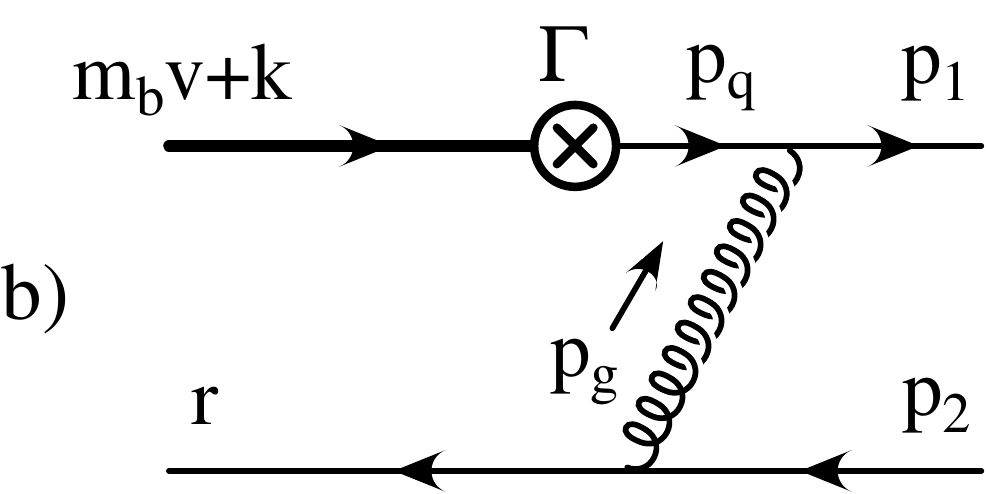}
\end{center}
\caption{Diagrams for the $B\to\pi$ form factor in QCD.}
\label{fig4}
\end{figure}

The soft-collinear effective theory (SCET) based on the collinear
factorization is formulated in the framework of OPE 
\cite{bfl,bfps,cbis,bpssoft}. The matching at
different scales involved in $B$ meson decays has been carefully
handled in SCET. Take the simple $B\to\pi$ transition form factor in
Fig.~\ref{fig4} as an example. The soft spectator in the $B$ meson
carries the momentum $r\sim O(\Lambda_{\rm QCD})$, because it is
dominated by soft dynamics. If the spectator in the energetic pion
carries the momentum $p_2 \sim O(m_b)$, the virtual gluon in
Fig.~\ref{fig4} is off-shell by $p_g^2=(p_2-r)^2=-2p_2\cdot r \sim
O(m_b\Lambda_{\rm QCD})$. Then the virtual quark in
Figs.~\ref{fig4}(a) is off-shell by $(m_bv+k+p_g)^2-m_b^2\sim
O(m_b^2)$, where $v$ is the $b$ quark velocity and $k\sim
O(\Lambda_{\rm QCD})$ denotes the Fermi motion of the $b$ quark.
Hence, $B$ meson decays contain three scales below $m_W$: $m_b$,
$\sqrt{m_b\Lambda_{\rm QCD}}$, and $\Lambda_{\rm QCD}$.

\begin{figure}
\begin{center}
\includegraphics[width=.3\textwidth]{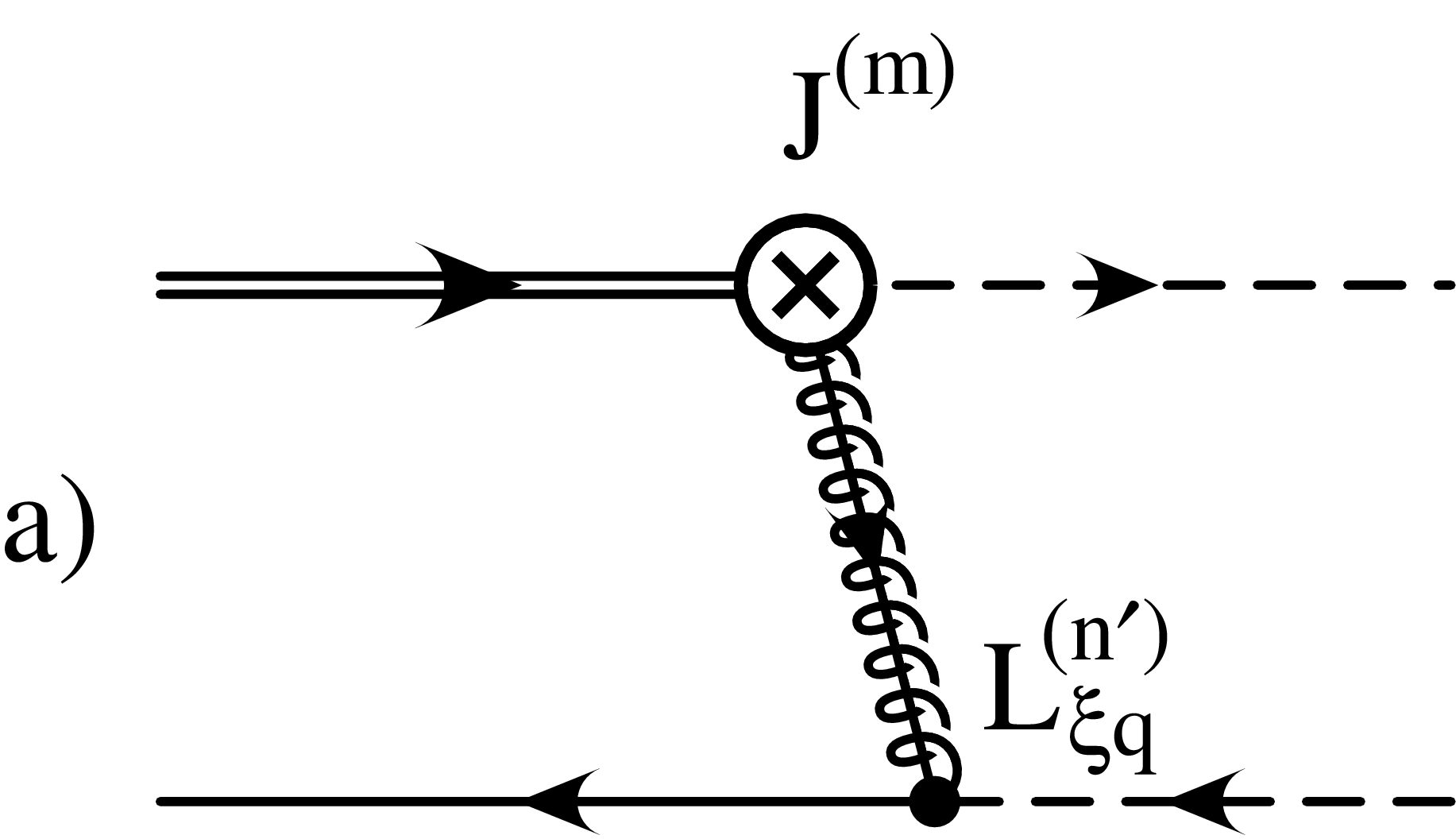}\hspace{1.0cm}
\includegraphics[width=.3\textwidth]{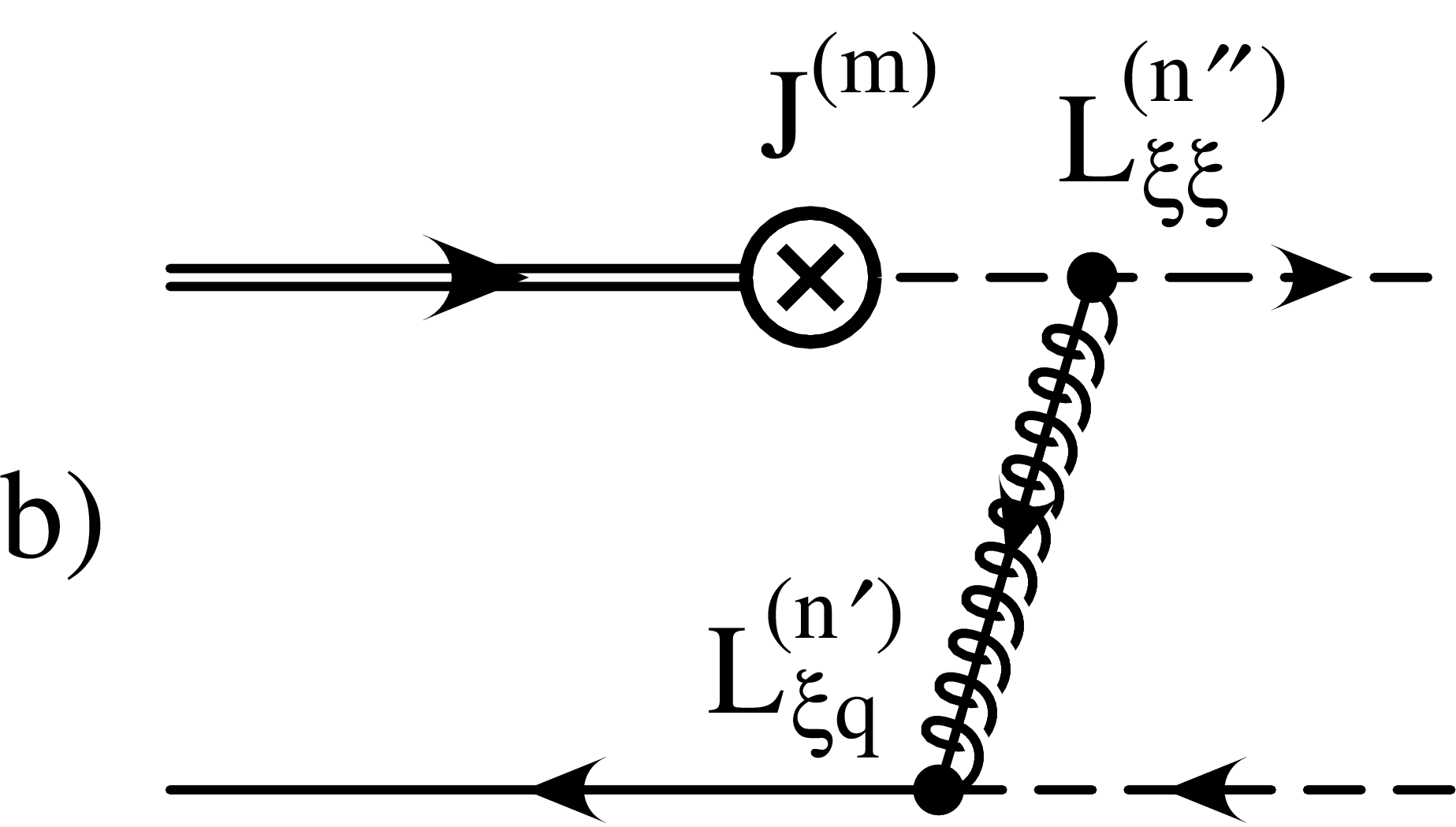}
\end{center}
\caption{Diagrams for the $B\to\pi$ form factor in SCET$_{\rm I}$.}
\label{fig5}
\end{figure}

\begin{figure}
\begin{center}
\includegraphics[width=.35\textwidth]{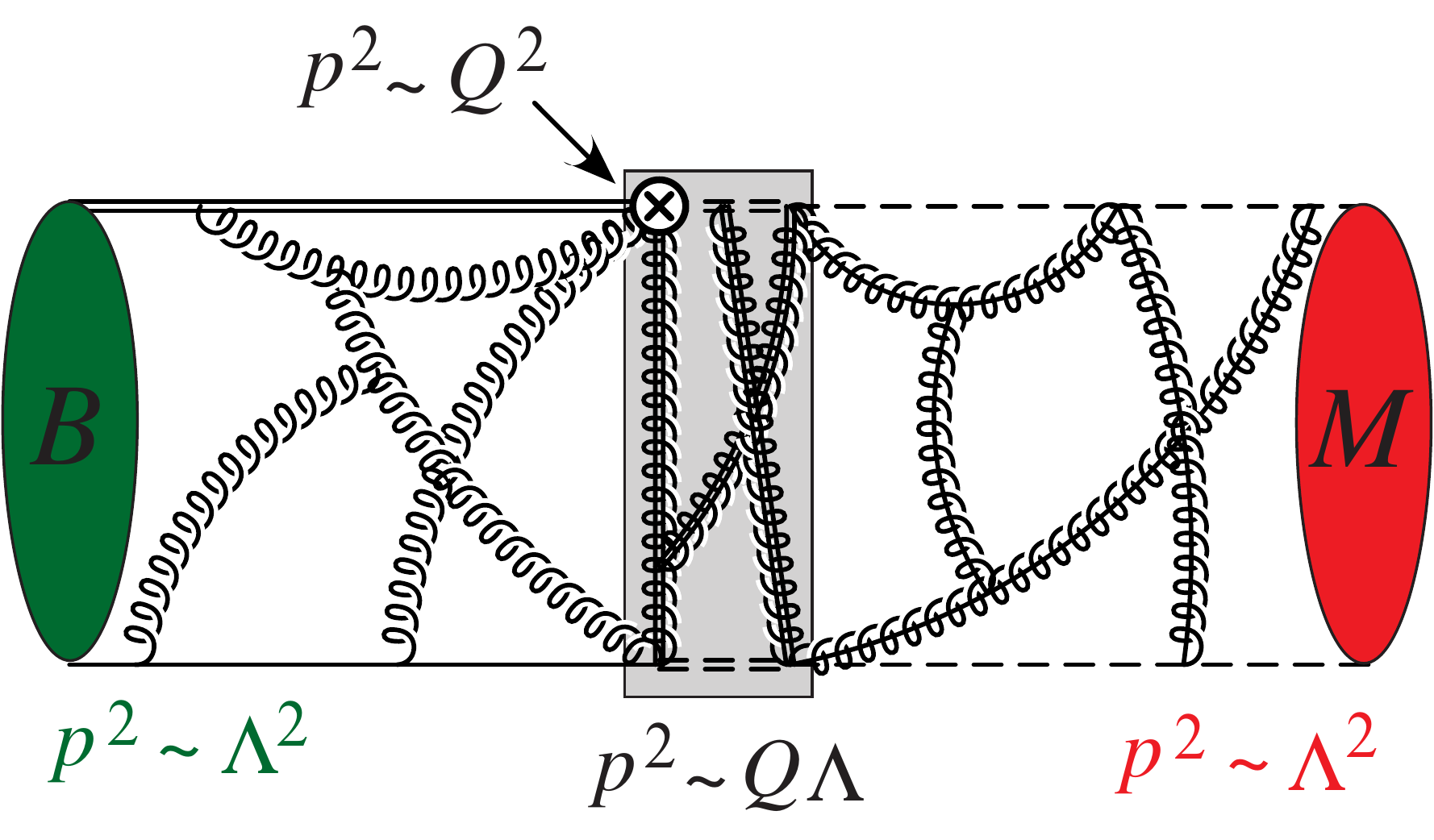}\hspace{1.0cm}
\includegraphics[width=.35\textwidth]{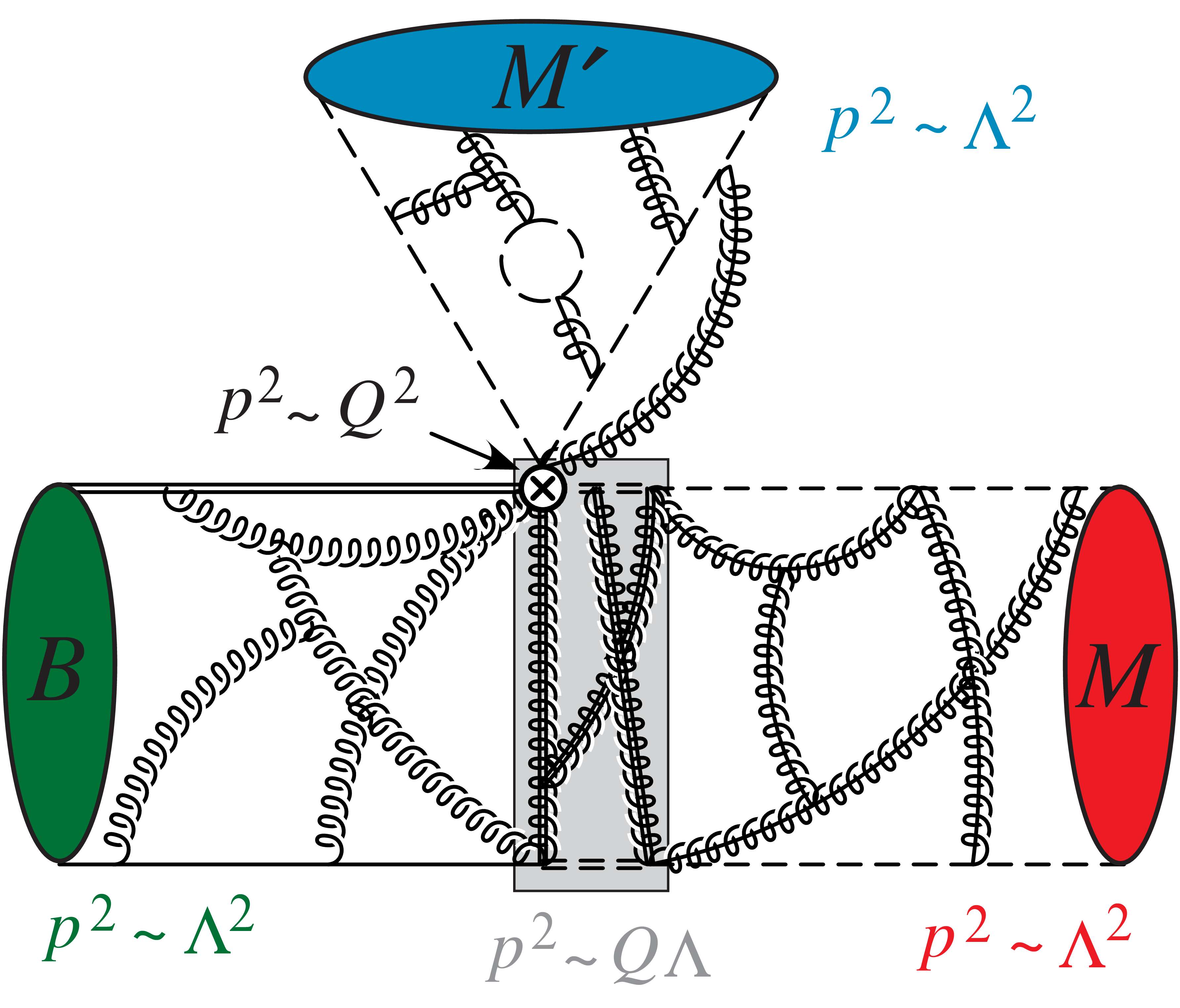}
\end{center}
\caption{Factorization of the $B\to\pi$ form factor and of the $B\to
M_1M_2$ decay in SCET.} \label{fig6}
\end{figure}

The separate matching at the two scales $m_b$ and
$\sqrt{m_b\Lambda_{\rm QCD}}$ is briefly explained below
\cite{BPS03}. The first step is to integrate out the lines off-shell
by $m_b^2$ in QCD, and the resultant effective theory is called
SCET$_{\rm I}$. One then derives the zeroth-order effective current
$J^{(0)}$ from the $b\to u$ weak vertex, and the first-order
effective current $J^{(1)}$ by shrinking the virtual $b$ quark line
in Fig.~\ref{fig4}(a). The next step is to integrate out the lines
off-shell by $m_b\Lambda_{\rm QCD}$ in SCET$_{\rm I}$, arriving at
SCET$_{\rm II}$. The relevant diagrams to start with are displayed
in Fig.~\ref{fig5}. Shrinking all the lines off-shell by
$m_b\Lambda_{\rm QCD}$, one derives the corresponding Wilson
coefficients, i.e., the jet functions, and the effective
four-fermion operators. Sandwiching these four-fermion operators by
the initial $B$ meson state and the final pion state leads to the
$B$ meson and pion distribution amplitudes. The $B\to\pi$ transition
form factor is then factorized as depicted in Fig.~\ref{fig6}. The
factorization of two-body hadronic $B$ meson decays is constructed
in a similar way, and the result is also shown in Fig.~\ref{fig6}.

At leading power in $1/m_b$, there is no large source of strong
phases in SCET (the annihilation contribution is parametrically
power-suppressed). To acquire strong phases, it has been argued that
$c\bar c$ (charming) penguins could give long-distance effects at
leading power \cite{BPRS04}. This contribution is nonperturbative,
so it must be parameterized as an arbitrary amplitude $A^{c\bar c}$.
Including the charming penguin, SCET has been applied as an
QCD-improved parametrization, and $A^{c\bar c}$ is determined
together with other hadronic inputs from data. It should be
mentioned that the long-distance charming-penguin contribution is
power-suppressed according to QCDF, PQCD and light-cone sum rules
\cite{KMM03}.

\subsection{Puzzles in $B$ Physics}

Before concluding, I review the long-standing puzzles in hadronic
two-body $B$ meson decays, which have not yet been fully resolved
so far. According to a naive estimate of the color-suppressed tree
amplitude, the hierarchy of the branching ratios
$B(B^0\to\pi^0\pi^0)\sim O(\lambda^2)B(B^0\to\pi^\mp\pi^\pm)$ with
the CKM parameter $\lambda\approx 0.2$ is
expected. However, the data \cite{HFAG}
\begin{eqnarray}
B(B^0\to\pi^\mp\pi^\pm)&=&(5.10\pm 0.19)\times 10^{-6}\;,\nonumber\\
B(B^0\to\pi^0\pi^0)&=&(1.91^{+0.22}_{-0.23})\times 10^{-6}\;,\label{data}
\end{eqnarray}
imply $B(B^0\to\pi^0\pi^0)\sim O(\lambda)B(B^0\to\pi^\mp\pi^\pm)$,
giving rise to the $B\to\pi\pi$ puzzle. As observed in
\cite{LM06}, the NLO corrections, despite of increasing the
color-suppressed tree amplitude significantly, are not enough to
enhance the $B^0\to\pi^0\pi^0$ branching ratio to the measured
value. A much larger color-suppressed tree amplitude, about the
same order as the color-allowed tree amplitude, must be obtained
in order to resolve the puzzle \cite{Charng2,P06}. To make sure
that the above NLO effects are reasonable, the PQCD formalism has
been applied to the $B\to\rho\rho$ decays \cite{LM06}, which also
receive the color-suppressed tree contribution. It was observed that
the NLO PQCD predictions are in agreement with the data
$B(B^0\to\rho^0\rho^0)=(0.73^{+0.27}_{-0.28})\times 10^{-6}$ \cite{HFAG}.
One concludes that it is unlikely to accommodate the measured
$B^0\to\pi^0\pi^0$ and $\rho^0\rho^0$ branching ratios
simultaneously in PQCD, and that the $B\to\pi\pi$ puzzle remains.

It has been claimed that the $B\to\pi\pi$ puzzle has been resolved in
the QCDF approach \cite{BBNS} with an input from SCET
\cite{BY05,BJ05,BJ052}: the inclusion of the NLO jet function, the
hard coefficient of SCET$_{\rm II}$, into the QCDF formula for the
color-suppressed tree amplitude gives sufficient enhancement of
the $B^0\to\pi^0\pi^0$ branching ratio, if adopting the parameter
scenario ''S4" \cite{BN}. It is necessary to investigate whether
the proposed new mechanism deteriorates the consistency of
theoretical results with other data. The formalism in \cite{BY05}
has been extended to the $B\to\rho\rho$ decays as a check
\cite{LM06}. It was found that the NLO jet function overshoots the
observed $B^0\to\rho^0\rho^0$ branching ratio very much as
adopting ''S4". That is, it is also unlikely to accommodate the
$B\to\pi\pi$ and $\rho\rho$ data simultaneously in QCDF.

\begin{table}[ht]
\begin{center}
\caption{Polarization fractions in the penguin-dominated $B\to VV$
decays. }
\begin{tabular}{c|c|c}\hline\hline
Mode&BABAR&Belle \\
\hline
$\phi K^{*+}$&$0.49\pm 0.05\pm 0.03 $&$0.52\pm 0.08\pm 0.03$ \\
\hline$K^{*+}\rho^0$&$0.78\pm 0.12\pm 0.03$& \\
\hline$K^{*0}\rho^+$&$0.52\pm 0.10\pm 0.04$&$0.43\pm 0.11^{+0.05}_{-0.02}$ \\
\hline$K^{*+}K^{*0}$&$0.75^{+0.16}_{-0.26}\pm 0.03$& \\
\hline\hline
\end{tabular}
\label{tab:tab1}
\end{center}
\end{table}

For penguin-dominated $B\to VV$ decays, such as those listed in
Table~\ref{tab:tab1} \cite{HFAG}, the polarization fractions
deviate from the naive counting rules based on kinematics
\cite{CKL2}. This is the so-called the $B\to\phi K^*$ puzzle. Many
attempts to resolve the $B\to \phi K^*$ polarizations have been
made \cite{LM04}, which include new physics
\cite{G03,YWL,KDY,CG0504,HKW}, the annihilation contribution
\cite{AK,BRY06} in the QCDF approach, the charming penguin in SCET
\cite{BPRS}, the rescattering effect \cite{CDP,LLNS,CCS}, and the
$b\to sg$ (the magnetic penguin) \cite{HN} and $b\to s\gamma$
\cite{BRY} transitions. The annihilation contribution from the
scalar penguin operators improves the consistency with the data,
because it is of the same order for all the three final helicity
states, and could enhance the transverse polarization fractions
\cite{CKL2}. However, the PQCD analysis of the scalar penguin
annihilation amplitudes indicates that the $B\to\phi K^*$ puzzle
cannot be resolved completely \cite{LM04}. A reduction of the
$B\to K^*$ form factor $A_0$, which is associated with the
longitudinal polarization, further helps accommodating the data
\cite{L04}.  

The penguin-dominated $B\to K^*\rho$ decays are expected to
exhibit similar polarization fractions. This is the reason why the
longitudinal polarization fraction in the $B^+\to K^{*0}\rho^+$
decay, which contains only the penguin contribution, is close to
$f_L(\phi K^*)\sim 0.5$ as listed in Table~\ref{tab:tab1}. Another
mode $B^+\to K^{*+}\rho^0$, nevertheless, exhibits a large
longitudinal polarization fraction around 0.8. This mode
involves tree amplitudes, which are subdominant, and should not
cause a significant deviation from $f_L\sim 0.5$. Though the data
of $f_L(K^{*0}\rho^+)$ from BABAR still suffer a large error,
the different longitudinal polarization
fractions, $f_L(K^{*+}\rho^0)\not=f_L(K^{*0}\rho^+)$, call for a
deeper understanding. The $B^+\to K^{*+}K^{*0}$ decay shows a longitudinal
polarization fraction smaller than unity, but larger than 0.5.
A more thorough study of the $B\to K^*K^*$ decays can help discriminating
the various resolutions for the $B\to\phi K^*$ puzzle \cite{L04,DGLN}.

The $B^0\to K^\pm\pi^\mp$ decays depend on the tree amplitude $T$
and the QCD penguin amplitude $P$. The data of the direct CP
asymmetry $A_{CP}(B^0\to K^\pm\pi^\mp)\approx -10\%$ then imply a
sizable relative strong phase between $T$ and $P$, which
verifies the LO PQCD prediction made years ago \cite{KLS}:
the scalar penguin annihilation provides an important source
of strong phases. The PQCD predictions for significant penguin
annihilation have been confirmed by the recent measurement of the pure
annihilation mode, $B(B_s\to\pi^+\pi^-)=(0.73\pm 0.14)\times 10^{-6}$, which is
consistent with $0.57\times 10^{-6}$ obtained in the LO PQCD approach
\cite{Ali07}. The $B^\pm\to K^\pm\pi^0$ decays contain the additional
color-suppressed tree amplitude $C$ and electroweak penguin
amplitude $P_{ew}$. Since both $C$ and $P_{ew}$ are
subdominant, the approximate equality for the direct CP
asymmetries $A_{CP}(B^\pm\to K^\pm\pi^0)\approx A_{CP}(B^0\to
K^\pm\pi^\mp)$ is expected. However, this naive expectation is in
conflict with the data \cite{HFAG},
\begin{eqnarray}
A_{CP}(B^0\to K^\pm\pi^\mp)&=&-0.086\pm 0.007\nonumber\\
A_{CP}(B^\pm\to K^\pm\pi^0)&=&0.040\pm 0.021,
\end{eqnarray}
making the $B\to K\pi$ puzzle.

While LO PQCD gives a negligible $C$ \cite{KLS,LUY}, it is
possible that this supposedly tiny amplitude receives a
significant subleading correction. Note that the small $C$ is
attributed to the accidental cancellation between the Wilson
coefficients $C_1$ and $C_2/N_c$ at the scale of $m_b$.
In \cite{LMS05} the important NLO contributions to the
$B\to K\pi$ decays from the vertex corrections, the quark loops,
and the magnetic penguins were calculated. It was observed that
the vertex corrections increase $C$ by a factor of 3, and induce
a large phase about $-80^o$ relative to $T$. The large and
imaginary $C$ renders the total tree amplitude $T+C$ more
or less parallel to the total penguin amplitude $P+P_{ew}$ in
the $B^\pm\to K^\pm\pi^0$ decays, leading to nearly vanishing
$A_{CP}(B^\pm\to K^\pm\pi^0)=(-1^{+3}_{-6})\%$ at NLO (it is about
-8\% at LO). One concludes that the $B\to K\pi$ puzzle has been
alleviated, but not yet gone away completely. Whether new physics
effects \cite{BFRS,BL07} are needed will become clear when the
data get precise. More detailed discussion on this subject can be
found in \cite{G07}.

\section{Summary}

Despite of nonperturbative nature of QCD, theoretical frameworks with
predictive power can be developed. They are based on the factorization
theorems, in which nonperturbative dynamics is absorbed into PDFs, and
the remaining infrared finite contributions go to hard kernels.
A PDF is universal (process-independent) and can be extracted from data,
while a hard kernel is calculable in perturbation theory.
Both the collinear and $k_T$ factorization theorems are the fundamental
tools of pQCD. The collinear factorization theorem is a simpler version,
and has been intensively studied and widely applied.
The $k_T$ factorization theorem is more complicated, and many of its
aspects have not been completely explored.

Sophisticated evolution equations and resummation techniques have been
developed in pQCD, which enhance predictive power, and increase 
theoretical precision. All the known single- and
double-logarithm summations, including their unifications, have been
explained in the CSS resummation formalism. The point is the treatment of
real gluon emissions under different kinematic orderings, and the resultant
logarithmic summations are summarized in Table~\ref{T2}.
The $k_T$ and threshold resummations, and the DGLAP and BFKL equations
have been applied to various QCD processes.

\begin{table}[ht]
\begin{center}
\caption{Single- and double-logarithmic summations under different kinematic
orderings.}
\begin{tabular}{ccccccc}
\hline\hline
              &$|$& small $x$ &$|$& intermediate $x$ &$|$& large $x$  \\
\hline
rapidity ordering &$|$& BFKL equation&$|$& $k_T$ resummation &$|$&         \\
\hline
$k_T$ ordering &$|$&   &$|$& DGLAP equation &$|$& threshold resummation \\
\hline
angular ordering    &$|$& CCFM      & & equation; \hspace{1.0cm}joint & &  resummation \\
\hline\hline\label{T2}
\end{tabular}
\end{center}
\end{table}

Experimental and theoretical studies of jet physics have been reviewed.
Especially, it was pointed out that jet substructures could be calculated
in pQCD: starting with the jet function definition, applying the factorization
theorem and the resummation technique, one can predict observables, which are
consistent with data. Because fixed-order calculations are not reliable at
small jet invariant mass, and event generators have ambiguities, pQCD provides an
alternative approach, that resolves the above difficulties. The pQCD formalism
will improve the jet identification and new particle search at the LHC.

We have been able to go beyond the factorization assumption fr hadronic
two-body heavy-quark decays by including QCD corrections. Different
approaches have been discussed and commented: in QCDF the high-power corrections must
be parameterized due to the existence of the endpoint singularities. There
are no endpoint singularities in PQCD, which is based on the $k_T$
factorization theorem, and in SCET, which employs the zero-bin subtraction
\cite{MS06}. A major difference arises from the treatment of the annihilation
contribution, which is parameterized in QCDF and neglected in SCET, but
is the main source of strong phases in PQCD.

Many subtle subjects on pQCD deserve more exploration, including the legitimate
definition of TMDs, the gauge invariance of the $k_T$ factorization,
resummations of other types of logarithms, such as rapidity logarithms,
non-global logarithms, and etc., jet substructures of boosted heavy particles,
and the long-standing puzzles in $B$ physics. pQCD remains as one of the most
challenging research fields in high-energy physics.

\section*{Acknowledgment}
This work was supported in part by the National Science Council of
R.O.C. under Grant No. NSC-101-2112-M-001-006-MY3, and by the National
Center for Theoretical Sciences of R.O.C.. The author acknowledges the
hospitality of the organizers during the First Asia-Europe-Pacific
School of High-energy Physics at Fukuoka, Japan in Oct., 2012.

\end{document}